\documentclass[12pt,preprint]{aastex}

\usepackage{epsfig}
\usepackage{color}
\usepackage{lscape}

\usepackage{color,epsfig}
\definecolor{brown}{rgb}{0.6,0.4,0.2}
\definecolor{purple}{rgb}{0.5,0,0.5}

\newcommand{\spitzer}{\textit{Spitzer}}

\newcommand{\mic}{$\mu$m}
\def\us{\char`\_}

\def\gteleven{G311.5-0.3}

\def\gtfe{G348.5-0.0}

\shorttitle{X-ray Properties of {\it Spitzer}-Detected Supernova Remnants}
\shortauthors{Pannuti et al.}

\begin{document}
\title{The X-ray Properties of Five Galactic Supernova Remnants Detected by the
{\it Spitzer} GLIMPSE Survey} 

\author{Thomas G. Pannuti}
\affil{Space Science Center, Department of Earth and Space Sciences, Morehead
State University, 235 Martindale Drive, Morehead, KY 40351}
\email{t.pannuti@moreheadstate.edu}

\author{Jeonghee Rho}
\affil{SETI Institute and SOFIA Science Center, NASA Ames Research Center, MS 211-3, 
Mountain View, CA 94035}
\email{jrho@sofia.usra.edu}

\author{Craig O. Heinke}
\affil{Department of Physics, CCIS 4-183, University of Alberta, Edmonton, AB T6G 2E1, Canada}
\email{heinke@ualberta.ca}

\and

\author{William P. Moffitt}
\affil{Space Science Center, Department of Earth and Space Sciences, Morehead State University,
235 Martindale Drive, Morehead, KY 40351}
\email{w.moffitt@moreheadstate.edu} 

\begin{abstract}
We present a study of the X-ray properties of five Galactic supernova remnants (SNRs)
--  Kes 17 (G304.6$+$0.1), G311.5$-$0.3, G346.6$-$0.2, CTB 37A (G348.5$+$0.1) 
and G348.5$-$0.0 -- that were detected in the infrared by \citet{Reach06} in an analysis of data from
the Galactic Legacy Infrared Mid-Plane Survey Extraordinaire (GLIMPSE) that was conducted by the 
{\it Spitzer} Space Telescope. We present and analyze archival {\it ASCA} observations of Kes 17, 
G311.5$-$0.3 and G346.6$-$0.2, archival {\it XMM-Newton} observations of Kes 17, CTB 37A and 
G348.5$-$0.0 and an archival {\it Chandra} observation of CTB 37A. All of the SNRs are clearly 
detected in the X-ray possibly except for G348.5$-$0.0. Our study reveals that the four detected 
SNRs all feature center-filled X-ray morphologies and that the observed emission from these sources 
is thermal in all cases. We argue that these SNRs should be classified as mixed-morphology SNRs
(MM SNRs): our study strengthens the
correlation between MM SNRs and SNRs interacting with molecular clouds
and suggests that the origin of mixed-morphology SNRs may be due to the interactions between these
SNRs and adjacent clouds. Our {\it ASCA} analysis of G311.5$-$0.3 reveals for the first time X-ray 
emission from this SNR: the X-ray emission is center-filled within the radio and infrared shells and 
thermal in nature ($kT$ $\sim$ 0.98 keV), thus motivating its classification as an MM SNR. 
We find considerable spectral variations in the properties associated with the plasmas of the other 
X-ray-detected SNRs, such as a possible overabundance of magnesium in the plasma of Kes 17. 
Our new results
also include the first detailed spatially-resolved spectroscopic study of CTB 37A using {\it Chandra} 
as well as a spectroscopic study of the discrete X-ray source CXOU J171428.5$-$383601, which may 
be a neutron star associated with CTB 37A.  Finally, we also estimate 
such properties as electron density $n$$_e$, radiative age $t$$_{rad}$ and swept-up mass 
$M$$_X$ for each of the four X-ray-detected SNRs. Each of these values are comparable
to archetypal mixed-morphology SNRs like 3C 391 and W44. In an analysis of the spectrum of
Kes 17, we did not find 
evidence of over-ionization unlike other archetypal mixed-morphology SNRs like W28 and W44.
\end{abstract}

\keywords{ISM: individual objects (SNR 304.6+0.1, SNR 311.5-0.3, SNR 346.6-0.2, SNR 348.5+0.1,
SNR 348.5-0.0) -- ISM: supernova remnants -- X-rays: ISM}

\section{Introduction}

\citet{Reach06} presented a search for infrared counterparts
to Galactic supernova remnants (SNRs) that were sampled by the Galactic Legacy Infrared
Mid-Plane Survey Extraordinaire (GLIMPSE) Legacy survey \citep{Benjamin03}. The 
GLIMPSE survey imaged the inner Galactic plane -- specifically the Galactic latitude and
longitude ranges of $|$$b$$|$$<$1$^{\circ}$ and 10$^{\circ}$$<|$$l$$|<$65$^{\circ}$, respectively --
at four wavelength bands (3.6, 4.5, 5.8 and 8.0 $\mu$m) with the Infrared Array Camera (IRAC)
\citep{Fazio04} aboard the {\it Spitzer Space Telescope} \citep{Werner04}. The results presented 
by \citet{Reach06} were very intriguing: of the 95 SNRs known (at the time) that were sampled by 
GLIMPSE, eighteen SNRs were detected (many for the first time) and the infrared emission from these 
sources typically coincided (at least partially) with known radio structures. \citet{Reach06} also
presented simple models to generate template colors in the IRAC bands for the primary 
emission mechanisms expected from SNRs. These primary emission mechanisms include the
following: (i) {\it molecular} line colors that were estimated using a three temperature component 
H$_2$ excitation model based on numerous H$_2$ lines detected toward IC 443 \citep{Rho01}, 
(ii) {\it ionized} gas line colors that were estimated based on hydrogen recombination as well as
the dominant fine-structure lines of iron and argon (assuming solar abundances) and (iii) 
interstellar medium (ISM) polycyclic aromatic hydrocarbon (PAH) colors that were estimated 
using a spectrum of the reflection nebula NGC 7023.
In their analysis, \citet{Reach06} found that the spectra of nine SNRs, three SNRs and four SNRs
were dominated by molecular lines, fine-structure lines and PAHs, respectively. The spectra of
the remaining SNRs indicated a mixture of ionic and molecular shocks, suggesting a combination
of shock types. 
\par
Interestingly, many of the SNRs detected with IRAC also appear to be interacting 
with relatively dense gas. Evidence for such interactions is the excess emission detected from these
SNRs at a wavelength of 4.5 $\mu$m: such an excess is likely produced by H$_2$ and CO line 
emission and such line emission is associated with molecular shocks. 
Furthermore, follow-up spectroscopy observations conducted with the Infrared Spectrograph (IRS)
aboard \spitzer\ \citep{Houck04} revealed H$_2$ emission from all of the SNRs with IRAC by the
survey conducted by \citet{Reach06}, suggesting
that each of these SNRs are interacting with molecular clouds \citep{Hewitt09, Andersen11}.
We shall refer to these SNRs as molecular SNRs for the remainder of the paper.
\par
Several molecular SNRs are detected in the X-ray (see \citet{Vink12} for a review of X-ray
emission from SNRs) and all of them belong to the class of sources known as mixed-morphology 
SNRs (MM SNRs). The distinguishing characteristics of MM SNRs are a center-filled thermal X-ray
morphology paired with a well-defined radio shell \citep{Rho98}.  Examples of well-known MM SNRs
include W28, W44 and 3C 391: each of these SNRs was observed in the infrared with the Long 
Wavelength Spectrometer \citep[LWS;][]{Clegg96} aboard the {\it Infrared Space Observatory} 
(ISO) \citep{Kessler96} and H$_2$ emission was detected from 
all three SNRs \citep{Reach00}. Many of the molecular SNRs detected by \citet{Reach06} have not 
been well-studied in the X-ray, probably because the Galactic column densities toward these sources 
are elevated ($N$$_H$ $\sim$ 10$^{22}$ cm$^{-2}$), causing them to appear to be faint in the
X-ray. In this paper
we describe the X-ray properties of five molecular SNRs detected by \citet{Reach06}: Kes 17
(G304.6$+$0.1), G311.5$-$0.3, G346.6$-$0.2, CTB 37A (G348.5$+$0.1) and G348.5$-$0.0. For this  
study, we use data from archival observations made with $\it{ASCA}$, $\it{Chandra}$ and 
$\it{XMM-Newton}$.  These SNRs had been studied almost exclusively in the radio before the
infrared observations made with {\it Spitzer}: they have not been detected in the optical most likely
because of the high extinction toward these sources. We note that three of these SNRs (Kes 17,
G346.6$-$0.2 and CTB 37A) have been the subjects of pointed {\it Chandra}, {\it XMM-Newton} and 
{\it Suzaku} observations that have been described and analyzed in separate publications 
\citep{Aharonian08,Combi10,Sezer11a,Sezer11b,Gok12,Yamauchi13}. Serendipitous observations
made of G346.6$-$0.2, CTB 37A and G348.5$-$0.0 with {\it ASCA} have also been described by 
\citet{Yamauchi08}. In this paper,
we present new analyses of the {\it ASCA} observations of Kes 17 and G311.5$-$0.3 (including
the first reported detection of X-ray emission from this SNR) and a spatially-resolved spectroscopic
analysis of the {\it Chandra} observation of CTB 37A. We will compare the results
of our analyses with the results for all of the SNRs with the previously-published papers mentioned 
above. 
\par
The organization of this paper may be described as follows: the observations as conducted 
by each observatory and the subsequent data analyses are described in Section \ref{ObsSection}. 
In Section \ref{PropsXrayImageSpecSection} we present a description of the 
five SNRs in our sample and the X-ray results -- both imaging 
and spectroscopic -- for
each SNR. A summary of the plasma conditions of these sources as inferred from these 
X-ray observations is presented in Section \ref{PlasmaSection}. In Section
\ref{OverionizationSection} we discuss the phenomenon of overionization in mixed-morphology SNRs 
and comment on its applicability to the SNRs in our sample. In Section \ref{ComparisonSection} we 
compare the properties of the SNRs in our sample and discuss the implications for SNR evolution.
Finally, the conclusions reached by this work are presented in Section \ref{ConclusionSection}.
\par
In Table \ref{SNRPropsTable} we list general properties -- including RA (J2000.0),
Dec (J2000.0), angular diameter, distance, physical size, column density $N$$_H$ and 
optical extinction $A$$_V$ -- of the five SNRs considered
in the present paper. 

\section{Observations and Data Reduction}
\label{ObsSection}

\subsection{X-ray Observations}

\subsubsection{{\it ASCA} Observations and Data Reduction}

All five of the SNRs studied in the present paper were
sampled (either directly by pointed observations or indirectly by pointed observations
that targeted other sources but included the SNR within the field of view) by the 
Advanced Satellite for Cosmology and Astrophysics ({\it ASCA} -- \citet{Tanaka94}). 
The instruments carried by {\it ASCA} included two Solid-State Imaging Spectrometers
denoted as SIS0 and SIS1 \citep{Burke91} and two Gas Imaging Spectrometers
denoted as GIS2 and GIS3 \citep{Makishima96, Ohashi96}: these instruments were
located at the focal planes of four thin-foil X-ray telescopes (XRT -- \citet{Serlemitsos95}).
A single GIS unit sampled a field of view $\sim$50$\arcmin$ in diameter while a single
SIS unit sampled a field of view approximately 44$\arcmin$$\times$44$\arcmin$ in size.
The nominal full-width at half-maximum (FWHM) angular resolution of both the
GIS and the SIS at 1 keV was approximately 1 arcminute.

\par

Data for all of the {\it ASCA} observations considered in the present paper were
extracted from the archive located at the High Energy Astrophysics
Science Archive Research Center (HEASARC\footnote{HEASARC is a
service of the Laboratory for High Energy Astrophysics (LHEA) at 
the National Aeronautics and Space Administration Goddard Space
Flight Center (NASA/GSFC) and the High Energy Astrophysics
Division of the Smithsonian Astrophysical Observatory (SAO).
For more information on HEASARC, please see
http://heasarc.gsfc.nasa.gov.}) and reduced using standard tools in the FTOOLS
package\footnote{http://heasarc.gsfc.nasa.gov/docs/software/floods/ftools\_menu.html}.
However, only the observations of the Kes 17, G311.5$-$0.3 and G346.6$-$0.2 yielded
enough counts from the particular SNRs for detailed spectral fitting. In the cases of the
SNRs CTB 37A and G348.5$-$0.0 which appear adjacent to each other in the sky, these 
particular sources were located too close to the edges of the GIS fields of view to extract useful 
spectra. In addition, as noted by \citet{Yamauchi08}, contaminating stray flux from bright X-ray
sources located out of the field of view and south of these two SNRs made it difficult to isolate 
their X-ray emission (see Section \ref{XMMG348Section} for more
details, particularly about our attempt to extract X-ray spectra for G348.5$-$0.0 from the
{\it XMM-Newton} observation, which also contained contaminating stray X-ray flux).  
The angular extents of the three remaining SNRs for which we extracted {\it ASCA} spectra
(that is, Kes 17, G311.5$-$0.3 and G346.6$-$0.2) were compact enough to fit within the fields 
of view of the GIS2 and the GIS3: in the case of Kes 17, the SNR was also compact enough 
to fit entirely within the fields of view of the SIS0 and the SIS1. Therefore, we also extracted SIS
spectra of this SNR as well. Unfortunately the signal-to-noise of the extracted SIS1 source 
spectrum of Kes 17 was too poor for any detailed spectral fitting, so we do not consider it further 
in this paper.  The data reduction
was accomplished using another program available from the HEASARC site, 
{\tt xselect} (Version 2.4b): standard REV2 screening was applied to both the GIS and
SIS datasets. We used {\tt xselect} to extract GIS2 and GIS3 spectra for Kes 17, G311.5$-$0.3
and G346.6$-$0.2: in the case of Kes 17, we also used {\tt xselect} to extract an SIS0 spectrum
as well. For spectral analysis, the standard GIS2 and GIS3 response matrix files (RMFs)
were used while the FTOOL {\tt sisrmg} was used to generate an RMF for the SIS0 spectrum of
Kes 17. Lastly, the FTOOL {\tt ascaarf} was used to create ancillary response files (ARFs)
for each extracted GIS and SIS spectra. In Table \ref{ASCAObsSummaryTable}, we present
a summary of the {\it ASCA} observations made of the SNRs that we consider in this paper.

\subsubsection{{\it XMM-Newton} Observations and Data 
Reduction\label{XMMSubsection}}

While all five of the SNRs considered in the present paper were
sampled (either directly by pointed observations or indirectly by pointed observations
that targeted other sources but included the SNR within the field of view) by the 
{\it XMM-Newton} Observatory \citep{Jansen01}, in the cases of two SNRs -- G311.5$-$0.3 and
G346.6$-$0.2 -- the source was located so far from the centers of the fields of view
and the exposure times of the observations were so short that no useful spectral data
could be extracted for these SNRs by these observations. We therefore only consider
the observations made by {\it XMM-Newton} of the other three SNRs in our study, 
namely Kes 17, CTB 37A and G348.5$-$0.0. The main science instruments aboard
{\it XMM-Newton} are the pn-CCD camera \citep{Struder01} and the Multi-Object
Spectrometer cameras \citep{Turner01}: we will refer to these instruments as the PN,
MOS1 and MOS2 for the remainder of the paper and note that both of these instruments
comprise the European Photon Imaging Camera (EPIC). The fields of view of the PN,
MOS1 and MOS2 cameras are approximately 30 arcminutes and the nominal angular
FWHM at 1 keV of all three cameras is $\sim$6$\arcsec$.

\par

Data for the observations was downloaded from the on-line {\it XMM-Newton}
data archive\footnote{http://xmm.esac.esa.int/xsa/}. The observations were reduced 
using standard tools in the Science Analysis System (SAS -- \citet{Gabriel04}) software 
package Version 11.0.0. The tool {\tt emchain} was used to process data from the observations
made by the MOS1 and the MOS2 cameras and the tool {\tt mos-filter} was used to filter out
background flares during the observation and to identify good time intervals. Similarly, the tool 
{\tt epchain} was used to process data from the observations made by the PN camera and the
tool {\tt pn-filter} was used to identify background flares during the observation and to identify
good time intervals. To extract MOS1, MOS2 and PN spectra 
for both source and background regions, the tool {\tt evselect} was used and the extracted
spectra were grouped to a minimum of 15 counts per channel. Subsequently, the 
tool {\tt backscale} was run to compute the proper value for the BACKSCAL keyword in the
extracted spectra. Finally, the tools {\tt rmfgen} and {\tt arfgen} were used to generate the
RMFs and ARFs (respectively) for the spectra which are necessary for spectral fitting.
In Table \ref{XMMObsSummaryTable}, we present a summary of the {\it XMM-Newton} observations
made of the SNRs that we consider in this paper. 

\subsubsection{{\it Chandra} Observations and Data 
Reduction\label{ChandraSubsection}} 

One of the SNRs studied in the present paper  -- CTB 37A -- was the target of
a pointed observation made with the {\it Chandra} X-ray Observatory 
\citep{Weisskopf02}: the corresponding ObsID of this observation is 6721
and data for this observation was downloaded from the
on-line {\it Chandra} data archive\footnote{http://cda.haravrd.edu/chaser/.}. The observation 
of CTB 37A was conducted in VERY FAINT Mode on 2006 October 7
with the Advanced CCD Imaging Spectrometer (ACIS) at a focal plane temperature
of $-$120$^{\circ}$ C. The entire angular extent of the X-ray emitting plasma associated
with SNR fit within the field of view of the ACIS-I array and the geometric center of
this angular extent aligned roughly with the aimpoint of the observation. 
Properties of the ACIS chips
(including the ACIS-I array) have been described by \citet{Garmire03}: 
to summarize, the ACIS-I array is composed of four front-illuminated CCD chips arranged
in a 2$\times$2 array. Each chip has a field of view of 8.3$^{\arcmin}$ $\times$ 8.3$^{\arcmin}$:
thus, the entire array has a field of view of 17$^{\arcmin}$ $\times$ 17$^{\arcmin}$.
The chips are sensitive to photons with energies ranging from 0.4 to 10.0 keV and the 
full-width at half-maximum (FWHM) angular resolution of each chip at 1 keV is 
1$^{\arcsec}$. The maximum effective collecting area of each chip is 525 cm$^{2}$: this
maximum corresponds to photons with energies of 1.5  keV. Finally, the spectral resolution
of each chip at 1 keV is 56. 
\par
The dataset from this observation was  
reduced using the {\it Chandra} Interactive Analysis of Observations 
(CIAO\footnote{See \citet{Fruscione06} and http://cxc.harvard.edu/ciao/.})
software package Version 4.3 (CALDB Version 4.4.1). Standard processing was applied
to this dataset: to generate a new event file, the CIAO tool ${\tt acis\us process\us events}$ was
used. This tool applies the latest temperature-dependent charge transfer
inefficiency (CTI) correction, the latest time-dependent gain adjustment as well as the
latest gain map: ${\tt acis\us process\us events}$ also flags bad pixels and applies background 
cleaning for datasets created in VERY FAINT mode, which is the case of the observation
considered here. The new event file was further filtered for bad grades (that is, events
which had GRADE values of 1, 5 or 7 were excluded) and proper status columns:
in addition, the good time interval supplied by the pipeline was also applied. Finally,
the event file was also screened for background flare activity using the CIAO tool {\tt chips}
\footnote{http://cxc.harvard.edu/chips/index.html.}: background flares did
not affect the observation at a significant level. The resulting total effective exposure time
of the observation was 19.7 ks: in Table \ref{ChandraObsTable} we present a summary of the
parameters of this observation.
\par
The CIAO tool {\tt specextract} was used to extract spectra from different
regions of the X-ray emitting plasma associated with CTB 37A (along with selected background
regions located outside of the X-ray emitting plasma). In addition to extracting spectra for
source and background regions, {\tt specextract} also creates weighted ARFs and RMFs
by initiating the CIAO tools {\tt mkwarf}
and {\tt mkrmf}, respectively. Finally, {\tt specextract} also initiates the CIAO tool {\tt dmgroup}
to group the source and background spectra: in the case of our spectral analysis, the
spectra were all grouped to a minimum of 15 counts per channel. 
We used the CIAO wavelet detection routine $\tt{wavdetect}$ \citep{Freeman02} to identify
unrelated discrete X-ray sources detected along the line of sight toward CTB 37A by the
{\it Chandra} observation: such sources may contribute undesired contaminating flux that
interferes with spectral analyses of the SNR itself. When regions of spectral extraction
were defined, care was taken to exclude flux from the positions of these detected sources
so that the analyzed spectra more faithfully represented the true emission from the SNR. 
The locations of these source and background regions as well as the 
analysis of these spectra are presented in Section \ref{CTB37AXraySubsection}.

\subsection{Infrared Observations\label{InfraredSubSection}}
We used \spitzer\  IRAC and MIPS images from \citet{Reach06} and \citet{Andersen11},
respectively, of the five SNRs considered in this paper to compare the infrared morphologies
of these sources with their X-ray and radio morphologies. For details on the IRAC and MIPS
observations of these SNRs and the subsequent data processing and analyses of the images,
the reader is referred to those papers. 

\subsection{Radio Observations\label{RadioSubSection}}

As part of our study, we also included radio maps of each SNR that were made 
at a frequency of 843 MHz by the Molonglo
Observatory Synthesis Telescope \citep[MOST\footnote{The MOST is operated by The 
University of Sydney with support from the Australian Research Council and the Science
Foundation for Physics within The University of Sydney.},][]{Mills81,Robertson91}: these 
maps have been consolidated into the MOST Supernova Remnant Catalogue 
(MSC\footnote{http://www.physics.usyd.edu.au/sifa/Main/MSC.}). We 
obtained publicly-available reduced radio maps for each SNR from the MGPS Web 
Page\footnote{http://www.astrop.physics.usyd.edu.au/MGPS/.} 
for the purposes of the present research. The MSC 
contains known southern Galactic SNRs located within the area 
245$^{\circ}$ $\leq$ $\ell$ $\leq$ 355$^{\circ}$, $|$$b$$|$ $\leq$
1.5$^{\circ}$ that were observed with the MOST during a survey conducted at the
frequency of 843 MHz of the southern Galactic plane 
\citep{Whiteoak96}. The angular resolution attained by
these observations was 43$\arcsec$ $\times$ 43$\arcsec$ csc $\delta$,
the sensitivity was approximately 1-2 mJy/beam
(1 $\sigma$) and the positional accuracy was 1-2$\arcsec$.  
For more information about MOST, MGPS and these observations the reader is
referred to the above references.

\section{Results and Discussion}
\label{PropsXrayImageSpecSection}

We now present an analysis of the X-ray properties of each SNR. All spectra were
analyzed using the {\it XSPEC} software package \citep{Arnaud96} Version 12.6.0q.

\subsection{Kes 17 (G304.6$+$0.1)}

Kes 17 (G304.6$+$0.1) was first identified as a discrete radio source by the survey 
conducted by \citet{Kesteven68a} at a frequency of 408 MHz with the Molonglo Cross
Telescope. It was later classified as a radio SNR by \citet{Kesteven68b} based on its
measured non-thermal radio spectral index ($\alpha$=-0.77\footnote{Here and for the
remainder of the paper, we will adopt the convention for the measured radio spectral index
$\alpha$ to be defined as $S$$_{\nu}$ $\propto$ $\nu$$^{\alpha}$.}). An 843 MHz map 
of this SNR presented by \citet{Whiteoak96} based on observations made with the MOST reveal 
an incomplete shell-like morphology with bright southern and northwestern rims: a gap in 
the radio shell is seen toward the northeastern boundary. \citet{Whiteoak96} measured the
angular extent of Kes 17 to be 8 arcminutes and reported an integrated flux density at 843 MHz 
for the SNR of 18 Jy.  \citet{Frail96} reported on the detection of 
hydroxyl (OH) maser emission at 1720 MHz from Kes 17 based on observations made with 
the Parkes 64-meter telescope. Based on HI absorption measurements
along the line of sight to this source, \citet{Caswell75} estimated a minimum distance to
Kes 17 of 9.7 kpc: we will adopt this distance to this SNR for the remainder of this paper.
Previously published analyses of X-ray observations of Kes 17 have been presented by 
\citet{Combi10} and \citet{Gok12} (who examined {\it XMM-Newton} and {\it Suzaku} observations,
respectively): we compare our analyses with the results from those two papers in Section
\ref{Kes17DiscussSubSection}. Finally, \citet{Wu11} reported the detection of $\gamma$-ray 
emission from Kes 17 based on observations made by the Large Area Telescope (LAT) aboard the 
$\it{Fermi}$ $\gamma$-ray Space Telescope \citep{Atwood09}.

\subsubsection{{\it ASCA} and {\it XMM-Newton} Observations of 
Kes 17 (G304.6$+$0.1)\label{Kes17XraySubSection}}

\begin{figure}
\plotone{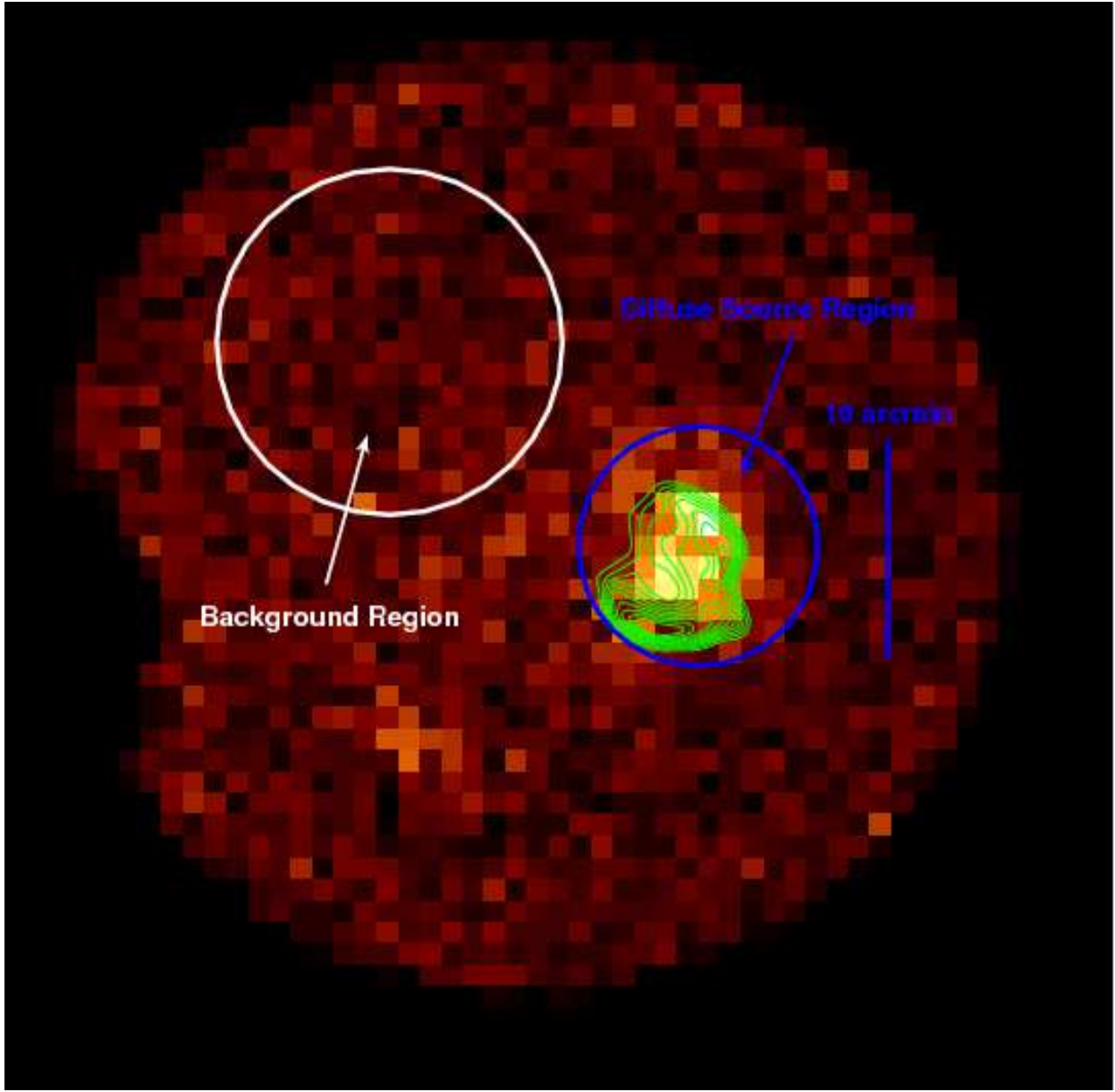}
\caption{{\it ASCA} GIS2  X-ray image of Kes 17 (in color) for the
energy range 0.7 to 10.0 keV. Radio emission (as detected with the MOST at a frequency
of 843 MHz) is overlaid on the X-ray emission in green contours. The contour levels 
correspond to 0.10, 0.15, 0.20, 0.25, 0.30, 0.35, 0.40, 0.45, 0.50 and 0.55 Jy/beam. The 
region of spectral extraction (namely the entire diffuse region) is indicated. See Section
\ref{Kes17XraySubSection}.}
\label{Kes17_ASCA_GIS2GIS3_MOST}
\end{figure}

\begin{figure}
\plotone{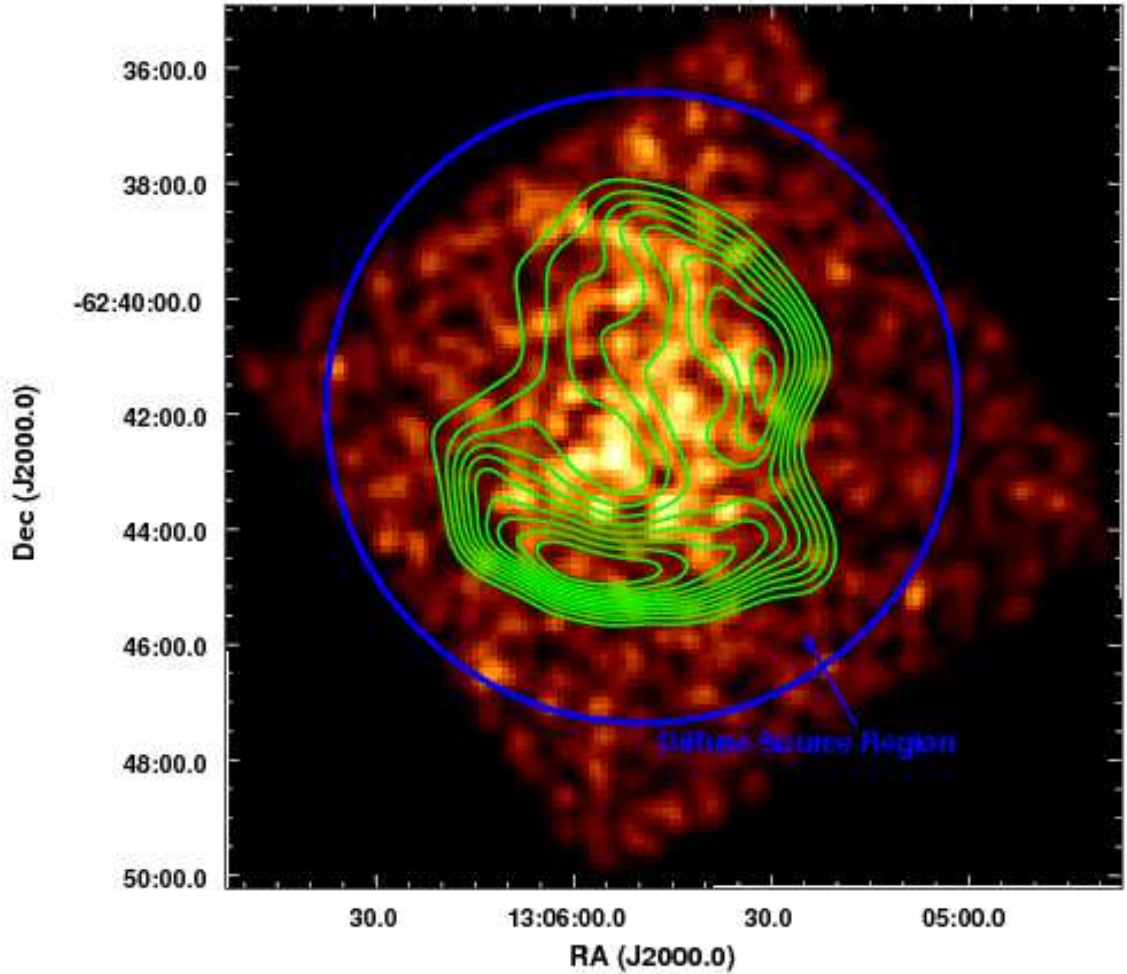}
\caption{{\it ASCA} SIS0 image of Kes 17 (in color) for the energy range 0.7 to 10.0 keV.
Radio emission (as detected with the MOST at a frequency of 843 MHz) is overlaid on the
X-ray emission in green contours: the contour levels are the same as those depicted in
Figure \ref{Kes17_ASCA_GIS2GIS3_MOST}. The region of spectral extraction for the source
spectrum (namely the entire diffuse region) is indicated. A background spectrum was extracted
from an SIS0 blank sky observation. See Section \ref{Kes17XraySubSection}.}
\label{Kes17_ASCA_SIS_MOST}
\end{figure}

In Figure \ref{Kes17_ASCA_GIS2GIS3_MOST} we present an {\it ASCA} 
GIS2 image of Kes 17 over the energy range from 0.7 to 10 keV. Kes 17
was offset by approximately seven arcminutes from the center of the field of view of the GIS
detectors when the observation was conducted.  
Radio emission (as detected by the MOST at 843 MHz) is plotted as 
contours overlaying the X-ray emission. The observed emission is broadly contained
within the radio shell but the poor angular resolution of the X-ray observation prevents
a detailed description of the morphology of the X-ray emission. In Figure 
\ref{Kes17_ASCA_SIS_MOST} we present an {\it ASCA} SIS0 image of Kes 17 (again,
the radio emission from the SNR is shown with contours): in the GIS and SIS images, it is clear
to see that the X-ray emission is contained within the radio shell and that the bright radio rims
(as seen toward the southern and western portions of the SNR)
are X-ray faint. We extracted GIS2, GIS3 and SIS0 spectra for Kes 17 using the
extraction regions indicated with blue circles in Figures \ref{Kes17_ASCA_GIS2GIS3_MOST}
and \ref{Kes17_ASCA_SIS_MOST}. For the GIS2 and GIS3, we used a source region 
approximately 5.5 arcmin radius and a background region approximately 9 arcmin in radius from a 
portion of the field of view away from the location of Kes 17 itself. For the SIS0, we 
extracted a source spectrum from a region approximately 5.5 arcmin in radius: because this
source region filled virtually the entire field of view of the SIS0 chip, we extracted a background
spectrum from a SIS0 blank sky observation. We fit the extracted source spectra
with the thermal models APEC \citep[Astrophysical Plasma Emission Code -- see][]{Foster12} and 
NEI \citep[Non-Equilibrium Ionization -- see][]{Borkowski01}: satisfactory fits were obtained using 
these models.
Because a statistically acceptable fit was obtained with the APEC model (which describes
a plasma in collisional ionization equilibrium -- CIE) and because the ionization 
timescale parameter $\tau$ derived for the NEI fit is rather high ($\tau$ $\sim$ 10$^{13}$ s 
cm$^{-3}$ with lower limits of approximately $\tau$ $\sim$ 10$^{10}$ - 10$^{11}$ s cm$^{-3}$) 
we conclude that the X-ray-emitting plasma associated with Kes 17 is in CIE. For the spectral
fits for all datasets considered in this paper, the photoelectric absorption model PHABS will
be used to account for Galactic photoelectric absorption along the lines of sight to the SNRs. 
\par
The quality of 
the fits (as measured by the $\chi$$_{\nu}^2$ statistic) were improved in both cases 
if we allowed the abundance parameter to vary: in each case, the derived abundance was
sub-solar (approximately 0.2). To investigate the possibility of variations in the abundances of
particular elements, we also fitted the extracted spectra with the thermal model VAPEC -- this
model has as free parameters the abundances of particular elements. We generated two
acceptable fits with the VAPEC model: in the first fit, we found marginal evidence for enhancement
in the abundance of magnesium (Mg=2.38$^{+4.32}_{-1.58}$ -- all quoted errors in the present paper
for fit parameters correspond to 90\% confidence levels) while the abundances of
other elements remained frozen to solar values. In the second fit, we found evidence for
subsolar abundances for silicon and sulfur (Si=0.47$^{+0.27}_{-0.19}$ and S=0.21($<$0.56))
while the abundances of the other elements were left frozen to solar values. For all of our
fits, both the column density $N$$_H$ and the temperature $kT$ parameters remained broadly
consistent (approximately $N$$_H$ $\sim$ 3-4$\times$10$^{22}$ cm$^{-2}$ and $kT$ $\sim$
0.5-0.7 keV). 
\par
\begin{figure}
\psfig{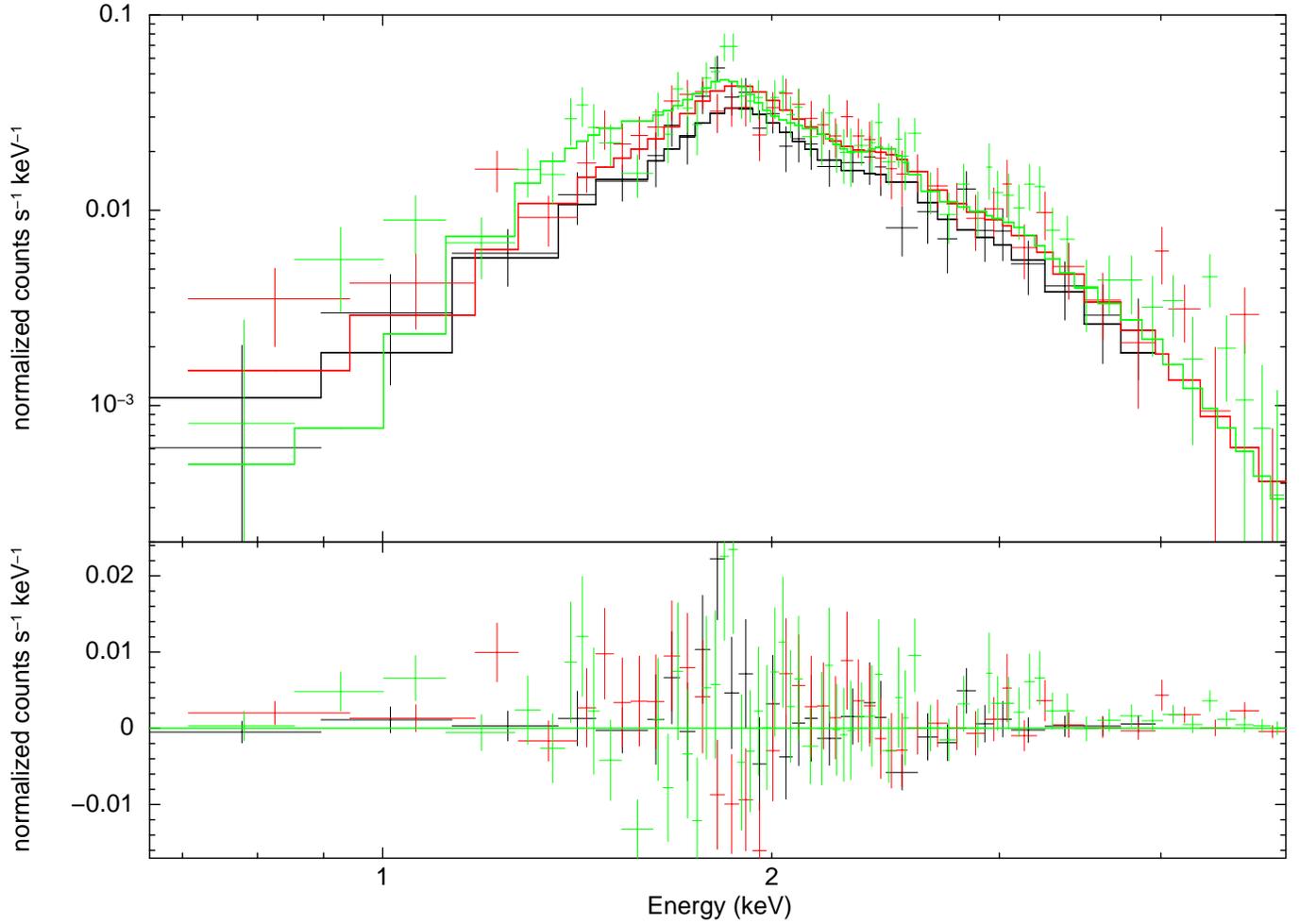}
\caption{{\it ASCA} GIS2, GIS3 and SIS0 (in black, red and green, respectively) spectra for 
Kes 17 as fit using a PHABS$\times$APEC model with variable abundance.
See Table \ref{GISSISSpectraSummaryKes17} and Section \ref{Kes17XraySubSection}.
\label{KES17ASCASpectra}}
\end{figure}
\begin{figure}
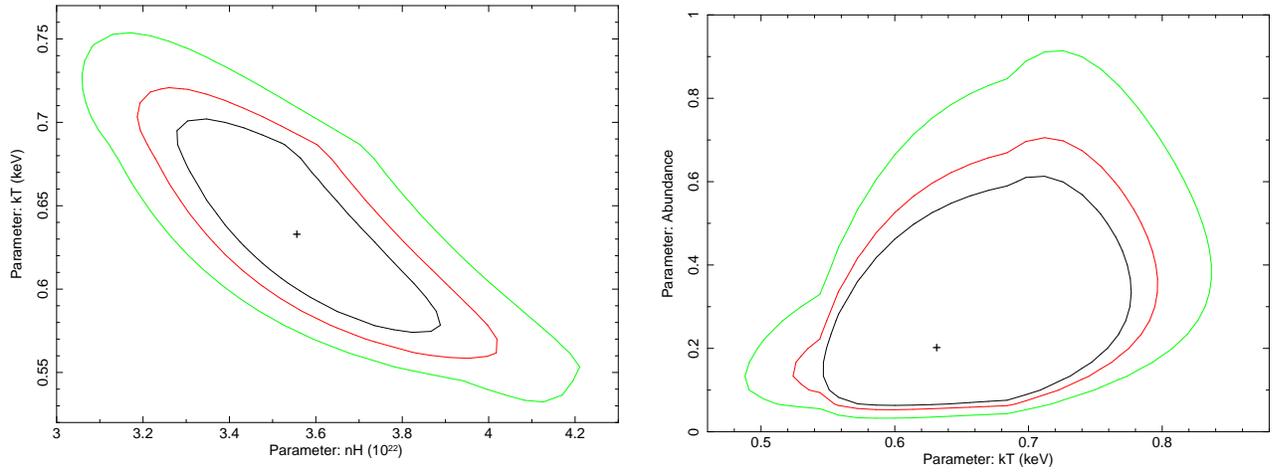

\hbox{
\psfig{figure=KES17_GIS2GIS3SIS0_PHABSAPEC_VABUND_NHKT.ps,angle=270,scale=0.35}
\psfig{figure=KES17_GIS2GIS3SIS0_PHABSAPEC_VABUND_KTABUND.ps,angle=270,scale=0.35}
}
\caption{Confidence contour plots for column density $N$$_H$ versus temperature $kT$ 
(left) and abundance versus temperature $kT$ (right) for the fit generated using the 
PHABS$\times$APEC model with variable abundances to the {\it ASCA} GIS2, GIS3 and SIS0 
spectra of Kes 17. See Figure \ref{KES17ASCASpectra}, Table \ref{GISSISSpectraSummaryKes17} 
and Section \ref{Kes17XraySubSection}.\label{KES17ASCASpectralPlots}}
\end{figure}

In Figures \ref{KES17ASCASpectra} we present the extracted GIS2, GIS3 and
SIS0 spectra of Kes 17 as fit with the PHABS$\times$APEC model with variable abundance
and in Figure \ref{KES17ASCASpectralPlots} we present confidence contour plots (namely
$N$$_H$ versus $kT$ and $kT$ versus abundance) for this fit. In Table
\ref{GISSISSpectraSummaryKes17} we present a summary of all of the spectral fits to
the extracted {\it ASCA} spectra of Kes 17. Confidence contour plots for the $N$$_H$, $kT$ and
the abundance parameters for the PHABS$\times$APEC fit are presented in Figure
\ref{KES17ASCASpectralPlots}.
\par
\begin{figure}
\plotone{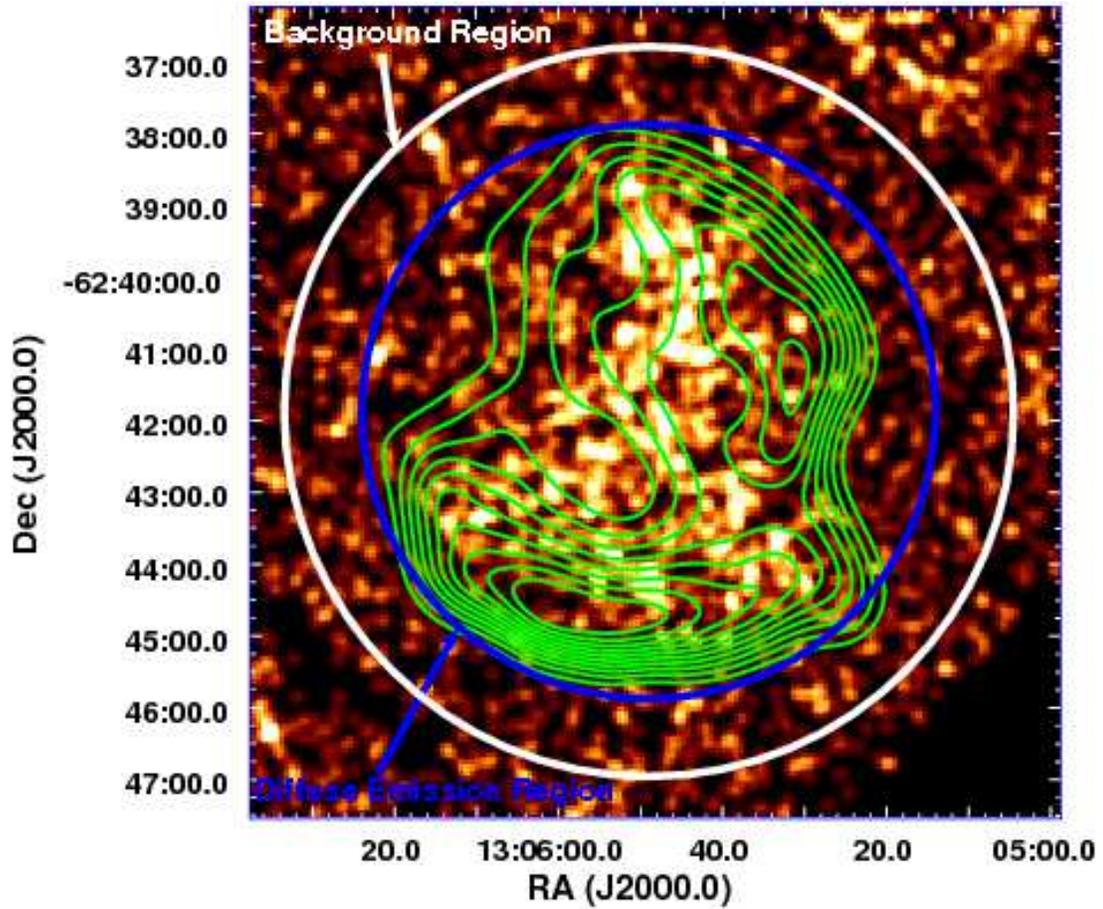}
\caption{{\it XMM-Newton} MOS1 image of Kes 17 (in color) for the energy range 0.7 to 10.0
keV. Radio emission (as detected with the MOST at a frequency of 843 MHz) is overlaid on
the X-ray emission in green contours: the contour levels are the same as those depicted in
Figure \ref{Kes17_ASCA_GIS2GIS3_MOST}. Regions of spectral extraction (namely the entire 
diffuse region and the background region) are indicated. See Section \ref{Kes17XraySubSection}.}
\label{Kes17_MOS1_MOST}
\end{figure}

In Figure \ref{Kes17_MOS1_MOST} we present an {\it XMM-Newton}
MOS1 image of Kes 17: once again, radio emission from the SNR is plotted with contours.
With an angular resolution of approximately 6$\arcsec$, 
the {\it XMM-Newton} observation of Kes 17 has yielded the highest angular resolution
image ever produced of this SNR. Inspection of this image clearly shows that the diffuse
X-ray emitting plasma is enclosed by the radio shell and that the bright radio rims are
X-ray dim. Such a contrasting X-ray and radio morphology of this SNR strongly indicates
that it belongs to the well-known class of SNRs known as the mixed-morphology SNRs
\citep{Rho98}. We extracted MOS1, MOS2 and PN spectra of Kes 17 using
a source extraction region approximately 4 arcminutes in radius and centered on the nominal
center of the X-ray-emitting plasma: a background spectrum was extracted from an annulus
surrounding the source extraction region and with a thickness of approximately 1 arcminute
(the locations of these extraction regions are indicated in Figure \ref{Kes17_MOS1_MOST}).
We then fit the extracted spectra with the same thermal models used in analyzing the
extracted {\it ASCA} spectra: once again, statistically-acceptable fits were derived with
the APEC and NEI models. In agreement with the analysis of the {\it ASCA} spectra, the
derivation of statistically-acceptable fits with the APEC model and a high derived ionization
timescale with the NEI model ($\tau$ = 3.70$^{+6.30}_{-1.70}$$\times$10$^{11}$ s cm$^{-3}$) 
imply that
the X-ray emitting plasma is close to CIE. In contrast to our fits derived in analyzing the
{\it ASCA} spectra, thawing the elemental abundance as a free parameter in the APEC model
still returns a value of the abundance which is broadly consistent with a solar value 
(0.53$^{+0.77}_{-0.23}$) and did not improve the quality of the fit. 
\par
Extending our prior analysis of the {\it ASCA} spectra of Kes 17, in analyzing the extracted
{\it XMM-Newton} spectra we also searched for variations in the abundances of particular elements 
by using the VAPEC model. We again first fit the spectra with the magnesium abundance thawed
and the abundances of the other elements frozen to solar values. Separately, we fit the spectra
again with the VAPEC model but this time with the abundances of silicon and sulfur thawed
and the abundances of the other elements frozen to solar values. In the case of the latter
fit, the abundances of silicon and sulfur were both broadly consistent with solar values
(Si=0.79$^{+0.51}_{-0.27}$ and S=0.71$^{+0.49}_{-0.29}$) but in the former case we again find
marginal evidence for an overabundance of magnesium (Mg=2.75$^{+2.20}_{-1.70}$). We also 
note that for all of our fits with thermal models, the values for both $N$$_H$ 
and $kT$ remain broadly similar (that is, $N$$_H$$\sim$3$\times$10$^{22}$
cm$^{-2}$ and $kT$$\sim$0.80-1.10 keV). Finally, we fit the extracted spectra using the VAPEC
model with the abundances of magnesium, silicon and sulfur all thawed: this new fit did not
yield a significant improvement in the value of $\chi$$^2_{\nu}$.
If the detection of an overabundance of magnesium
in the X-ray emitting plasma associated with Kes 17 is verified, this SNR may be the newest
member of the class of mixed-morphology SNRs which feature ejecta-dominated emission:
another example of such a source is the mixed morphology SNR CTB 1, which was shown to
feature an overabundance of oxygen in spectra extracted from a {\it Chandra} observation of
the SNR \citep{Pannuti10}.
\par
\begin{figure}
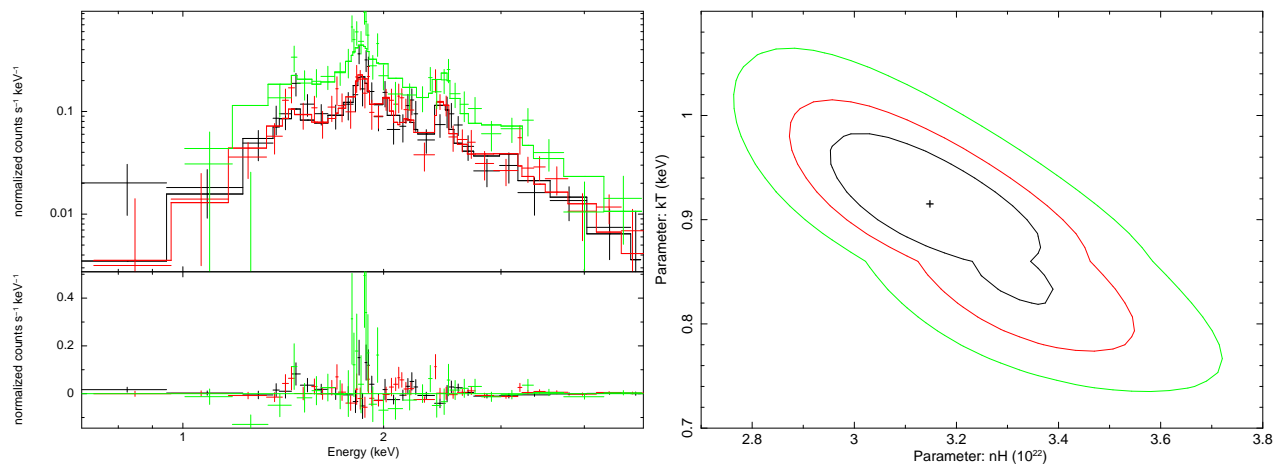

\hbox{
\psfig{figure=KES17_MOS1MOS2PN_PHABSAPEC_SPECTRA.ps,angle=270,scale=0.35}
\psfig{figure=KES17_MOS1MOS2PN_PHABSAPEC_NHKT.ps,angle=270,scale=0.35}
}
\caption{(left) {\it XMM-Newton} MOS1, MOS2 and PN (in black, red and green, 
respectively) spectra of diffuse emission from Kes 17 as fit with the PHABS$\times$APEC
model. (right) Confidence contour plots for for column density $N$$_H$ versus
temperature $kT$ for the PHABS$\times$APEC fit. The contours are plotted at the
1$\sigma$, 2$\sigma$ and 3$\sigma$ levels. See Table \ref{XMMSpectraSummaryKes17} 
and Section \ref{Kes17XraySubSection}.} 
\label{Kes17_MOS1MOS2PN_SPECTRA}
\end{figure}
\begin{figure}
\epsscale{0.8}
\plotone{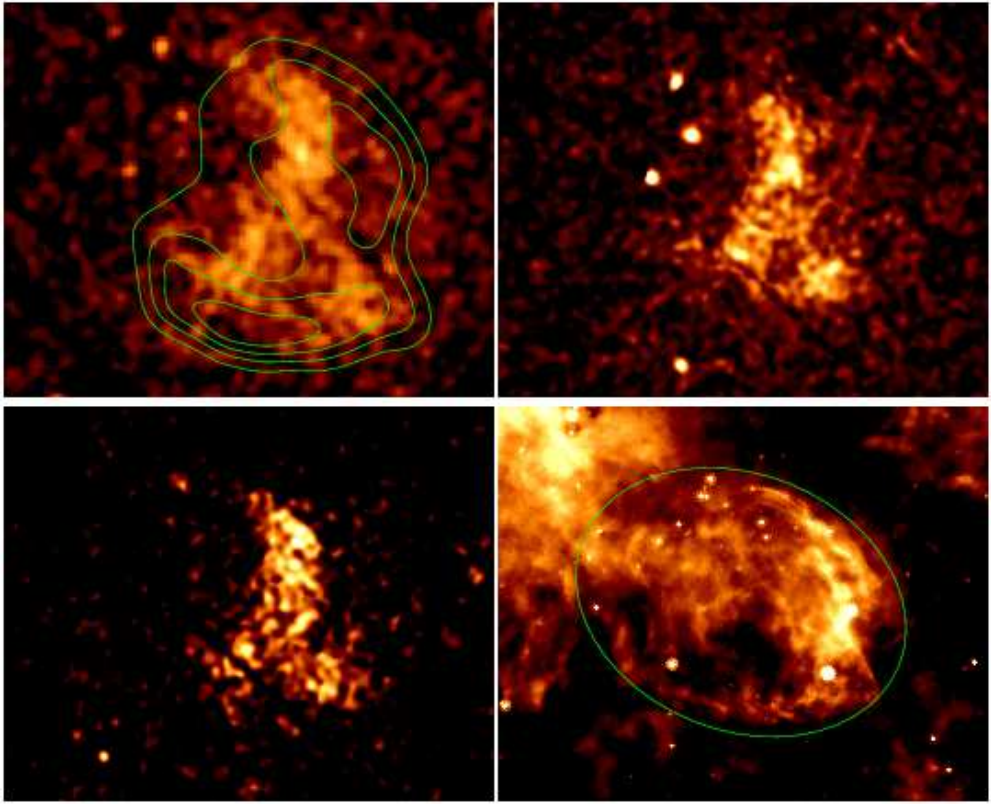}
\caption{(Top panel) Multiwavelength images of Kes 17. (a-upper left), (b-upper right) and (c--lower 
left): {\it XMM-Newton} X-ray images of Kes 17 in the (a) total band (with radio contours overlaid), (b) 
soft band (0.5-2 keV), and (c) hard band (2-6 keV). (d - lower right) {\it Spitzer} MIPS 24\mic\ image 
with a green ellipse to demarcate the emission that is plausibly associated with the SNR. 
The image is centered on RA (J2000.0) 13$^h$ 05$^m$ 58$^s$.3, Dec (J2000.0) 
$-$62$^{\circ}$ 42$'$ 08$''$.8 and the field of view is 12$'$. See Section \ref{Kes17XraySubSection}.}
\label{kes17xmmimage}
\end{figure}
\begin{figure}
\epsfig{figure=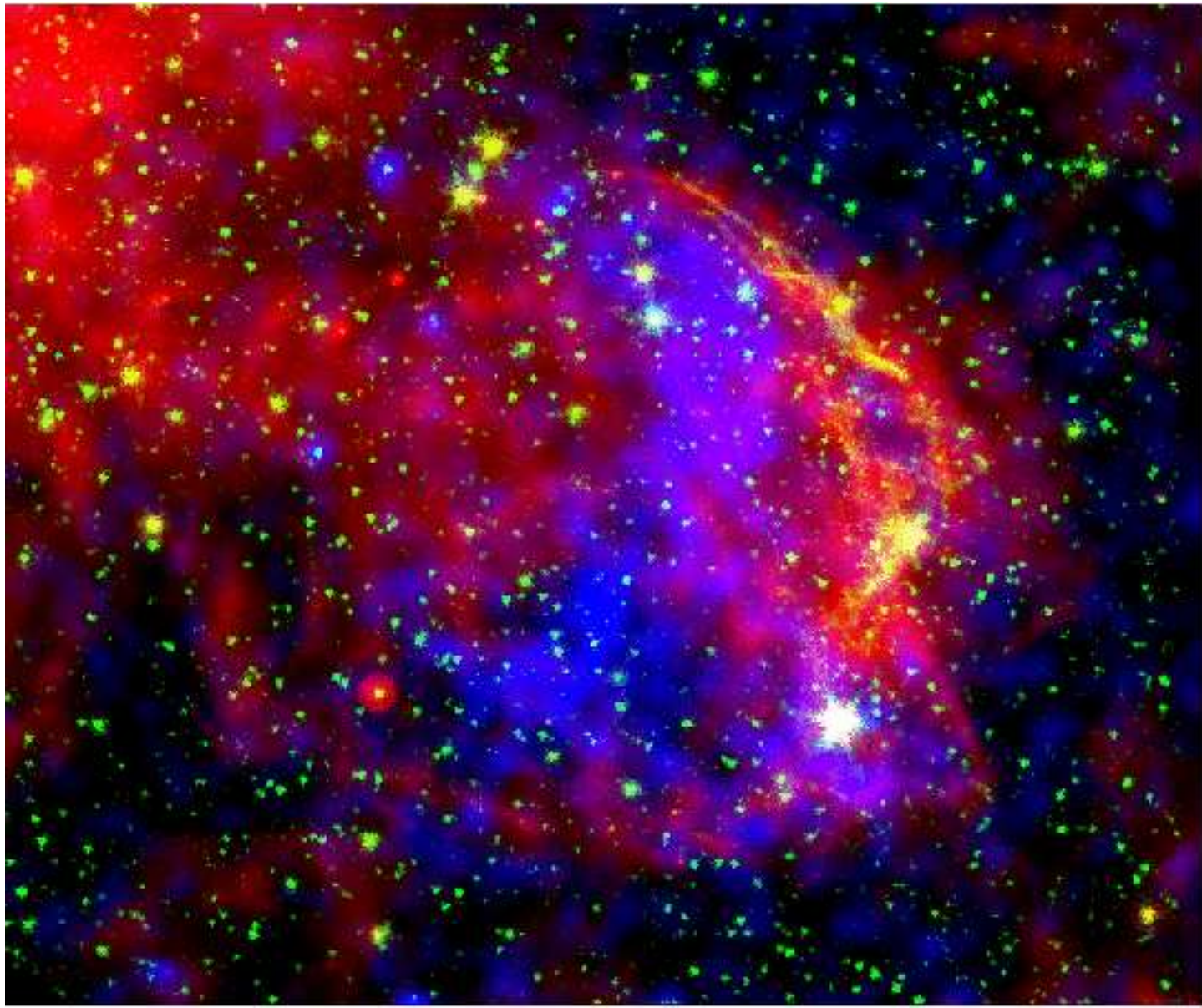}
\caption{Mosaicked three color images of Kes 17. Blue, green and red represent  {\it XMM-Newton} 
X-ray (total band), {\it Spitzer} MIPS at 24$\mu$m and IRAC 4.5$\mu$m images (from \citet{Reach06} 
and \citet{Andersen11}, respectively). Note that the observed X-ray emission may be described as
center-filled within infrared radiative shells. See Section
\ref{Kes17XraySubSection}.} 
\label{kes17rayinfrared}
\end{figure}

In Figure \ref{Kes17_MOS1MOS2PN_SPECTRA} we present the
MOS1, MOS2 and PN spectra of Kes 17 as fitted with the PHABS$\times$APEC model
as well as a confidence contour plot for column density $N$$_H$ and temperature
$kT$ for this fit. In Table \ref{XMMSpectraSummaryKes17} 
we provide a summary of the fits to the extracted MOS1, MOS2 and PN spectra of Kes 17.
In Figures \ref{kes17xmmimage} and \ref{kes17rayinfrared} we present multi-wavelength images 
-- X-ray, radio and infrared (namely {\it Spitzer} IRAC and MIPS images) -- of this SNR.

\subsubsection{Discussion of X-ray Properties of Kes 17\label{Kes17DiscussSubSection}}

We first describe prior analyses made of the X-ray properties of Kes 17. 
\citet{Combi10} presented a study of a pointed {\it XMM-Newton} observation of Kes 17
that included an analysis of MOS1, MOS2 and PN spectra of this SNR: we have analyzed this
same observation in the present paper. Those authors divided 
the SNR into three regions and extracted spectra for each region while we have considered just the 
spectra of the entire diffuse emission. \citet{Combi10} fit the extracted spectra using 
the absorption model WABS \citep{Morrison83} coupled with the non-equilibrium ionization model 
with constant temperature PSHOCK\citep{Mazzotta98} and a power law model with a photon index 
$\Gamma$ (defined such that $E$$^{-\Gamma}$). In addition, \citet{Combi10} allowed the
abundance of the PSHOCK model to vary. The values for $N$$_H$, $kT$, abundances, $\tau$,
and $\Gamma$ obtained in the fits presented by \citet{Combi10} show some variation, which was
interpreted by those authors to indicate spectral variations from one portion of the SNR to the other.
We adopted mean values for these parameters and fit our extracted MOS1, MOS2 and PN
spectra with these mean values in a WABS$\times$(PSHOCK+POWER LAW) model: we present 
the results of the fit  using these parameters for Table \ref{XMMSpectraSummaryKes17}. 
We note that we obtain a 
marginally statistically acceptable fit to the MOS1+MOS2+PN spectra using this model with these fit 
parameters ($\chi$$^2_{\nu}$ = 1.21).
The values for the parameters of the thermal model and the column density derived by \citet{Combi10}
are comparable to those that we obtained. Like \citet{Combi10}, we conclude that the X-ray emitting 
plasma associated with Kes 17 is in CIE (as indicated by their fitted value for $\tau$), but unlike 
\citet{Combi10}, we do not require an additional 
power law component to obtain an acceptable fit to the extracted spectra.
When we added the power law, the value of $\chi^2_{\nu}$ was not improved (as listed in Table 
\ref{XMMSpectraSummaryKes17}).
The differences between our results and those published by \citet{Combi10} 
may originate from the choice of background subtraction (we have chosen an annular region for the
extraction of a background spectrum while \citet{Combi10} extracted background spectra ``from a
region where no X-ray emission was detected"). After careful background subtraction, the detection of
emission above 5 keV was marginal in the spectra.
\par
\citet{Gok12} presented an analysis of observations
made of Kes 17 with the X-ray Imaging Spectrometers (XISs) aboard {\it Suzaku}. Those authors
fitted the extracted spectra using the WABS absorption model coupled with the variable abundance
thermal model VMEKAL \citep{Mewe85, Mewe86, Liedahl95}: the abundances of magnesium,
silicon and sulfur were thawed by those authors during the fitting while the abundances of the other
elements were frozen to solar values. In addition, \citet{Gok12} used an additional 
power law component in their spectral fitting like \citet{Combi10}. 
To compare our results with those presented by \citet{Gok12}, we fitted our extracted 
MOS1+MOS2+PN spectra with the fit parameters published in that paper.
We present the results of this fitting in Table \ref{XMMSpectraSummaryKes17}:
we obtain a statistically acceptable fit using these fit parameters ($\chi$$^2$$_{\nu}$ = 1.11) and again
the values for the parameters $kT$ and $N$$_H$ derived by \citet{Gok12} are similar to those
obtained by \citet{Combi10} and the present work. \citet{Gok12} also concluded that the X-ray emitting 
plasma associated with Kes 17 is in CIE. We also attempted to include a power law component
in the spectral fitting: although the improvement of the fit was not significant, 
the temperature did become lower (from 0.90 to 0.84 keV). The temperatures that we obtained were 
comparable to the temperatures published in those two papers. Since the qualities of the fits of 
the thermal model with or without a 
power law in the \citet{Combi10} and \citet{Gok12} were not compared, the essentiality of the power 
law is not clear in the X-ray energy range.
Inspection of our high
energy {\it XMM-Newton} image of Kes 17 (see Panel (c) of Figure \ref{kes17xmmimage}) does
suggest that some high energy X-ray emission is indeed associated with Kes 17 and the origin
of this emission is uncertain. Possible origins for this particular emission include high energy 
thermal emission from the SNR
or synchrotron emission from an as yet undiscovered neutron star associated with Kes 17. 
Finally, we comment that \citet{Combi10},
\citet{Gok12} and our present work all agree that the contrasting center-filled X-ray morphology and
shell-like radio morphology of Kes 17 motivate its classification as a mixed-morphology SNR. We
will return to the study of the MOS1+MOS2+PN spectra of Kes 17 when we discuss overionization
in mixed-morphology SNRs in Section \ref{OverionizationSection}.




\subsection{G311.5$-$0.3}

G311.5$-$0.3 was first discovered
during surveys made with the Parkes 64-meter radio telescope at 5000 MHz
\citep{Shaver70a} and the Molonglo radio telescope at 408 MHz
\citep{Shaver70b}. It was first classified as a SNR by \citet{Shaver70c}
based on its nonthermal spectral index ($\alpha$=$-$0.49). Those authors also measured 
an angular 
diameter of 3.9 arcminutes for G311.5$-$0.3 and extrapolated a flux
density at 1 GHz of 3.7 Jy from measured flux densities at 408 MHz and
5000 MHz. Using the flux density at 1 GHz, the measured angular extent and 
the surface-brightness diameter ($\Sigma$-D) relation for SNRs derived by 
\citet{Kesteven68b},\footnote{$\Sigma$ $\propto$ $D$$^{-4.5}$ when $\alpha$=$-$0.5
which is approximately the same index as the one measured for 
G311.5$-$0.3} \citet{Shaver70c} calculated a distance of 12.5 kpc to
this SNR and a corresponding diameter of 14 pc: we adopt this distance to G311.5$-$0.3
in this paper. We also note that \citet{Caswell75} placed a lower limit of 6.6 kpc on the
distance to this SNR based on its HI absorption profile. Additional radio observations 
made of a field including this SNR at the same two frequencies as 
\citet{Shaver70a} and \citet{Shaver70b} were described by 
\citet{Caswell85}: the values for the angular extent, flux densities
and spectral index measured by those authors for the SNR were in
good agreement with the values given by \citet{Shaver70c}. A shell-like 
radio morphology for this SNR was first suggested by \citet{Retallack80}:
in addition, \citet{Whiteoak96} described observations made of 
G311.5$-$0.3 at the frequency of 843 MHz using the MOST: they also
reported a shell-like morphology for the source and
measured a flux density of 2.9 Jy at 843 MHz for
this SNR. Finally, \citet{Cohen01} compared MOST observations made
at this frequency with observations made at 8.3 $\mu$m with the 
Midcourse Space Experiment \citep[MSX,][]{Mill94} and reported 
that emission from G311.5$-$0.3 at this wavelength was not detected 
by MSX down to a limiting flux of 1.0$\times$10$^{-6}$ W m$^{-2}$ 
sr$^{-1}$ = 2.35$\times$10$^{-13}$ ergs cm$^{-2}$ sec$^{-1}$. \citet{Frail96}
included G311.5$-$0.3 among a set of Galactic SNRs that were observed with
the intent of detecting maser emission from the OH satellite line at 1720.5 MHz 
with the Parkes 64-meter radio telescope, but no such emission was detected from G311.5$-$0.3.
There have been no previously-published analyses of X-ray emission from this SNR. 

\subsubsection{{\it ASCA} Observations of G311.5$-$0.3\label{G311Results}}

\begin{figure}
\epsfig{figure=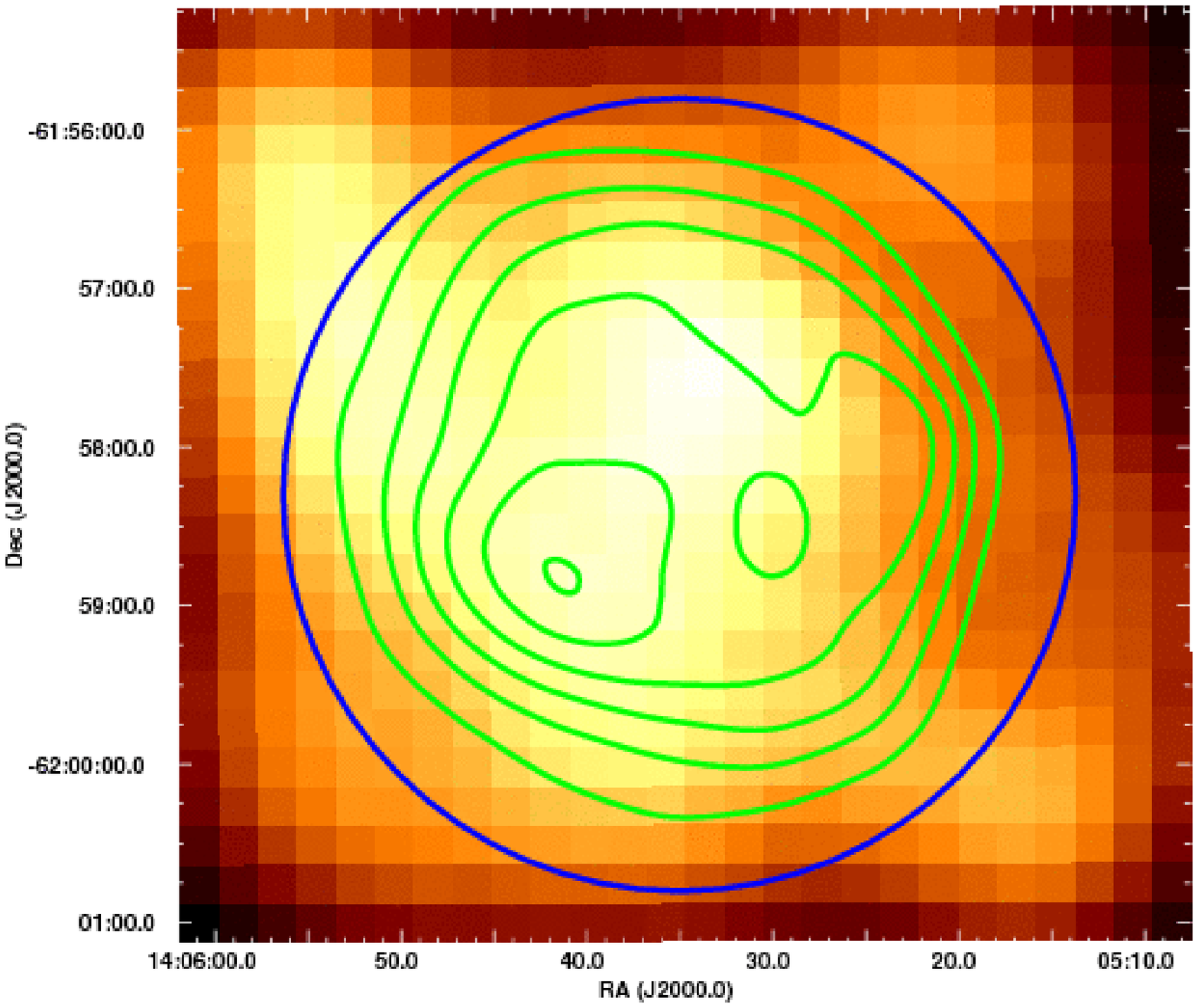}
\caption{{\it ASCA} GIS2+GIS3 co-added X-ray image of G311.5$-$0.3 (in color) for the
energy range 0.6 to 10.0 keV. Radio emission
(as detected with the MOST at a frequency of 843 MHz) is overlaid on the X-ray emission in
green contours. The pixel range of the X-ray emission is 0.01 to 0.08 counts arcmin$^{-2}$
and the radio contours are at the levels of of 0.05, 0.10, 0.15, 0.20, 0.25 and 0.29 Jy/beam.
The blue circle (with a radius of 2.5 arcminutes) indicates the region of spectral extraction 
of the GIS2 and GIS3 spectra. See Section \ref{G311Results}.}
\label{g311ascaimage}
\end{figure}

In Figure \ref{g311ascaimage} we present a co-added {\it ASCA}
GIS2+GIS3 image of G311.5$-$0.3: the SNR was offset by approximately 7 arcminutes
from the nominal center of the field of view during the observation. Radio emission
(as detected at the frequency of 843 MHz) is plotted as contours overlaying the X-ray
emission. The work we describe here represents the first detection and analysis of
X-ray emission from G311.5$-$0.3. The X-ray emission in fact is faint and -- coupled
with the poor angular resolution of the GIS instruments -- prevents us from making 
very definitive statements about the X-ray morphology. To the limit of the existing data
it appears that the emission is centrally-concentrated, in contrast to the shell-like radio
morphology of the SNR. As shown below, our spectral analysis strongly indicates that
the X-ray emission from G311.5$-$0.3 is thermal in origin. A centrally-concentrated
thermal X-ray morphology coupled with a shell-like radio morphology strongly 
suggests that G311.5$-$0.3 is in fact an MM SNR, though additional X-ray
observations with improved sensitivity and angular resolution are needed to investigate
the true nature of the X-ray morphology of G311.5$-$0.3 more thoroughly.

\par

\begin{figure}
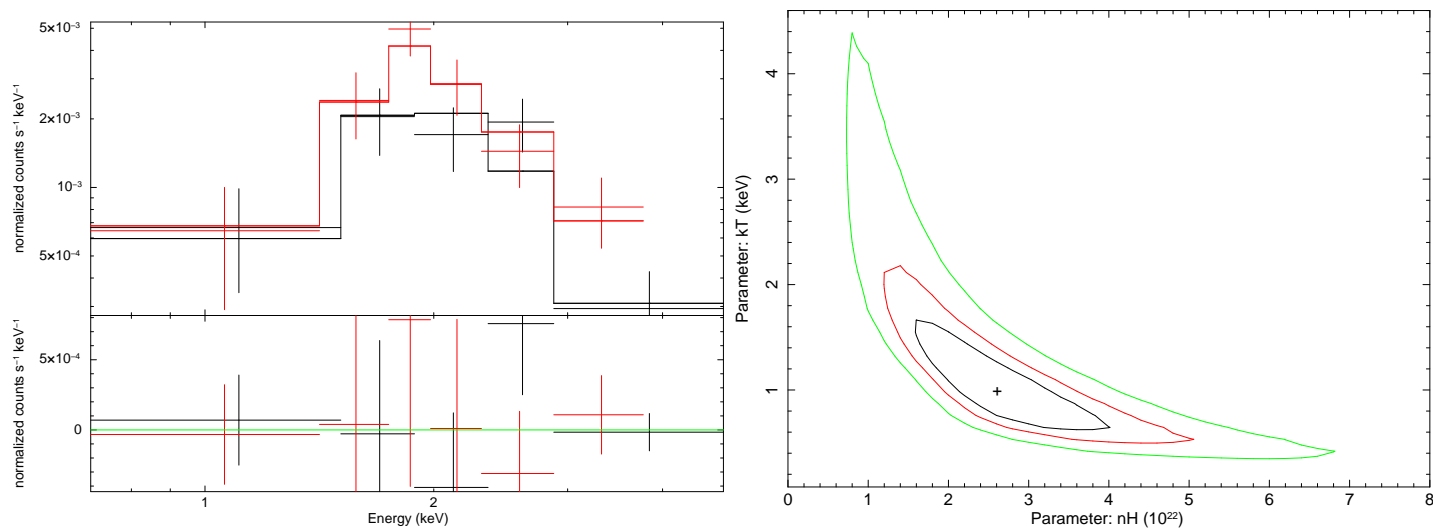
                                                                           
\hbox{
\epsfig{figure=G311_GIS2GIS3_PHABSAPEC_SPECTRA_071911.ps,angle=270,width=9.5truecm}
\epsfig{figure=G311_GIS2GIS3_PHABSAPEC_NHKT_071911.ps,angle=270,scale=0.4}
}
\caption{(left) {\it ASCA} GIS2 and GIS3 (in black and red, respectively) spectra for
G311.5$-$0.3 as fit using the PHABS$\times$APEC model. (right) Confidence 
contour plot for column density $N$$_H$ versus temperature $kT$ for the 
fit generated using the PHABS$\times$APEC model. See Table \ref{GISSpectraSummaryG311}
and see Section \ref{G311Results}.}                                                   
\label{g311ascaspec}
\end{figure}

GIS2 and GIS3 spectra were extracted for G311.5$-$0.3 using a
circular aperture 5 arcminutes in radius and centered at a position corresponding to the apparent
middle of the radio shell of this SNR. The corresponding
background GIS2 and GIS3 spectra required for fitting were generated using the FTOOL
{\tt mkgisbgd.} The number of detected counts for both spectra were very modest 
(75 and 111 counts for the GIS2 and GIS3 after background subtraction, respectively), 
so we only used simple models (namely
a power law model and the thermal plasma APEC model) to fit the spectra coupled with the
PHABS model to account for photoelectric absorption along the line of sight. A summary
of these fits is presented in Table \ref{GISSpectraSummaryG311}: we find that both the power 
law model and the APEC model adequately fit the extracted spectra, with each fit featuring
corresponding values for $\chi$$^2_{\nu}$ of $\sim$1.1. In addition, the values for
the column density $N$$_H$ returned by the two fits are comparable (ranging from 
$N$$_H$ $\sim$ 2.7$\times$10$^{22}$ cm$^{-2}$ to $N$$_H$ $\sim$ 3.3$\times$10$^{22}$
cm$^{-2}$). The value for $kT$ as returned by the fit with the PHABS$\times$APEC model is 
elevated but plausible for an SNR (that is, $kT$$\sim$0.92$^{+1.02}_{-0.22}$ keV) while 
the value for the photon index $\Gamma$ returned in the PHABS$\times$POWER LAW fit
is steep enough ($\Gamma$=3.10$^{+2.10}_{-1.20}$) to also suggest a 
thermal origin of the X-ray emission rather than (for example) synchrotron radiation. 
In Figure \ref{g311ascaspec} we present both
the {\it ASCA} GIS2 and GIS3 spectra as fit by the PHABS$\times$APEC model described
above (and summarized in Table \ref{GISSpectraSummaryG311}) and a confidence 
contour plot for this particular fit for the parameters {\it kT} and $N$$_H$. We conclude that
the {\it ASCA} spectra of G311.5$-$0.3 favor a thermal origin for the observed X-ray emission
from this source: follow-up observations made with higher spectral resolution with sufficient 
counts are required to confirm this interpretation. 
\par
\begin{figure}
\epsfig{figure=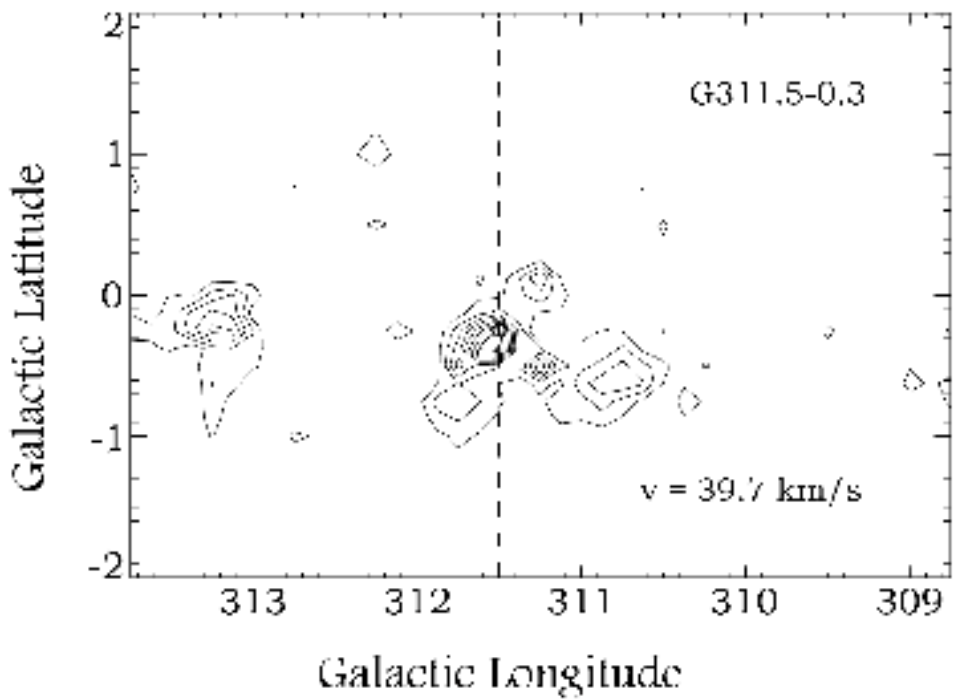}
\caption{CO map of \gteleven for the velocity $v$=39.7 km s$^{-1}$. The position of G311.5$-$0.3
is indicated by the intersection of the dashed line. Notice the giant molecular cloud (GMC) located
in proximity to G311.5$-$0.3. See Section \ref{G311Results}.}
\label{g311co}
\end{figure}
\begin{figure}
\epsfig{figure=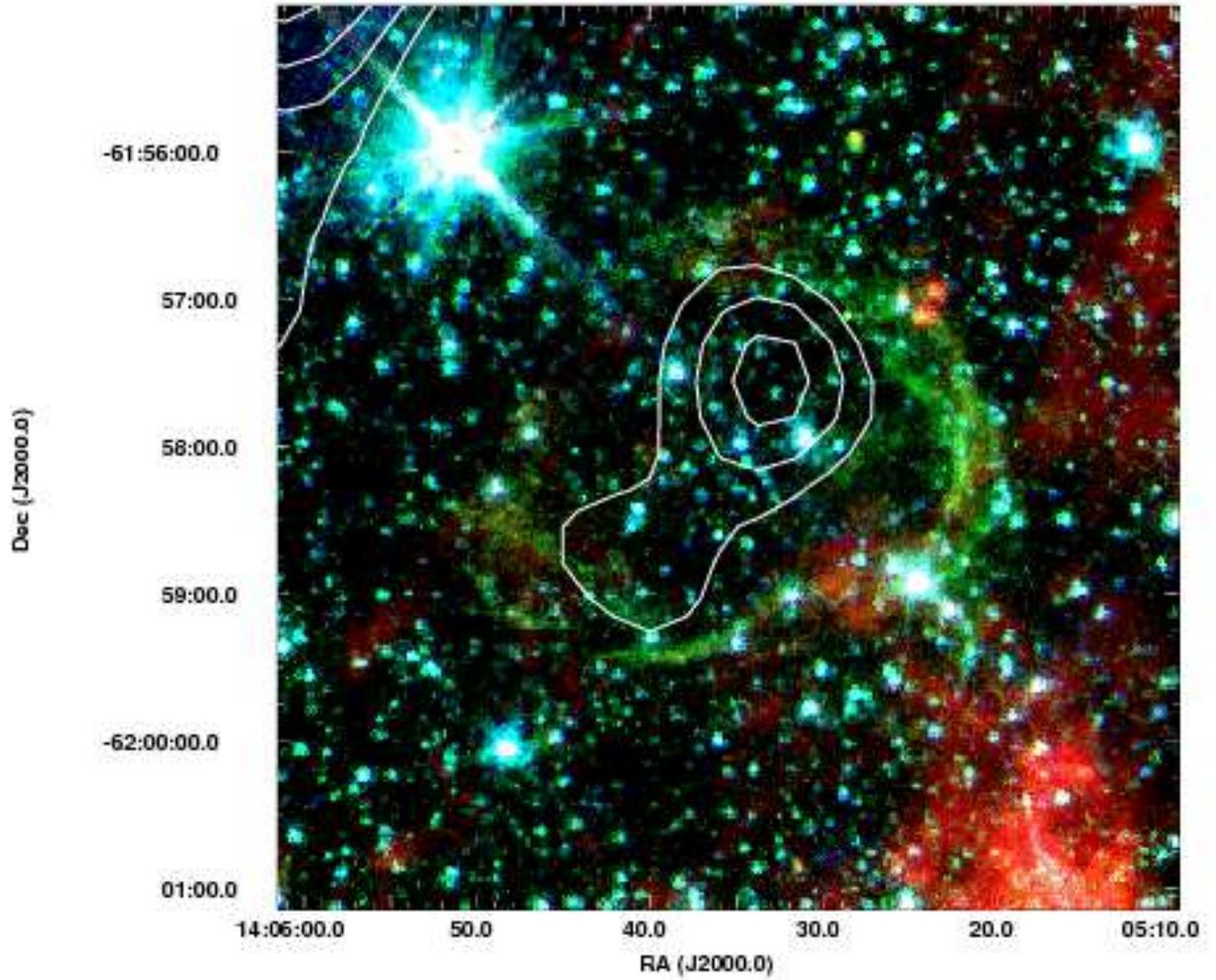}
\caption{{\it Spitzer} IRAC three color image of \gteleven\ (from \citet{Reach06}) with 
{\it ASCA} GIS X-ray contours overlaid. In this image, the red, green and blue correspond to emission 
detected at 8$\mu$m, 4.5 $\mu$m and 3.5$\mu$m, respectively. Notice the center-filled 
X-ray morphology which is characteristic of mixed-morphology SNRs. See Section \ref{G311Results}.}
\label{g311xrayinfrared}
\end{figure}

Our analysis of the {\it ASCA} observations of G311.5-0.3 represents the first detection of
X-ray emission from this SNR. As mentioned above, the distance
to this SNR is unknown: to help constrain the distance to this SNR more rigorously, we have examined 
CO data obtained from observations made toward G311.5$-$0.3. Data from a CO observation of
this source (corresponding to a velocity of 39.7 km s$^{-1}$) is presented in Figure \ref{g311co}:
we note the presence of a giant molecular cloud (GMC) detected in proximity to this SNR. 
\citet{Andersen11} have argued that G311.5$-$0.3 is interacting with molecular clouds based
on the detection of H$_2$ emission at infrared wavelengths. Based on this argument, we 
claim that G311.5$-$0.3 is associated with the GMC seen in projection toward G311.5$-$0.3: 
the particular velocity of 39.7 km s$^{-1}$ for the GMC corresponds to a distance of 12.5 kpc. 
This distance matches our adopted distance to this SNR. The fitted column density toward this 
source as determined from our X-ray spectral fitting (that is, $N$$_H$ = 2.7 $\times$
10$^{22}$ cm$^{-2}$) and our calculated X-ray luminosities have been determined using this
adopted distance. We also note that the distance toward G311.5$-$0.3 as adopted by 
\citet{Andersen11} 
is also based on this high column density as suggested by the velocity of 
the GMC detected in CO: the uncertainty of the distance estimate is not known. In Figure 
\ref{g311xrayinfrared} we present a {\it Spitzer} three-color IRAC image of G311.5$-$0.3 with X-ray 
contours (corresponding
to the emission detected by the {\it ASCA} GIS2+GIS3 detectors) overlaid. The nearly-complete
infrared emission with colors consistent with molecular shocks \citep[see][]{Andersen11} and the
center-filled thermal X-ray emission support the classification of this source as an MM SNR. 

\subsubsection{Discussion of X-ray Properties of G311.5$-$0.3\label{G311DiscussSubSection}}

Based on its center-filled thermal X-ray morphology and its shell-like radio morphology, we 
suggest that G311.5$-$0.3 belongs to the class of MM SNRs. 
As noted above, the infrared
properties of this SNR can be modeled with shocked molecular emission and we find evidence
(based on a CO observation) of a cloud seen in projection toward G311.5$-$0.3: the large
distance implied to this SNR based on the high column density (as derived from the spectral fitting)
matches the implied distance to this cloud. The infrared colors of G311.5$-$0.3 imply that the SNR 
is a molecular SNR \citep{Reach06} and IRS 
follow-up spectroscopy confirmed that the {\it Spitzer} IRAC 4.5$\mu$m band emission
shown in Figure \ref{g311xrayinfrared} is H$_2$ emission \citep{Hewitt09} 
which likely originates from the interaction between the SNR and dense clouds. 
Our {\it ASCA} analysis of G311.5$-$0.3 revealed for the first time X-ray emission from this SNR.
The X-ray morphology of the emission is center-filled -- lying within the radio and infrared shells
of emission -- and the X-ray emission is thermal with a temperature $kT$ $\sim$ 0.98 keV.
For these reasons, we suggest that G311.5$-$0.3 is an MM SNR, but because the number of
photons in the {\it ASCA} observation of G311.5$-$0.3 is limited, future observations of this SNR
with improved angular resolution and flux sensitivity are required to confirm this classification.
Such observations may also help strengthen the
correlation between the phenomenon of MM SNRs and SNRs interacting with adjacent molecular
clouds.

\subsection{G346.6$-$0.2}

Like G311.5$-$0.3, G346.6$-$0.2 was first detected as a radio source and 
classified as an
SNR based on surveys made with the Parkes 64-meter radio telescope at
5000 MHz and the Molonglo radio telescope at 408 MHz \citep{Clark73,Clark75}.
These observations -- along with more recent radio observations made of the SNR
by \citet{Dubner93} using the Very Large Array (VLA) at 1.465 GHz as well
as by \citet{Whiteoak96} using the MOST at 843 MHz -- reveal the SNR to be shell-like 
with a diameter of approximately 8.2 arcminutes. \citet{Dubner93}
measured a flux density of 8.1 Jy at 1.465 GHz and also estimated a spectral index of
$\alpha$ = -0.5 for this SNR. Those authors also commented that the shell of
radio emission from the SNR appeared to be flattened with a noticeable compression
along the eastern and northwestern borders. Based on observations made with the 
VLA of the OH ground state transition at 1720 MHz, \citet{Koralesky98} 
detected five masers along the southern rim of the SNR. \citet{Green97} also reports
the detection of OH maser emission at 1720 MHz from G346.6$-$0.2 as observed
with the Parkes 64-meter telescope. Interestingly, the radio
emission from the SNR is strongest from the northern, eastern and western portions
of the shell but is noticeably weaker along the southern rim, where the masers
are detected. \citet{Dubner93} estimated the distance to G346.6$-$0.2 to be $\sim$9 kpc
based on the $\Sigma$-$D$ relationship for SNRs that was presented by 
\citet{Huang85} while \citet{Koralesky98} provided lower and upper bounds on the
distance to this SNR of 5.5 kpc and 11 kpc respectively based on the estimates of the
rotational velocities of the molecular clouds argued to be interacting with G346.6$-$0.2
based on detected maser emission. We adopt a distance of 11 kpc to this SNR for
the remainder of this paper: we note that a larger distance estimate to G346.6$-$0.2
is supported by the enhanced column density seen toward this SNR based on the
results of our spectral fitting (see Section \ref{G346ResultsSection}). Previously-published 
X-ray studies of G346.6$-$0.2 have been presented by \citet{Yamauchi08} (who analyzed an {\it 
ASCA} observation of G346.6$-$0.2), \citet{Sezer11b} and \citet{Yamauchi13} (who both analyzed 
a {\it Suzaku} observation of this SNR). we compare our results
in the present paper with the results published in those papers in Section
\ref{G346DiscussSubSection}.

\subsubsection{{\it ASCA} Observations of G346.6$-$0.2\label{G346ResultsSection}}

\begin{figure}
\epsfig{figure=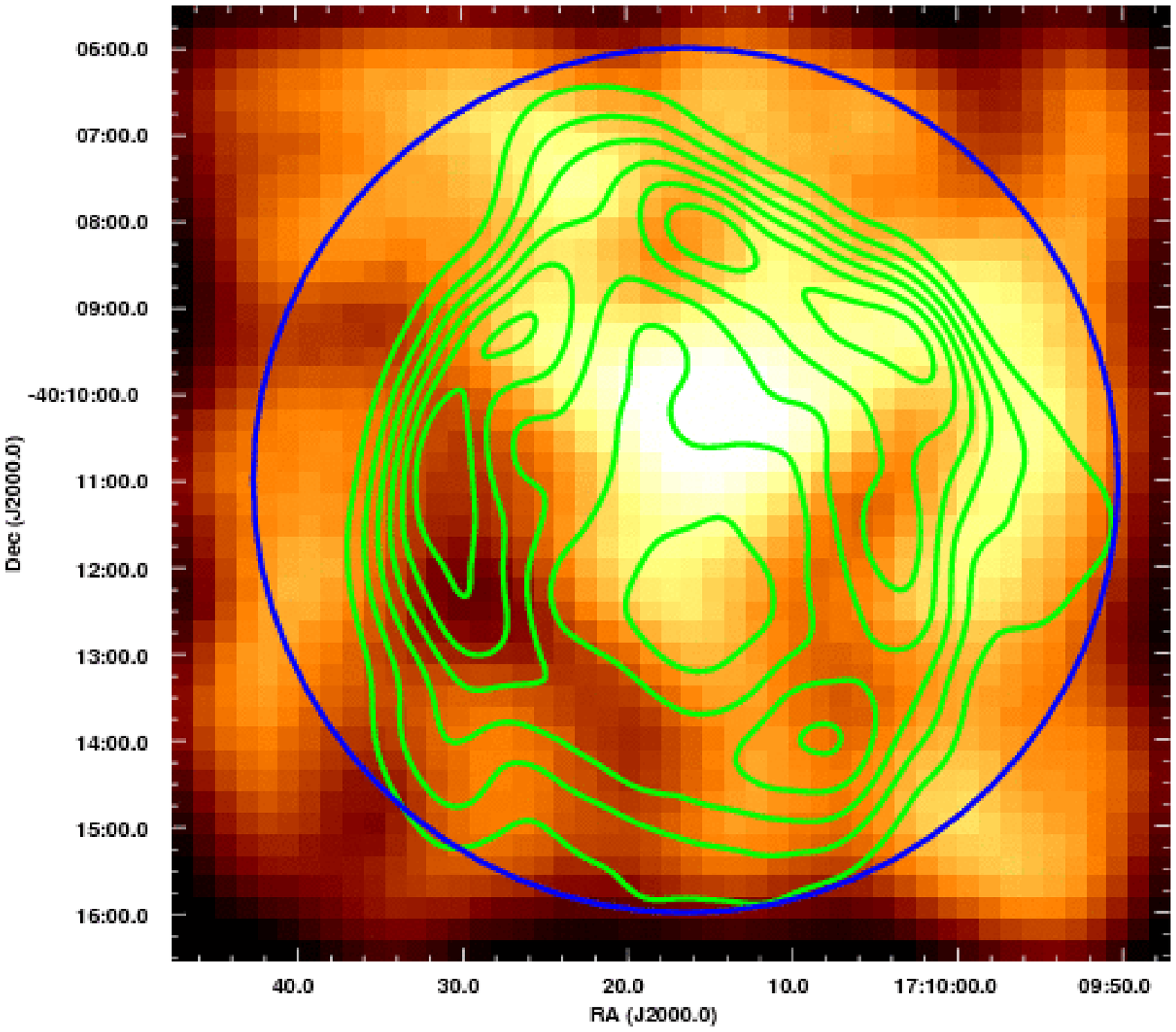}
\caption{{\it ASCA} GIS2+GIS3 co-added X-ray image of G346.6$-$0.2 (in color) for the
energy range 0.6 to 10.0 keV. Radio emission
(as detected with the MOST at a frequency of 843 MHz) is overlaid on the X-ray emission in
green contours. The pixel range of the X-ray emission is 0.02 to 0.10 counts arcmin$^{-2}$
and the radio contours are at the levels of of 0.04, 0.08, 0.12, 0.16, 0.20 and 0.24 Jy/beam.
The blue circle (with a radius of 5 arcminutes) indicates the region of spectral extraction 
of the GIS2 and GIS3 spectra. See Section \ref{G346ResultsSection}.}
\label{g346ascaimage}
\end{figure}

In Figure \ref{g346ascaimage} we present a co-added {\it ASCA} GIS2+GIS3 
image of G346.6$-$0.2: the center of the X-ray emission from the SNR was offset by approximately 
14 arcminutes from the nominal
center of the field of view during the observation. We have overlaid radio contours at
a frequency of 843 MHz to illustrate the radio morphology of the shell: similar to the
situation with G311.5$-$0.3 described previously, while G346.6$-$0.2 is only modestly
detected by {\it ASCA}, the X-ray morphology appears to be center-filled while the
radio emission is shell-like. Again, similar to the situation with G311.5$-$0.3, unfortunately the
poor angular resolution of the GIS2 and GIS3 instruments and the modest detection of 
the SNR prevent us from making more conclusive statements about the general X-ray morphology
of G346.6$-$0.2. As shown below, the X-ray emission detected from this SNR 
is predominately thermal in origin. The contrasting X-ray and radio morphologies of G346$-$0.2 
coupled with the thermal nature of its X-ray emission motivate the classification of this source
as a mixed-morphology SNR (which also parallels our results with G311.5$-$0.3). 
As mentioned previously, the detection of maser emission
from this SNR strongly suggests that G346.6$-$0.2 is currently interacting with a molecular
cloud: interactions with molecular clouds is another commonly-seen characteristic of
MM SNRs.

\par

GIS2 and GIS3 spectra were extracted for G346.6$-$0.2 using a circular 
aperture 5 arcminutes in radius and centered at a position corresponding to the apparent
middle of the radio shell of this SNR.
The region of
spectral extraction is indicated in Figure \ref{g346ascaimage}: similar to our work 
in analyzing the GIS2 and GIS3 spectra of G311.5$-$0.3, we used the FTOOL {\tt mkgisbgd} to
generate background spectra needed for spectral fitting. 
Inspection of
the extracted spectra revealed that there was a significantly greater amount of emission at higher
energies compared to the extracted spectra of Kes 17 and G311.5$-$0.3. 
Note that our re-analysis of {\it ASCA} data of G346.6-0.2 produces higher quality spectra than 
those presented by \citet{Yamauchi08}.
We first fitted the extracted spectra with a simple power law model: the photon index of the
power law fit was $\Gamma$ $\sim$ 2.5 but the fit was not statistically acceptable 
($\chi$$^2_{\nu}$=1.34). We then fitted the spectra with the plasma model APEC (with the abundance 
frozen to solar): the temperature of the fit was $kT$ $\sim$ 2.5 keV but once again the fit was not 
acceptable ($\chi$$^2_{\nu}$=1.45). We then fit the spectrum with the APEC model again,
this time with the abundance parameter thawed: the fit returned by this model was 
significantly improved ($\chi$$^2_{\nu}$=1.22) but still not statistically acceptable. The 
upper limit on the fitted elemental abundance was 0.4 (relative to solar) while the fitted temperature
was $kT$ $\sim$ 3.4 keV, which is higher than those typically measured for MMSNR SNRs.
If this estimated temperature for the SNR is indeed true, it may indicate that G346.6$-$0.2 is
a particularly young SNR. We also attempted to fit the 
spectra with the VAPEC model: we obtained a statistically-acceptable fit with the VAPEC model
($\chi$$^2_{\nu}$=1.14) though the only elemental abundance that was found
to exceed solar was neon, and in this case the fitted abundance was very high (with
a lower limit of 26 times solar). However, because the low signal-to-noise of the extracted GIS 
spectra made it too difficult to search for variations in the abundances of specific elements in a 
rigorous manner, we discarded this PHABS$\times$VAPEC fit. Next, we fit the spectrum with 
the non-equilibrium ionization plasma model NEI (with the abundance frozen to solar): an  
acceptable fit was obtained ($\chi$$^2_{\nu}$ of the fit was 1.07) but again the fitted 
temperature ($kT$ = 2.80$^{+0.80}_{-0.50}$ keV) was high for an SNR while 
the ionization timescale ($\tau$ = 7.27$^{+6.23}_{-3.77}$$\times$10$^{9}$ cm$^{-3}$ s) was lower
than other MM SNRs. This result
indicated that the X-ray-emitting plasma associated with G346.6$-$0.2 is not in CIE
and motivated the use of an NEI model with thawed abundance coupled with a power law to fit 
properly model the high energy X-ray emission. Due to the low signal-to-noise in the spectrum that 
would make calculating the error bounds of many individual parameters difficult, we attempted
fits where we froze values of $\tau$ and $\Gamma$ to physically realistic parameters (specifically
10$^{11}$ cm$^{-3}$ s for $\tau$ to reflect the X-ray-emitting plasma being outside of CIE and
0.5 and 2.0 for $\Gamma$ to reflect values obtained in fits of components of hard emission observed
toward other Galactic SNRs).
We attempted to analyze diagnostics of the NEI model using line ratios
such as He$\alpha$ over Ly$\alpha$ for Mg, Si or S, but none of them show clear lines so such
analysis was not possible.
In considering the fits with high $\tau$, all appear to be statistically
acceptable with approximately the same values of $\chi$$_{\nu}$$^2$: we believe that the fits made
using adopted values for $\Gamma$ and $\tau$ of 0.5 and 10$^{11}$ cm$^{-3}$ s 
respectively might be the most physically reasonable because these values for these fit parameters 
may be the most physically accurate. 
\par
\begin{figure}
\psfig{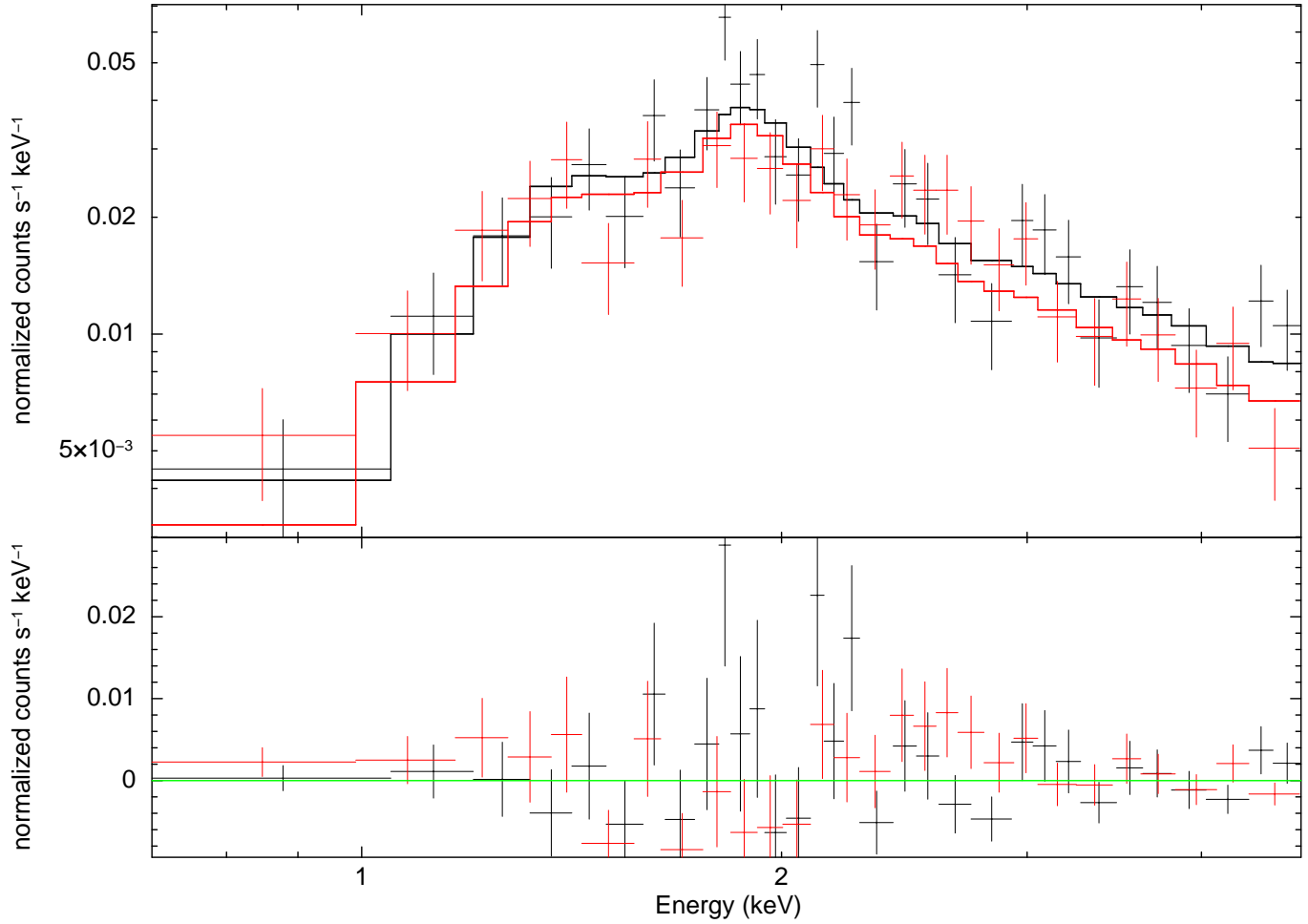}
\caption{{\it ASCA} GIS2 and GIS3 (in black and red, respectively) spectra for 
G346.6$-$0.2 as fit using a PHABS$\times$(NEI+Power Law) model. In this fit, the ionization 
timescale has been frozen to $\tau$ = 8$\times$10$^{10}$ cm$^{-3}$ s and the photon index has
been frozen to $\Gamma$=0.5. See Table \ref{GISSpectraSummaryG346} and Section 
\ref{G346DiscussSubSection}.
\label{G346Spectra}}
\end{figure}
\begin{figure}
\psfig{figure=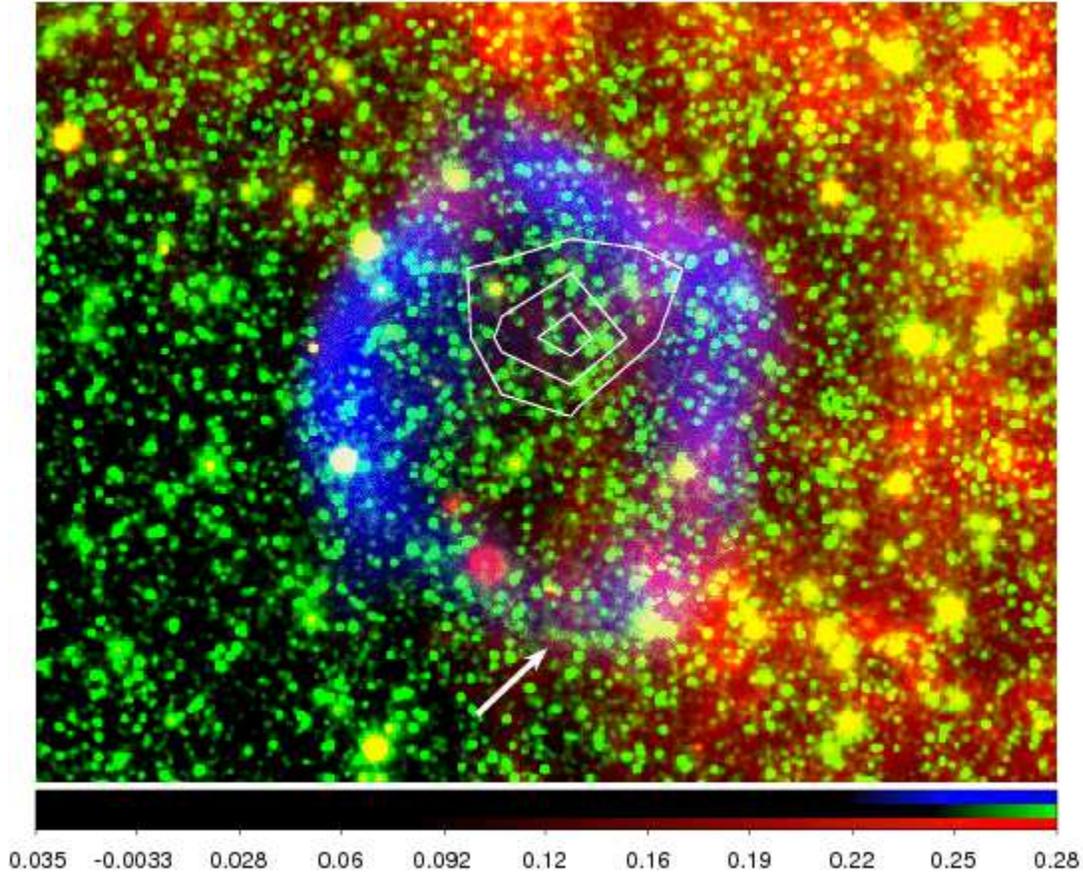,angle=0,scale=0.75}
\caption{Mosaicked three color images of G346.6-0.2 superposed on {\it ASCA} GIS2+GIS3 X-ray contours. 
Red, green and blue represent images of {\it Spitzer} 
MIPS at 24 $\mu$m, IRAC at 4.6 $\mu$m \citep{Reach06} and radio \citep{Green97}, respectively.
The arrow indicates the location where emission from molecular hydrogen is detected
\citep{Andersen11}. See Section \ref{G346ResultsSection}.}
\label{G346XrayInfraredRadioImage}
\end{figure}

We present a summary of these
fits in Table \ref{GISSpectraSummaryG346}. In Figure \ref{G346Spectra} we present
the {\it ASCA} GIS2 and GIS3 spectra as fit by the PHABS$\times$(NEI+Power Law) model described
above where we have frozen $\tau$ and $\Gamma$ to 8$\times$10$^{10}$ cm$^{-3}$ s and
0.5, respectively. The results of all of the fitting (including values for fit parameters) are  
listed in Table \ref{GISSpectraSummaryG346}.
Finally, in Figure \ref{G346XrayInfraredRadioImage} we present a 
multi-wavelength image -- X-ray, radio and infrared (namely {\it Spitzer} MIPS) -- of G346.6$-$0.2. 

\subsubsection{Discussion of X-ray Properties of G346.6$-$0.2\label{G346DiscussSubSection}}

In a previous analysis of extracted GIS spectra of this SNR, \citet{Yamauchi08} presented 
statistically-acceptable fits using a thermal component 
(WABS$\times$MEKAL, with fitted values of $N$$_H$=2.0$^{+3.5}_{-1.6}$$\times$10$^{22}$ 
cm$^{-2}$ and $kT$ = 1.6$^{+3.1}_{-1.2}$ keV) or a power law component (WABS$\times$Power 
Law, with fitted values of $N$$_H$=2.6$^{+8.2}_{-2.2}$$\times$10$^{22}$ cm$^{-2}$ and 
$\Gamma$=3.7($>$1.7). we find that we can obtain statistically acceptable fits to our extracted spectra 
using these models within the quoted error bounds. More recently, \citet{Sezer11b} described an 
analysis of an observation made of G346.6$-$0.2 with $\it{Suzaku}$: those authors presented spectral 
fits to the diffuse emission using several different thermal models (including VMEKAL, VNEI and the 
constant temperature plane parallel shock plasma model with variable elemental abundances 
VPSHOCK -- see \citet{Mazzotta98}) coupled with a power law model (the fitted values of the photon 
indices were $\Gamma$$\sim$0.5-0.6). The thermal models all featured
abundances of several elements (namely magnesium, silicon, sulfur, and iron) which were all
sub-solar while the abundance of calcium exceeded solar. We have fitted our extracted spectra
with the models and the associated fit parameters published by \citet{Sezer11b} and we present
our results in Table \ref{GISSpectraSummaryG346}. We obtain statistically acceptable fits to our
extracted spectra using these models with these associated parameters: like \citet{Sezer11b}, we
need a combination of a thermal component and a power law component to obtain a statistically
acceptable fit with physically realistic parameters. Unfortunately our extracted GIS spectra lack 
sufficient signal-to-noise for a detailed analysis of the elemental abundances of the X-ray-emitting
plasma as conducted by \citet{Sezer11b}. Using {\it ASCA} data, we confirm that
that the X-ray emitting plasma associated with G346.6$-$0.2 is thermal and that the X-ray 
morphology is center-filled  within a radio shell. 
These facts motivate the classification of G346.6$-$0.2 as an MMSNR. 
\citet{Yamauchi13} present a more detailed analysis of the {\it Suzaku} observations of G346.6$-$0.2
and argue that the X-ray spectrum of this SNR is best fit with a recombining plasma model,
while such spectral analysis was not possible with the {\it ASCA} data due to the limited number of 
counts and the poorer spectral resolution.
\par
Based on the measured elemental abundances in their fits to the extracted spectra, \citet{Sezer11b} 
argued that the X-ray emission from G346.6$-$0.2 is ejecta-dominated
and -- by analyzing the relative abundances of the elements -- that a Type Ia supernova
explosion produced this SNR. We note, however, that the H$_2$ emission detected from this
SNR by \citet{Hewitt09} indicates the presence of a fast J-shock (with a velocity of approximately
150 km s$^{-1}$): the spectral lines associated with shocked H$_2$ suggest a high preshock
density, such as the elevated preshock density associated with molecular clouds. We argue that
the presence of this shocked H$_2$ indicates an interaction between this SNR and the 
adjacent molecular cloud (as supported by the detection of maser emission from the perimeter of
this SNR by \citet{Koralesky98}). We therefore argue that G346.6$-$0.2 had a massive stellar
progenitor and that the type of supernova that created this SNR was a Type Ib/Ic/II supernova 
(that is, the supernova types that are all associated with massive stellar progenitors). Such 
massive stellar progenitors are usually associated with MM SNRs. 
Future {\it XMM-Newton} and {\it Chandra} observations of G346.6$-$0.2 would be critical
in confirming or refuting these conclusions.
 
\subsection{CTB 37A (G348.5$+$0.1)}

The CTB 37 radio complex was first cataloged as a radio source by a 960 MHz 
survey conducted by the Owens Valley Radio Observatory \citep{Wilson60, Wilson63}.
This complex was resolved into two discrete radio sources -- CTB 37A and CTB 37B --
by radio observations made at multiple frequencies by \citet{Milne69}, who measured
non-thermal radio spectral indices for both objects ($\alpha$=-0.55 in each case) and
thus classified the two sources as SNRs. Subsequent radio observations of CTB 37A
\citep{Milne75,Downes84,Kassim91} revealed this SNR to feature a shell-like morphology
with bright northern and eastern rims. \citet{Milne69} estimated an integrated flux density for
CTB 37A at 1410 MHz of approximately 76 Jy. Based on MOST observations of this SNR, 
\citet{Whiteoak96} commented on the shell-like morphology of CTB 37A by noting that this 
source possessed one of the highest mean surface brightnesses of any SNR in the MOST 
catalog. \citet{Whiteoak96} also estimated the angular extent of CTB 37A to be 19$\times$16
arcminutes and commented on the presence
of faint emission associated with this SNR extending into the southwest: such an extension
of emission is consistent with the expansion into a cavity. OH masers observed
at a frequency of 1720 MHz have been detected along the northern, western and southern
boundaries of this SNR as well as toward its center \citep{Frail96,Brogan00,Hewitt08}, though
a bimodal distribution of measured velocities was observed for these masers (specifically, 
masers with associated velocities of $-$65 km s$^{-1}$ were observed toward the center of the
SNR as well as the eastern and southern boundaries while masers with associated velocities
of $-$24 km s$^{-1}$ were observed toward the northern boundary). 
\citet{Caswell75} estimated a
distance of 10.2$\pm$3.5 kpc to CTB 37A based on its observed HI absorption profile while 
\citet{Reynoso00} derived a distance to the SNR of 11.3 kpc based on a putative interaction
between this SNR and adjacent molecular clouds, as observed by emission from the carbon
monoxide (CO) $J$=1 to 0 transition at approximately 115 GHz). More recently, \citet{Tian12}
bounded the distance to CTB 37A between 6.3 and 9.5 kpc based on the absorption profile
along the line of sight toward the SNR. Those authors concluded that CTB 37A lies in front of the 
far side of the molecular cloud seen toward this SNR. In the present paper, we adopt a distance
of 8 kpc toward this SNR, approximately the mean of the lower and upper distance bounds
proposed by \citet{Tian12}. We also note that \citet{Aharonian08} detected 
$\gamma$-ray emission from CTB 37A using the Cherenkov telescopes of the High
Energy Stereoscopic System (HESS -- see \citet{Bernlohr03}): specifically, those authors
associated the $\gamma$-ray source HESS J1713$-$385 with CTB 37A  (in particular
with the region of hard emission seen toward the northwestern rim of the SNR). Based
on fits to the spectrum of HESS J1713$-$385, \citet{Aharonian08} speculated that the
observed emission had a hadronic origin (that is, it is produced by the decay of pions that
are created by collisions between cosmic-ray protons and protons in an adjacent molecular
cloud). Regarding previous X-ray observations of CTB 37A, published work based on observations
made with {\it ASCA} \citep{Yamauchi08}, {\it Chandra}, {\it XMM} (both presented by 
\citet{Aharonian08}) and finally {\it Suzaku} \citep{Sezer11a} have all appeared previously in
the literature: we compare our results with the results presented in those papers in Section
\ref{CTB37ADiscussSubSection}.

\subsubsection{{\it XMM-Newton} and {\it Chandra} Observations of CTB 
37A\label{CTB37AXraySubsection}}

\begin{figure}
\psfig{figure=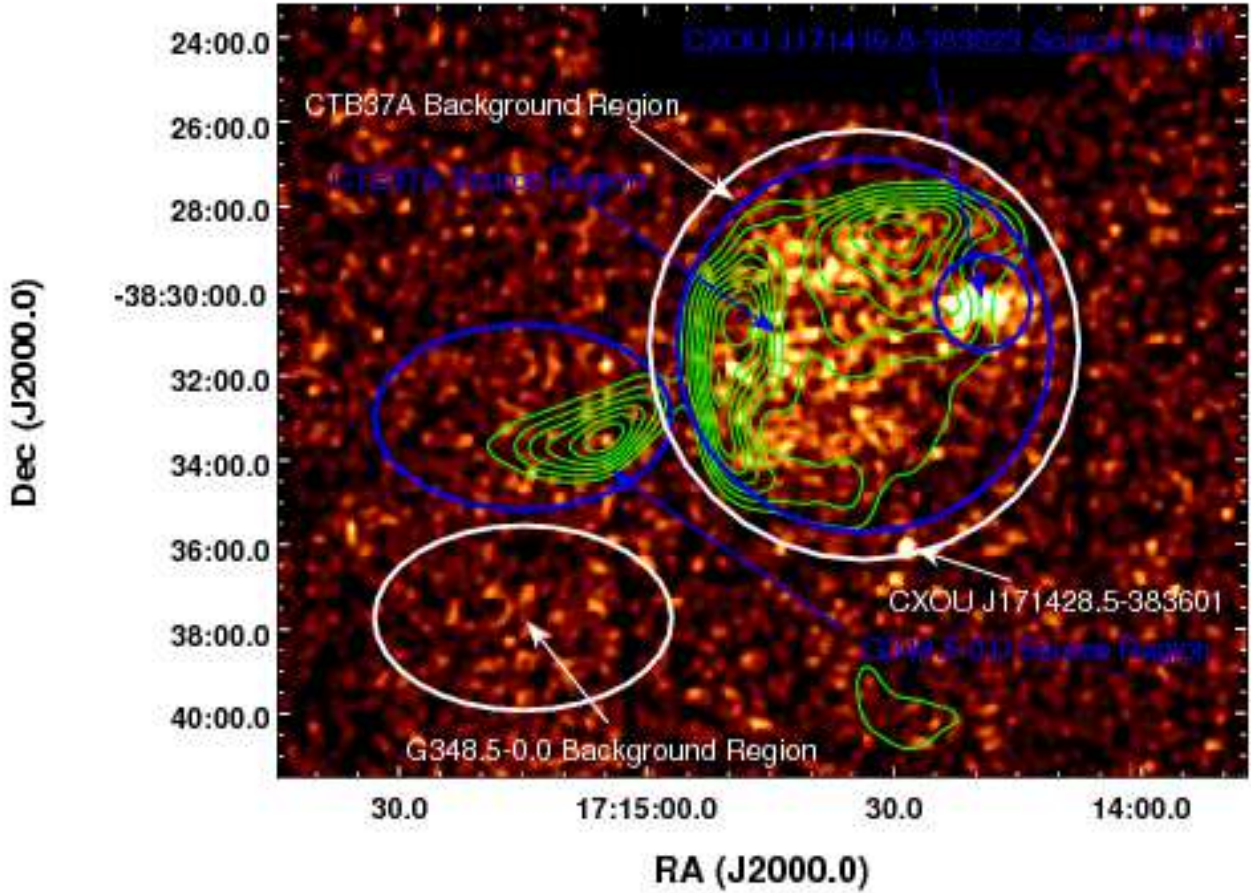}
\caption{{\it XMM-Newton} MOS1 image (in color) of CTB 37A and G348.5$-$0.0 for the energy 
range 0.7
to 10.0 keV. Radio emission (as detected with the MOST at a frequency of 843 MHz) is 
overlaid on the X-ray emission in white contours. The radio contours are at the level of 0.4, 0.6, 
0.8, 1.0, 1.2, 1.4, 1.6, 1.8 and 2.0 Jy/beam. Notice the crescent-shape filament of radio
emission located due east of CTB 37A: this filament corresponds to the separate SNR
G348.5$-$0.0. Regions of spectral extraction -- namely the entire source region for CTB 37A, the 
extended source CXOU J171419.8-383023 (associated with the northwestern rim of the SNR), the 
extended source region for G348.5$-$0.0 and the corresponding background regions -- are indicated.
See Sections \ref{CTB37AXraySubsection} and \ref{XMMG348Section}.
\label{CTB37A_MOS1figure}}
\end{figure}

\begin{figure}
\psfig{figure=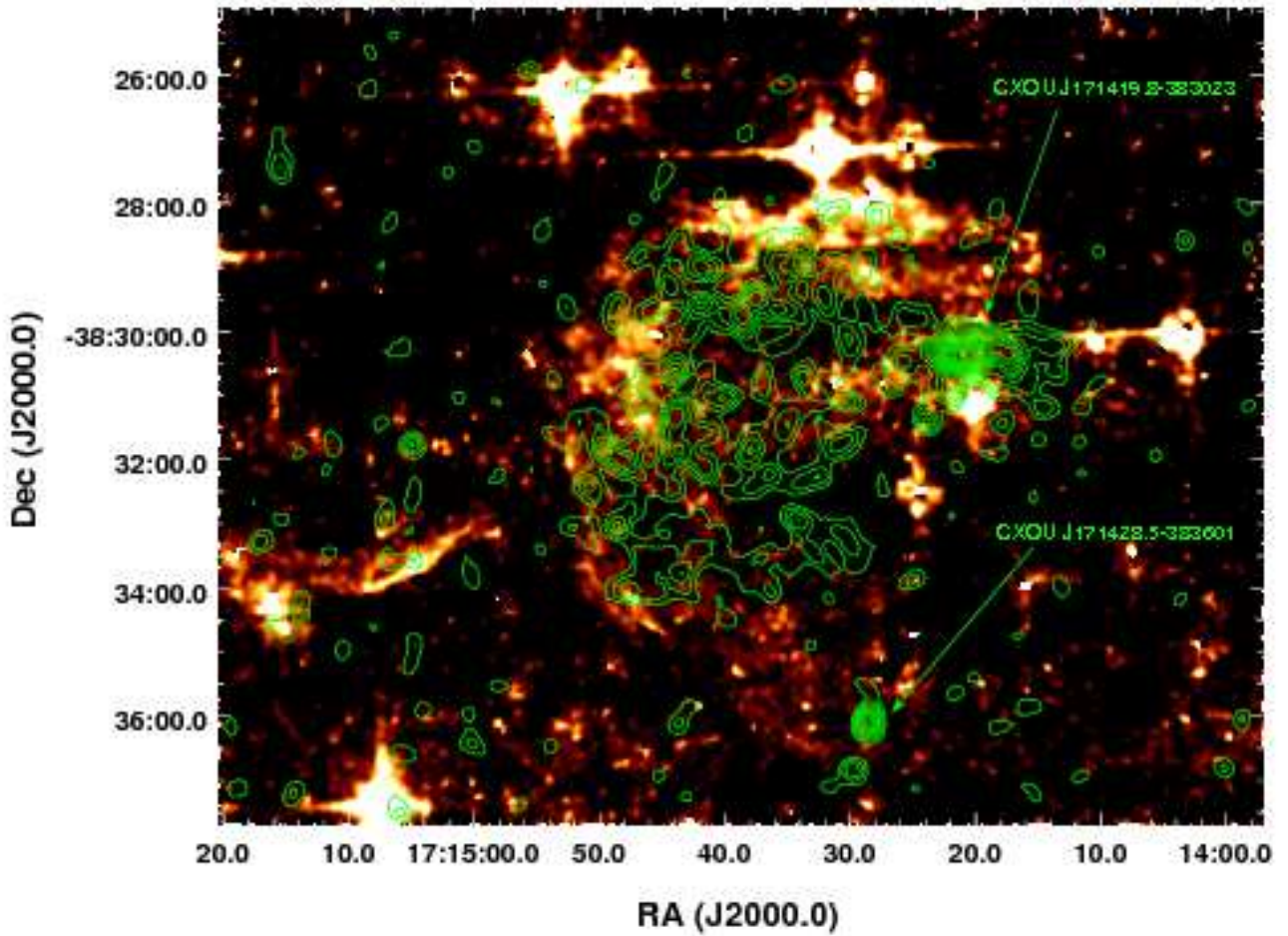}
\caption{Infrared {\it Spitzer} \label{CTB37AXMMMOSSpitzer} IRAC Channel 3 (5.8$\mu$m) color
image of CTB 37A with {\it XMM-Newton} MOS1 X-ray emission overlaid in contours. The contour 
levels are set at 0.12 to 0.80 counts per pixel (in steps of 0.04). The locations of the X-ray source 
CXOU J171419.8$-$383023 (associated with the northwestern rim of the SNR) and CXOU 
J171428.5$-$383601 (the southern discrete hard X-ray source) are both indicated. The crescent 
shape of infrared emission seen toward the east in the image is physically associated with the
SNR G348.5$-$0.0. See Sections \ref{CTB37AXraySubsection} and \ref{XMMG348Section}.}
\label{CTB37A_XMMSpitzerFigure}
\end{figure}

In Figure \ref{CTB37A_MOS1figure} we present an {\it XMM-Newton}
MOS1 image of CTB 37A over the energy range from 0.7 to 10.0 keV: contours depicting
the detected radio emission at 843 MHz as detected by MOST are overlaid.  
The X-ray emission detected by {\it XMM-Newton} is indeed extended and appears to lie 
interior to the radio shell: a similar description of the X-ray morphology of CTB 37A was
also presented by \citet{Aharonian08} and \citet{Sezer11a}. In Figure 
\ref{CTB37A_XMMSpitzerFigure} we present
a {\it Spitzer} IRAC 5.8 $\mu$m image of CTB 37A with {\it XMM-Newton} MOS1 X-ray emission
overlaid with contours. We have extracted MOS1, MOS2 and PN spectra for 
the entire diffuse emission (except for a region to the northwest which is known to feature 
a significantly harder spectrum than the diffuse emission from the rest of the SNR -- see 
the description below). This extraction region was approximately four arcminutes in diameter
and was centered on the nominal center of the observed diffuse X-ray emission. For the
purposes of spectral fitting, a background spectrum was extracted from an annular region
extending approximately 0.5 arcminutes in radius beyond the region of source extraction
and with the same center. These regions of extraction are depicted in Figure 
\ref{CTB37A_MOS1figure}. We also attempted to fit extracted MOS1, MOS2 and PN spectra
of the northwest region known to feature hard emission, but our extracted spectra lacked
a sufficient signal-to-noise ratio suitable for detailed spectral analysis. We will discuss
the properties of this source in more detail when we discuss an analysis of its {\it Chandra}
ACIS-I spectrum below.
\par
\begin{figure}
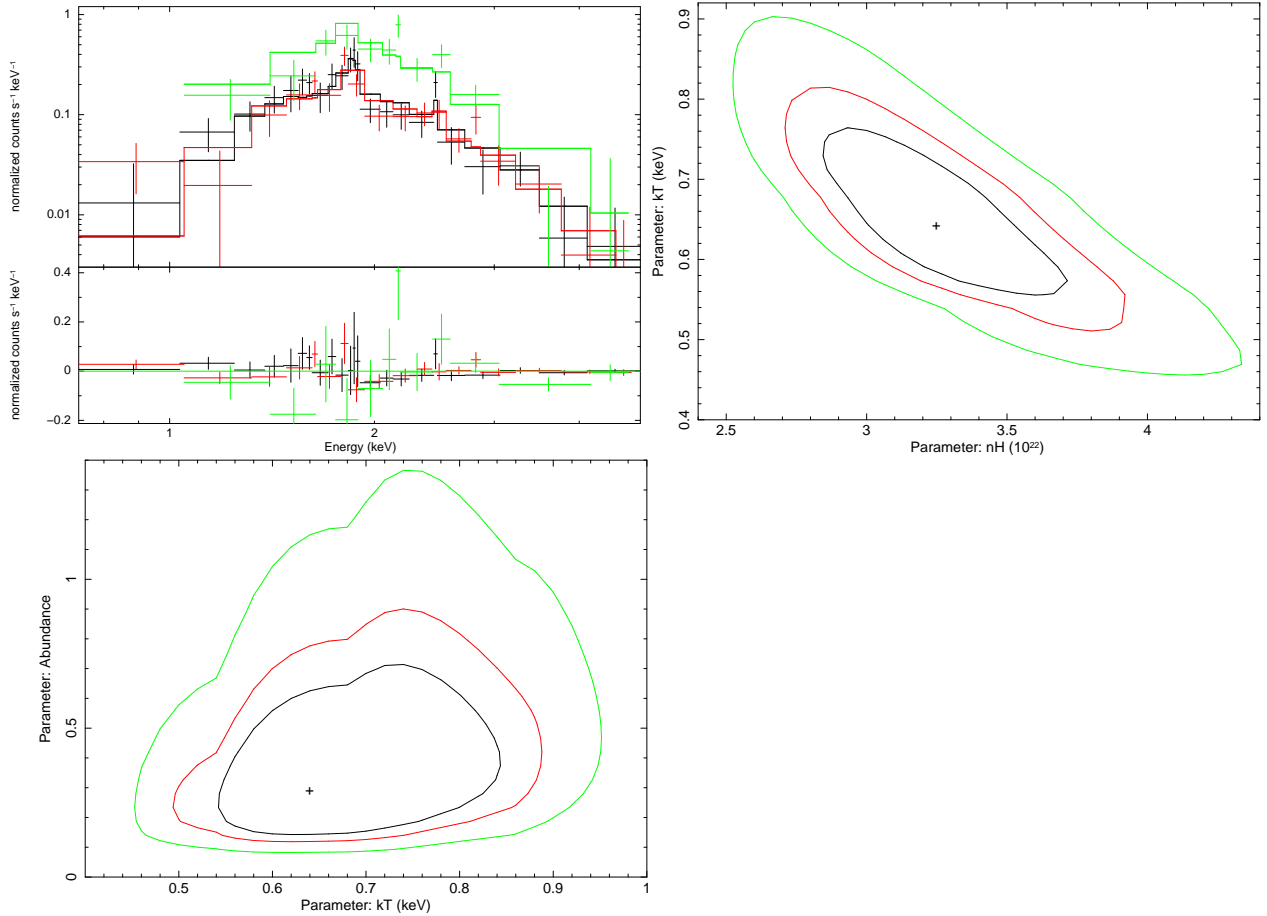

\hbox{
\psfig{figure=CTB37A_MOS1MOS2PN_PHABSAPEC_VABUND_SPECTRUM.ps,angle=270,scale=0.35}
\psfig{figure=CTB37A_MOS1MOS2PN_PHABSAPEC_VABUND_NHKT.ps,angle=270,scale=0.35}
}
\hbox{
\psfig{figure=CTB37A_MOS1MOS2PN_PHABSAPEC_VABUND_KTABUND.ps,angle=270,scale=0.35}
}
\caption{(top left) {\it XMM-Newton} MOS1, MOS2 and PN spectra of CTB 37A as fit
with a PHABS$\times$APEC model with variable abundance. (top right) Confidence contour 
plot for column density $N$$_H$ versus temperature $kT$. (bottom left) Confidence contour
plot for column density $N$$_H$ versus abundance. See Table \ref{XMMSpectraSummaryCTB37A}
and Section \ref{CTB37AXraySubsection}.\label{CTB37AXMMSpectralPlots}}
\end{figure}

The MOS1, MOS2 and PN spectra of the diffuse emission were
fitted with the APEC and NEI models: statistically acceptable fits were derived with both
models. Because the APEC model returned a statistically acceptable fit and because the
ionization timescale returned by the NEI fit was elevated ($\tau$ $\sim$ 10$^{13}$ s cm$^{-3}$
with a lower bound of $\sim$ 10$^{11}$ s cm$^{-3}$), we conclude that the X-ray emitting
plasma is close to thermal equilibrium. We also allowed the abundances of the APEC and NEI 
models to vary to
see if the quality of the fits could be improved. In fact, small improvements in the quality
of the fit (as measured by the $\chi$$^2_{\nu}$ parameter) were realized as sub-solar values for
the abundances were derived (approximately 0.3-0.4). We also tried fitting the spectra with
the VAPEC and VNEI models: only in the cases where silicon was varied were modest 
improvements in the value of $\chi$$^2_{\nu}$ attained. In those cases, the abundance of silicon 
was sub-solar (approximately
0.5-0.6). In Table \ref{XMMSpectraSummaryCTB37A} we present a summary of the
parameters for the different model fits to the MOS1, MOS2 and PN spectra of CTB 37A.
In Figure \ref{CTB37AXMMSpectralPlots} we present a plot of the MOS1, MOS2 and PN
spectra of CTB 37A as fit with the PHABS$\times$APEC model with variable abundances.
We also present in that figure confidence contour plots that depict $N$$_H$ versus $kT$
as well as $kT$ versus abundance.
\par
\begin{figure}
\psfig{figure=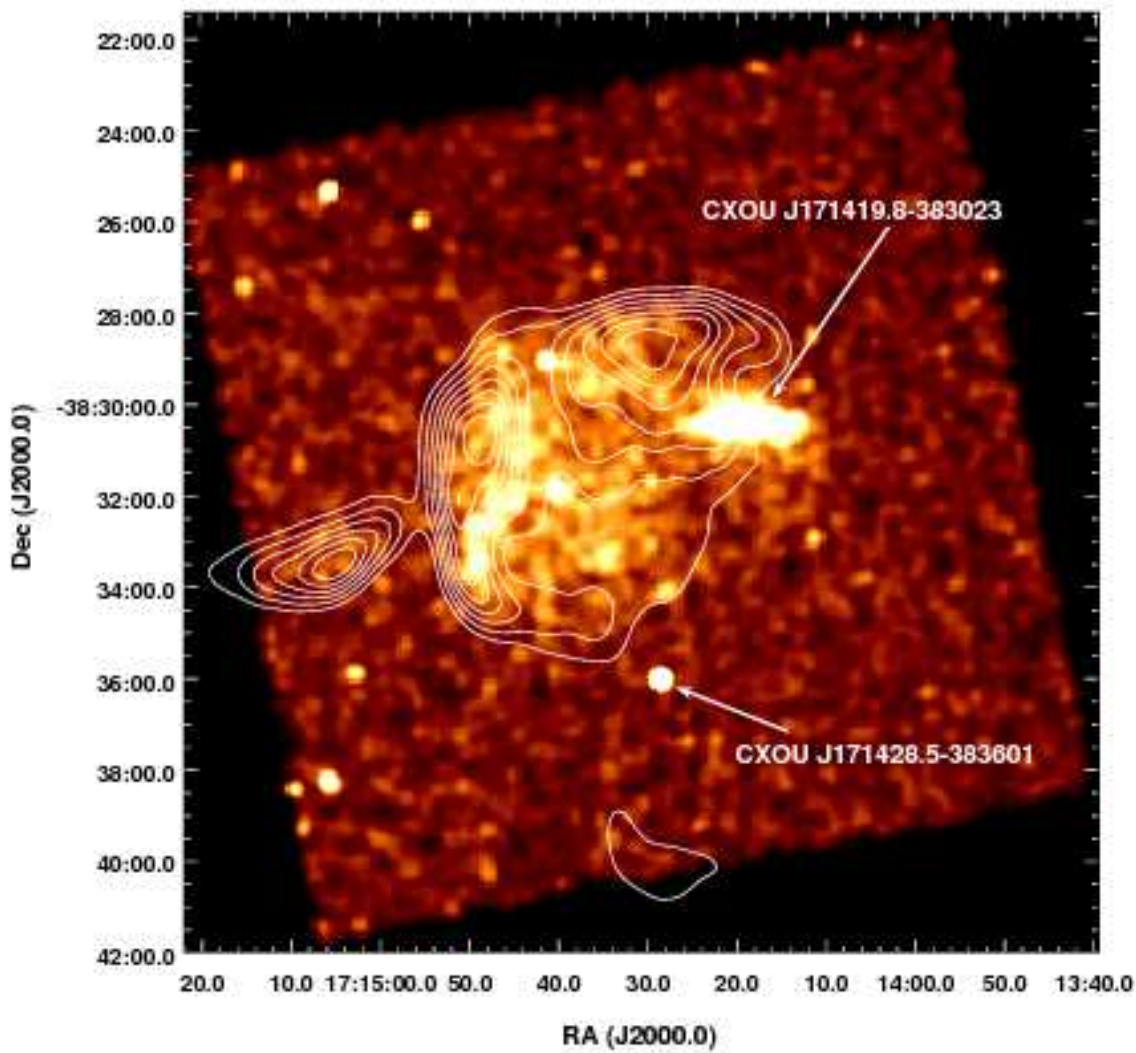}
\caption{Exposure-corrected {\it Chandra} ACIS-I image of CTB37A (in color) for the
energy range 0.5 to 10.0 keV. Radio emission (as detected with the MOST at a frequency of 843 MHz) 
is overlaid on the X-ray emission in white contours. The linear pixel range of the depicted X-ray 
emission is 0 to 4$\times$10$^{-7}$ photons cm$^{-2}$ s$^{-1}$ pixel$^{-1}$: the X-ray emission has 
also been smoothed by a Gaussian with a radius of 1.5 arcseconds. The radio contours are at the 
same levels as those depicted in Figure \ref{CTB37A_MOS1figure}. The locations of the X-ray source 
CXOU J171419.8$-$383023
(associated with the northwestern rim of the SNR) and CXOU J171428.5$-$383601 (the southern 
discrete hard X-ray source) are both indicated. Again, the crescent of radio emission seen to the
east is the SNR G348.5$-$0.0. See Section \ref{CTB37AXraySubsection}.
\label{ChandraMOSTCTB37A}} 
\end{figure}
\begin{figure}
\psfig{figure=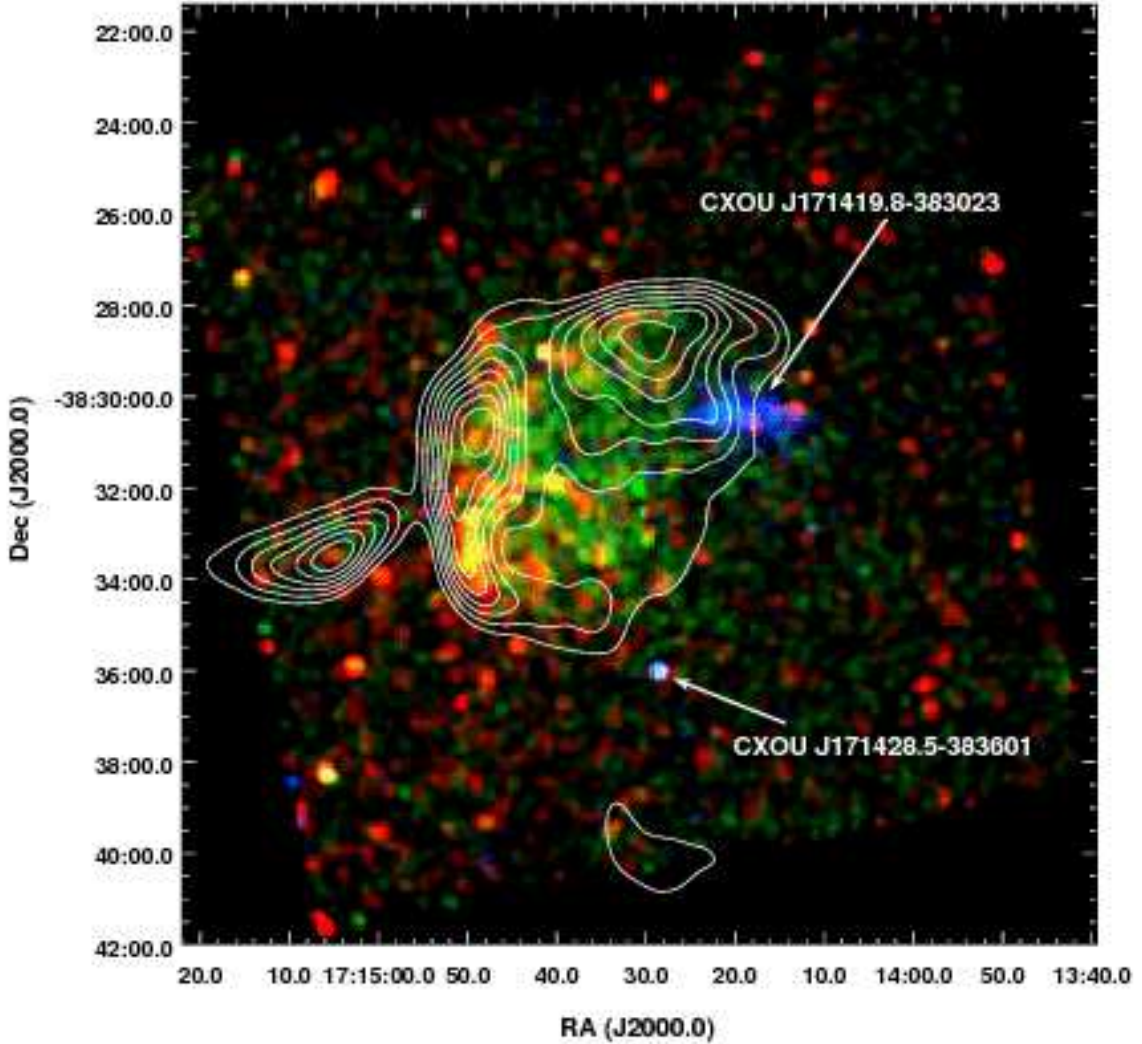}
\caption{Three-color exposure-corrected and smoothed (with a Gaussian with a radius
of 1.5 arcseconds) {\it Chandra} ACIS-I image of CTB 37A. This image has been made 
using monoenergetic exposure maps. Red, green and blue colors correspond to soft 
(0.5-1.5 keV), medium (1.5-2.5 keV) and hard (2.5-10.0 keV) emission, respectively. The 
linear pixel ranges of the depicted X-ray emission (in units of photons cm$^{-2}$ s$^{-1}$
pixel$^{-1}$) are as follows: 1$\times$10$^{-8}$ to
4$\times$10$^{-8}$ for the soft emission, 1$\times$10$^{-8}$ to 2$\times$10$^{-7}$ for the
medium emission and 1$\times$10$^{-9}$ to 5$\times$10$^{-9}$ for the hard emission. 
The white contours depict radio emission as observed with the MOST at a frequency of 843 MHz 
(the contour levels are the same as those depicted in Figure \ref{ChandraMOSTCTB37A}. The radio 
contours are at the level of 0.4, 0.6, 0.8, 1.0, 1.2, 1.4, 1.6, 1.8 and 2.0 Jy/beam. The locations of the 
X-ray source CXOU J171419.8$-$383023
(associated with the northwestern rim of the SNR) and CXOU J171428.5$-$383601 (the southern 
discrete hard X-ray source) are both indicated. Again, the crescent of radio emission seen to the
east is the SNR G348.5$-$0.0. See Section \ref{CTB37AXraySubsection}.
\label{ChandraCTB37AThreeColor}}.
\end{figure}
\begin{figure}
\psfig{figure=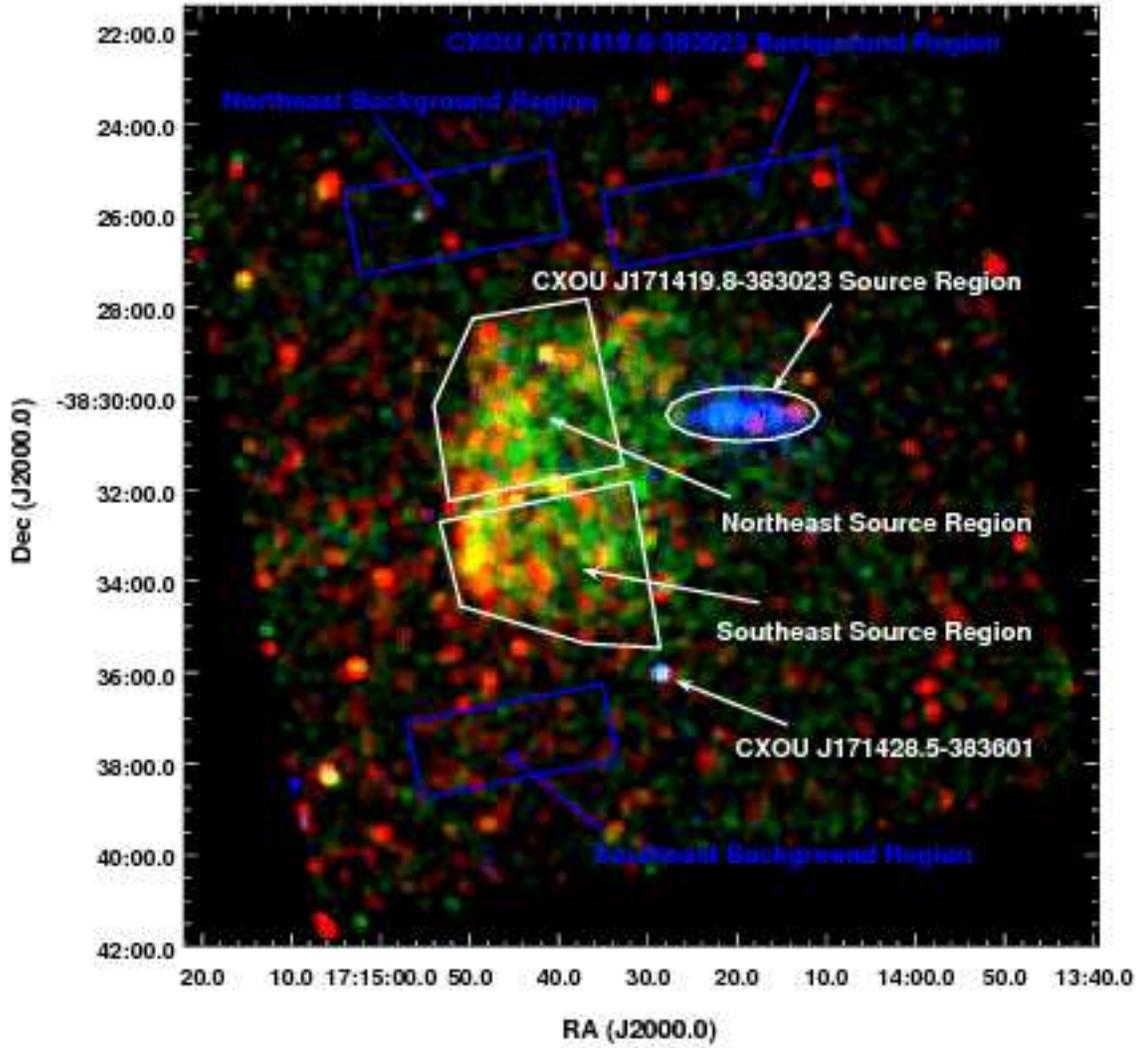}
\caption{Same as Figure \ref{ChandraCTB37AThreeColor} but with the regions of 
spectral extraction indicated (source regions are depicted in white and background
regions are depicted in blue). See Table \ref{ACISSpectraSummaryCTB37A} and Section
\ref{CTB37AXraySubsection}.\label{CTB37A_Chandra_SpecRegions}}
\end{figure}

In Figure \ref{ChandraMOSTCTB37A} we present an exposure-corrected {\it Chandra} 
ACIS-I image of CTB 37A over the energy range from 0.5 through 10.0 keV: contours 
depicting the detected radio emission at 843 MHz as detected by MOST are overlaid. Similar
to the morphology seen in Figure \ref{CTB37A_MOS1figure} where the X-ray emission
detected by the {\it XMM-Newton} MOS1 is detected, the X-ray emission from CTB 37A as
detected by the {\it Chandra} ACIS-I appears to lie interior to the well-defined radio shell: the
emission is brightest
toward the eastern half of the SNR. In Figure \ref{ChandraCTB37AThreeColor} we
present a three-color exposure-corrected {\it Chandra} ACIS-I image of CTB 37A
with the soft (0.5-1.5 keV), medium (1.5-2.5 keV) and hard (2.5-10.0 keV) emission
depicted in red, green and blue, respectively. We have extracted spectra from three
particular regions of CTB 37A: we denote these regions as the northwest region, the 
northeast region and the southeast region, respectively, and in Figure 
\ref{CTB37A_Chandra_SpecRegions} we show these regions of spectral extraction
(along with the accompanying regions where background spectra were extracted).
To the best of our knowledge, this is the first spatially-resolved spectroscopic study of the
X-ray emission from this SNR. Through this analysis (as described below), we have identified
spectral variations between the northeastern and southeastern regions of the SNR. 
\par
We first discuss our analysis of the northwest region: inspection of Figure 
\ref{CTB37A_Chandra_SpecRegions} clearly indicates that this source emits primarily
hard X-rays: we obtained an adequate fit
to this spectra using a simple power law with $\Gamma$ = 1.18$^{+0.42}_{-0.33}$ and 
$N$$_H$ = 4.08$^{+1.32}_{-0.88}$$\times$10$^{22}$ cm$^{-2}$.
These fit parameters are consistent with those obtained by \citet{Aharonian08} who
also extracted and fit the spectrum of this source (which they denoted as CXOU
J171419.8$-$383023 -- we shall use this designation for the remainder of the paper) with 
a PHABS$\times$POWER LAW model. \citet{Aharonian08} also speculated
that this hard X-ray source is associated with the $\gamma$-ray source HESS J1714$-$385
and -- regarding the true nature of this source -- considered such scenarios as the
high energy emission originating from an interaction between the SNR and the adjacent
molecular cloud complex or from a pulsar wind nebula, with the latter scenario seeming
more plausible. 
\par
\begin{figure}
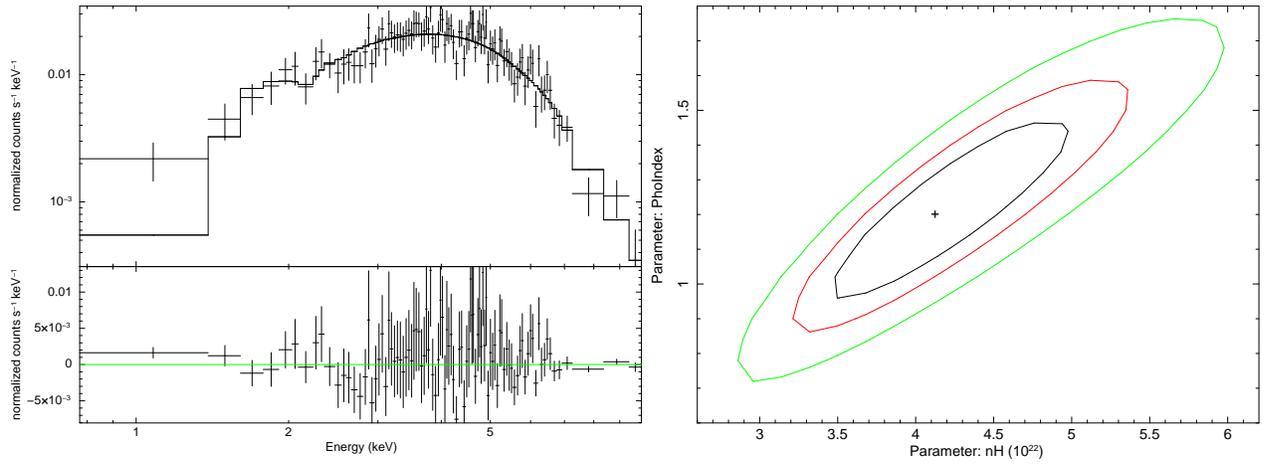

\hbox{
\psfig{figure=CTB37A_NWRIM_PHABSPOW_SPECTRUM.ps,angle=270,scale=0.35}
\psfig{figure=CTB37A_NWRIM_PHABSPOW_CONT.ps,angle=270,scale=0.35}
}
\caption{(left) {\it Chandra} ACIS-I spectrum of CXOU J171419.8-383023 as fit
with a PHABS$\times$POWER LAW model. (right) Confidence contour plot for column 
density $N$$_H$ versus photon index $\Gamma$. See Table 
\ref{ACISSpectraSummaryCTB37A} and Section \ref{CTB37AXraySubsection}. 
\label{CTB37ANWRimSpectralPlots}}
\end{figure}
\begin{figure}
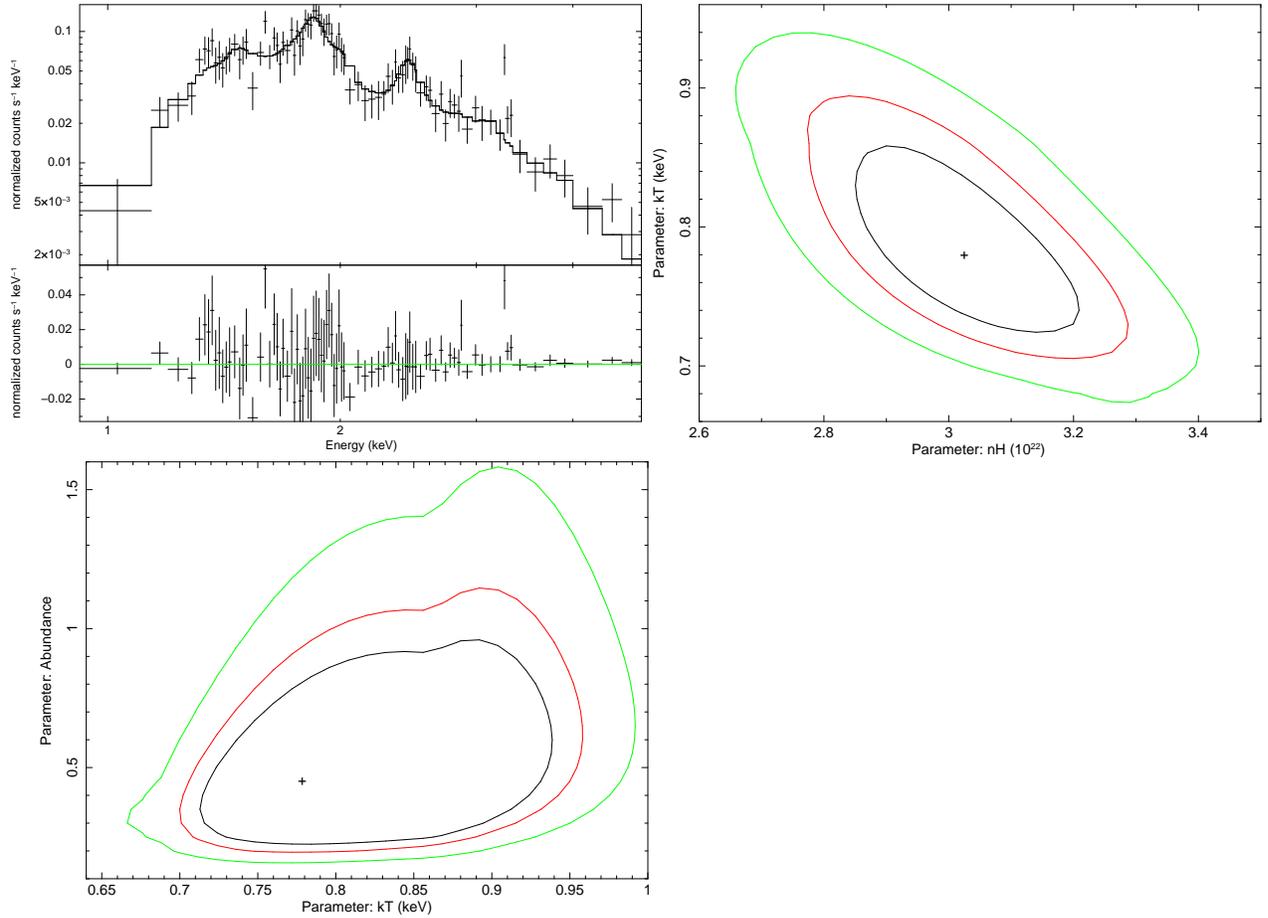

\hbox{
\psfig{figure=CTB37A_NEREGION_PHABSAPEC_VABUND_SPECTRUM.ps,angle=270,scale=0.35}
\psfig{figure=CTB37A_NEREGION_PHABSAPEC_VABUND_NHKT.ps,angle=270,scale=0.35}
}
\hbox{
\psfig{figure=CTB37A_NEREGION_PHABSAPEC_VABUND_ABUNDKT.ps,angle=270,scale=0.35}
}
\caption{(top left) {\it Chandra} ACIS-I spectrum of the northeastern region of CTB 37A as fit
with a PHABS$\times$APEC model with variable abundance. (top right) Confidence contour 
plot for column density $N$$_H$ versus temperature $kT$. See Table 
\ref{ACISSpectraSummaryCTB37A} and Section \ref{CTB37AXraySubsection}.
\label{CTB37ANERegionSpectralPlots}}
\end{figure}
\begin{figure}
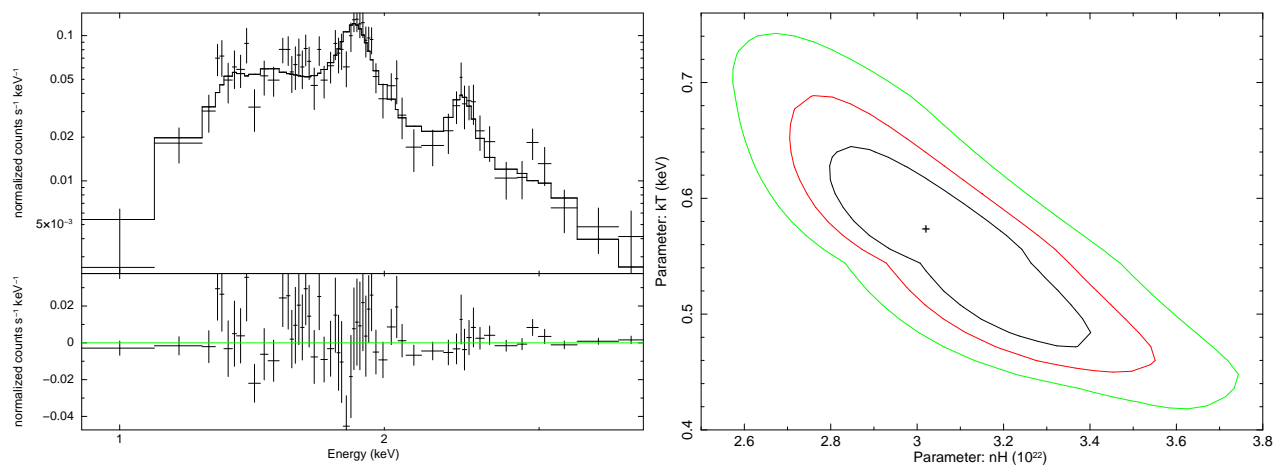

\hbox{
\psfig{figure=CTB37A_SEREGION_PHABSAPEC_SPECTRUM.ps,angle=270,scale=0.35}
\psfig{figure=CTB37A_SEREGION_PHABSAPEC_CONT_NHKT.ps,angle=270,scale=0.35}
}
\caption{(left) {\it Chandra} ACIS-I spectrum of the southeastern region of CTB 37A as
fit with a PHABS$\times$APEC model. (right) Confidence contour plot for column 
density $N$$_H$ versus temperature $kT$. See Table 
\ref{ACISSpectraSummaryCTB37A} and Section \ref{CTB37AXraySubsection}.
\label{CTB37ASERimSpectralPlots}}
\end{figure}

In the cases of the spectra of the northeast and southeast regions, we
obtained acceptable fits using the thermal models APEC and NEI. Statistically-acceptable
fits were obtained with both models: because acceptable fits were obtained with the
APEC model and because the ionization timescales derived by the NEI models were
high ($\tau$ $\sim$ 10$^{11}$ s cm$^{-3}$ with lower limits of 10$^{10}$ s cm$^{-3}$ or higher),
we conclude that the X-ray emitting plasma in both regions is close to ionization
equilibrium. The fitted column densities for both regions are comparable ($N$$_H$ $\sim$
3$\times$10$^{22}$ cm$^{-2}$). This fitted value is lower than the column density derived toward
CXOU J171419.8$-$383023 but -- within the error bounds -- the estimated column densities for all 
three regions are comparable. Interestingly, a difference in temperatures is seen, with the
fitted temperatures of the northeast region ($kT$ = 0.55-0.57 keV) being noticeably higher
than the southeast region ($kT$ = 0.78-0.83 keV). Again, compared to the fit to the extended
thermal emission from CTB 37A as derived by \citet{Aharonian08}, our derived column densities
for both regions are comparable and our derived temperature for the northeast region is quite
similar but our temperature for the southeast region is noticeably lower. We also find differences 
in the fitted values for the abundances of the two regions: while superior fits to the spectra of 
the northeast region are obtained with both the APEC and NEI models when the value of the
abundance is sub-solar ($\sim$0.4), the derived values of the abundances obtained from the 
fits to the spectra of the southeast region with both the APEC and NEI models are consistent 
with a solar abundance.
The detection of a reduced value of the abundance is consistent with the result seen previously
for the extracted {\it XMM-Newton} spectra of CTB 37A.
In Table \ref{ACISSpectraSummaryCTB37A} we present a summary of the fit parameters for 
the spectra of these three regions. In Figure \ref{CTB37ANWRimSpectralPlots} we present 
the extracted {\it Chandra} ACIS-I spectrum of CXOU J171418.8$-$383023 as fit with the 
PHABS$\times$POWER LAW
model and a confidence contour plot for this fit for the parameters $N$$_H$ and $\Gamma$.  
In Figure \ref{CTB37ANERegionSpectralPlots} we present the extracted {\it Chandra} ACIS-I
spectrum of the northeast region as fit with the PHABS$\times$APEC model with a thawed
abundance as well as confidence contour plots for the parameters $N$$_H$ and $kT$ as
well as abundance and $kT$. Lastly, in Figure \ref{CTB37ASERimSpectralPlots} we present the
extracted {\it Chandra} ACIS-I spectrum of the southeast region as fitted with the 
PHABS$\times$APEC model as well as a confidence contour plot for the parameters $N$$_H$ and 
$kT$.
\par
\begin{figure}
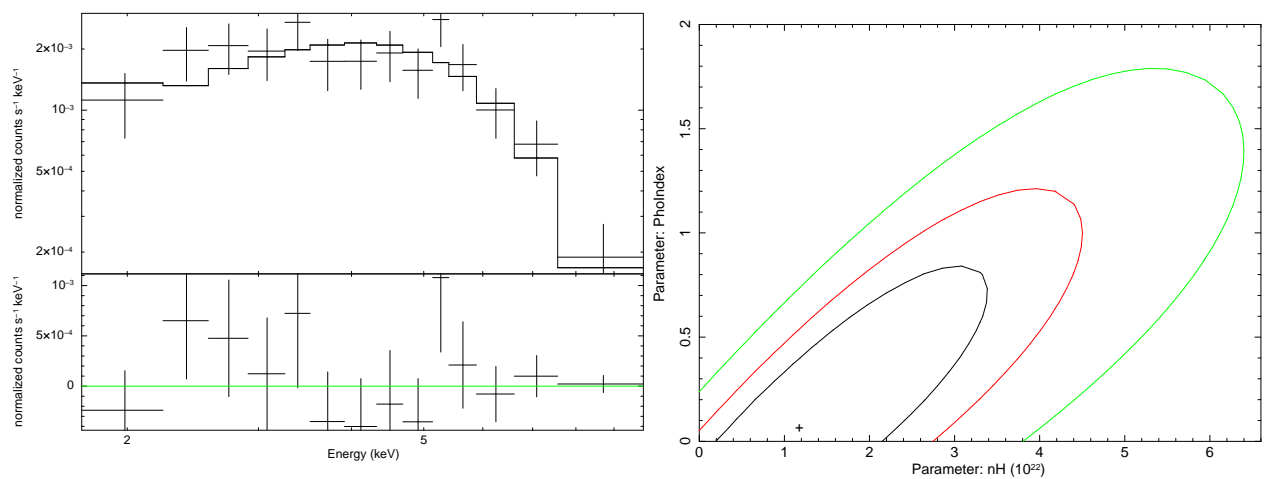

\hbox{
\psfig{figure=CTB37A_SOUTHERNHARDSOURCE_SPECTRUM.ps,angle=270,scale=0.35}
\psfig{figure=CTB37A_SOUTHERNHARDSOURCE_CONTOURS.ps,angle=270,scale=0.35}
}
\caption{(left) Extracted ACIS-I spectrum of the hard source CXOU J171428.5$-$383601 -- a
discrete source seen toward the southern edge of the angular extent of CTB 37A -- as fitted with
a PHABS$\times$Power Law model. (right) Confidence contour plot for the fit. See Section 
\ref{CTB37AXraySubsection}. \label{CTB37AHardSourcePlots} }
\end{figure}

We comment on the discrete X-ray source located at RA (J2000.0) 17$^h$ 14$^m$ 28.6$^s$, 
Dec (J2000.0) $-$38$^{\circ}$ 36' 01'': it is seen in both the {\it XMM} MOS1 and the 
{\it Chandra} ACIS-I images of  Figure \ref{ChandraMOSTCTB37A}. In the latter image, the source
appears to be blue, indicative of a hard spectrum for this source: it was also detected by
\citet{Aharonian08} and cataloged by those authors as CXOU J171428.5$-$383601. We will adopt 
that designation for the source for the remainder of the paper. It is possible that this source
may be the neutron star produced by the supernova explosion that created the SNR CTB 37A
itself. To investigate this possibility, we extracted an ACIS-I spectrum for this source and fitted it with a
power law coupled with the photoelectric absorption model PHABS: we obtained a 
statistically-acceptable fit ($\chi$$^2_{\nu}$ = 0.73) using this model.
The photon index of the fit is flat ($\Gamma$ $<$ 1.3) -- consistent with a spectral index expected
for a pulsar \citep{Weisskopf07} -- and the fitted column density ($N$$_H$ $<$ 
4.6$\times$10$^{22}$ cm$^{-2}$)
is consistent with the values for $N$$_H$ derived 
from thermal fits to extracted spectra for portions of the X-ray emitting plasma of CTB 37A. 
We present in Figure \ref{CTB37AHardSourcePlots} the extracted spectrum for CXOU 
J171428.5$-$383601 and the confidence contours for the PHABS$\times$Power Law fit
described here: it is possible that this source is the neutron star associated with CTB 37A but
additional observations are required to determine its true identity.
\par
\begin{figure*}
\vbox{
\epsfig{figure=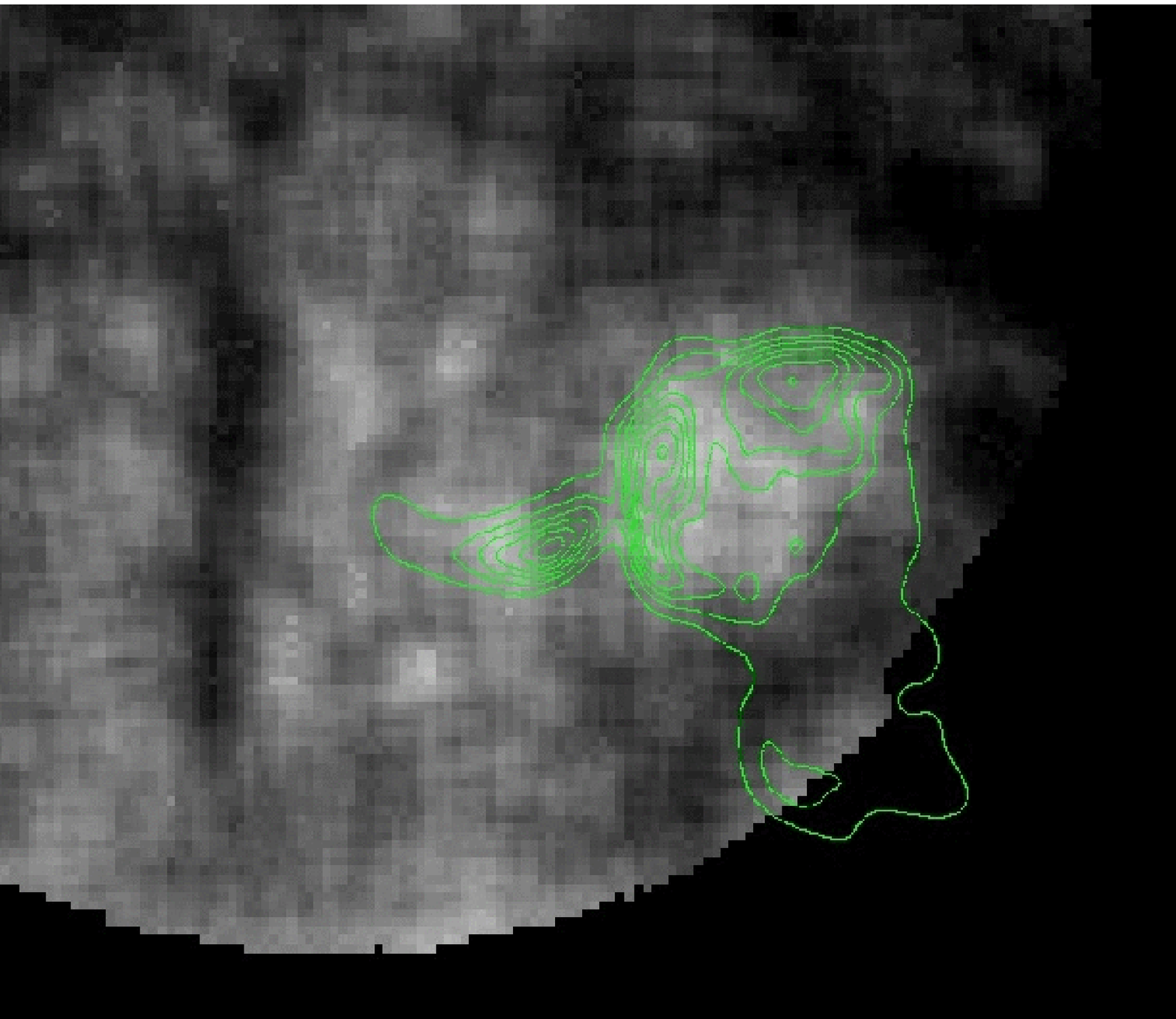,height=8truecm}
\epsfig{figure=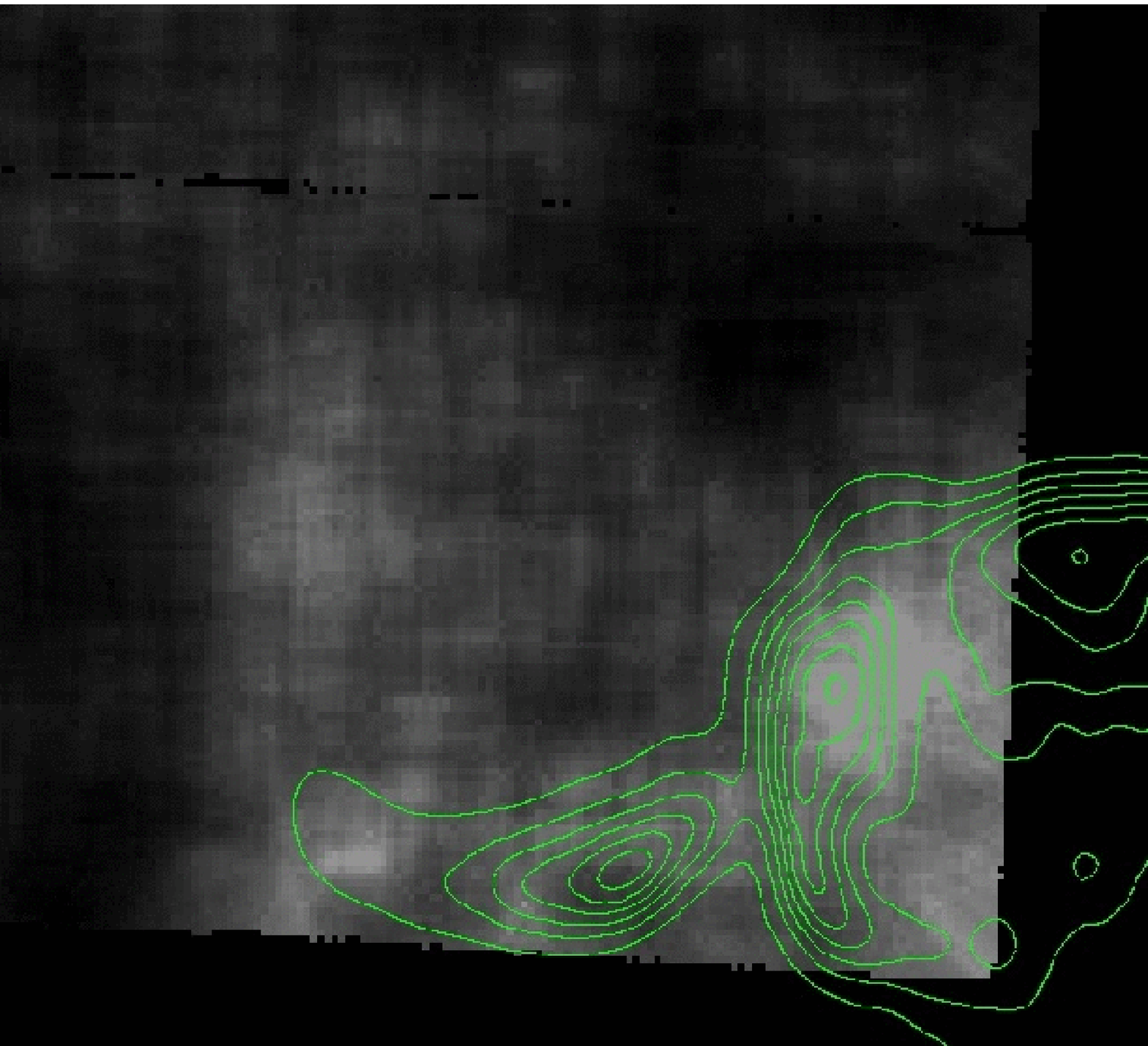,height=8truecm}
}
\caption{{\it ASCA} GIS (left) and SIS (right) images of CTB 37A and \gtfe\ with radio contours overlaid.
The circular radio structure seen to the west is CTB 37A (which is visible in X-ray and radio) while the 
radio filament extending eastward (which lacks an X-ray counterpart) is G348.5$-$0.0. See
Section \ref{CTB37AXraySubsection}.\label{CTB37AASCAimage}}
\end{figure*}

For the purposes of completeness, in Figure \ref{CTB37AASCAimage} we present
{\it ASCA} GIS and SIS images of CTB 37A and the adjacent SNR G348.0$-$0.0 (the latter object
is seen as a radio filament in these images and is discussed in more detail in Section 
\ref{XMMG348Section}). The image shows some hints of emission from both of these SNRs, but
because the signal-to-noise ratios of these images are poor, we have instead concentrated on
the more sensitive {\it XMM-Newton} and {\it Chandra} observations of these SNRs.

\subsubsection{Discussion of Physical Properties of CTB 37A\label{CTB37ADiscussSubSection}}

Previous X-ray spectral analyses of CTB 37A have been published in the literature based on
data from {\it ASCA} GIS observations \citep{Yamauchi08}, {\it Chandra} and {\it XMM}
\citep{Aharonian08} and finally {\it Suzaku} \citep{Sezer11a}. In the case of the {\it ASCA} 
observations, \citet{Yamauchi08} fit the extracted combined GIS spectrum (where the authors had
merged the GIS2 and GIS3 spectra to boost the signal-to-noise) with a WABS$\times$MEKAL model
with elemental abundances frozen to solar, a WABS$\times$MEKAL model with silicon abundance 
thawed and other elemental abundances thawed and a WABS$\times$Power Law model. 
Because the signal-to-noise of the fitted spectrum was poor, the fit parameters were poorly 
constrained: fitted values for $N$$_H$, $kT$ and $\Gamma$ ranged from 1.7$\times$10$^{22}$ 
to 2.4$\times$10$^{22}$, 2.2 keV to 2.8 keV and 2.4 to 3.6, respectively. We attribute the differences
between these fit results and the fit results presented in this paper to the poor signal-to-noise of
the merged GIS spectrum and significant contaminating flux from bright sources located outside
the field of view. \citet{Aharonian08} presented spectral fits to the extracted {\it Chandra} and 
{\it XMM-Newton} spectra of the entire diffuse emission associated with CTB 37A.
Those authors derived a column density $N$$_H$ = 3.15$^{+0.13}_{-0.12}$ $\times$ 10$^{22}$
cm$^{-2}$ and a temperature $kT$ = 0.81$\pm$0.04 keV. We find that while our derived
value for $N$$_H$ ($\sim$3$\times$10$^{22}$ cm$^{-2}$) is comparable to the value derived
by \citet{Aharonian08}, our derived values for the temperature ($kT$ $\sim$ 0.5-0.7 keV) are
slightly lower. We argue that our fit values are broadly consistent with the results published
by \citet{Aharonian08}: we also note that \citet{Yamauchi08} applied the fit parameters derived by
\citet{Aharonian08} to their extracted merged GIS spectrum and were able to obtain a 
statistically-acceptable fit.
\par
\citet{Sezer11a} presented a spectral analysis of the whole X-ray emission detected 
from CTB 37A using data obtained from a pointed observation made of the SNR by {\it Suzaku}. 
Those authors fit the extracted XIS spectra using a WABS$\times$(VMEKAL+Power Law) model:
their derived fit parameters were $N$$_H$$\sim$3$\times$10$^{22}$ cm$^{-2}$, $kT$$\sim$0.6 keV
and $\Gamma$=1.6: they obtained a statistically-acceptable fit with all elemental abundances frozen 
to solar values. We have fit our extracted {\it XMM-Newton} spectra with this model and present 
the results of this fit in Table \ref{XMMSpectraSummaryCTB37A}: we obtain a statistically-acceptable
fit (with a reduced $\chi$$^2$ value of 1.13) using the parameters derived by \citet{Sezer11a}. 
Regarding the additional power law component present in the \citet{Sezer11a} fit: we do not require
such a component to adequately fit the MOS1+MOS2+PN spectrum of CTB 37A. We attribute the
differences between the spectral fit that we derived and the spectral fit derived by \citet{Sezer11a}
to different selections of regions to extract background spectra. \citet{Sezer11a} chose a background
region which had the lowest X-ray flux available (see Figure 1 in their paper), not from a region 
accurately reflecting the Galactic X-ray background behind the SNR. We argue that our choice
of region for extraction of a background spectrum is more appropriate. 

\subsection{\gtfe\label{XMMG348Section}}

G348.5$-$0.0 was first identified as a ``jet"-like feature extending from the 
eastern edge of the bright radio rim of CTB 37A in early radio observations made by 
\citet{Milne79}. While those authors favored the scenario where the jet
was a radio feature generated by CTB 37A, subsequent work by \citet{Downes84}
instead argued that the jet was a separate SNR seen in projection toward CTB 37A.
The classification of G348.5$-$0.0 as a distinct SNR was confirmed by \citet{Kassim91}
based on radio observations made at the frequencies of 333 MHz, 1443 MHz and
4835 MHz. The high angular resolution radio maps provided by those authors of
G348.5$-$0.0 reveal that the morphology of the SNR may be described as a concave 
rim pointed downward with an angular extent of approximately 3.5 arcminutes. A similar
morphology for the SNR is described by \citet{Whiteoak96} based on their observations
made with the MOST of this SNR at a frequency of 843 MHz.  
\citet{Kassim91} estimated the flux density and the spectral index of the SNR to be 
6.2 Jy at 1443 MHz and $-$0.4$\pm$0.1, respectively: however, those authors stressed that their
estimated flux density at 1443 MHz (and at the other frequencies as well) must be interpreted
as a lower limit because observations of this source at frequencies greater than 1000 MHz
(including their own observation at 4835 MHz) resolve out much of the flux density of the
source. For the same reason, the calculated spectral index should be treated as a lower
limit as well (that is, $\alpha$$>$$-$0.4$\pm$0.1). 
\citet{Hewitt08} described observations made with the Green Bank Telescope (GBT) 
of a sample of Galactic SNRs (including G348.5$-$0.0) at the frequencies corresponding
to all four ground-state transitions of the OH radical (namely the frequencies of 1612.231,
1665.4018, 1667.359 and 1720.530 MHz). Those authors identified two masers 
that were spatially coincident with the SNR. \citet{Reynoso00} favored a distance of 13.5 kpc
to this SNR based on a putative association between a weak CO concentration observed
toward the west of the radio rim. However, \citet{Tian12} instead argued for an upper limit
on the distance to G348.5$-$0.0 of 6.3 kpc based on the HI absorption profile observed
along the line of sight toward this SNR. We therefore adopt a distance of 6.3 kpc to G348.5$-$0.0
for the remainder of this paper. 
\par
A clear detection of X-ray emission from \gtfe~(in particular, from the luminous southern
radio rim) has proven to be challenging with the datasets which are currently available 
for this SNR. In Figure \ref{CTB37AASCAimage} we present {\it ASCA} GIS and SIS images of 
G348.5$-$0.0: emission from this SNR is not clearly seen in either of these images. As noted
previously, \citet{Yamauchi08} stated that the {\it ASCA} datasets are badly confused by 
contaminating stray flux originating from a bright source located just south of the field of view.
For this reason, we were not able to detect clearly X-ray emission from the radio rim using the 
{\it ASCA} GIS data. While the pointed {\it Chandra} observation of CTB 37A which was conducted 
with the ACIS-I array also included the radio rim of \gtfe, unfortunately much of this rim fell into a 
gap between chips in the ACIS-I array (see Figures \ref{ChandraCTB37AThreeColor} and 
\ref{CTB37A_Chandra_SpecRegions}). This situation -- coupled with the short exposure time of the 
observation -- prevented the clear detection of X-ray emission from the radio-luminous limb of this 
SNR as well. 
\par
The pointed {\it XMM-Newton} observation made of CTB 37A also sampled G348.5$-$0.0
and we were able to extract MOS1+MOS2+PN spectra for the luminous radio rim
(though the short exposure time of the observation and the contaminating stray flux
severely limited the quality of the data). An elliptical region with radii
of 3.52 arcminutes $\times$ 2.19 arcminutes 
was used to extract X-ray spectra while 
a region with the same linear dimensions and position angle located approximately 5 arcminutes 
to the south was used to extract a background spectrum. The locations of these extraction
regions are shown in Figure \ref{CTB37A_MOS1figure}: we also note that in
Figure \ref{CTB37A_XMMSpitzerFigure} we depict X-ray emission as detected by MOS1 overlaid
in contours on a $\it{Spitzer}$ IRAC 5.8$\mu$m of a field containing both CTB 37A and
G348.5$-$0.0 (infrared emission from the latter SNR is visible as the concave-downward arc
visible at the eastern edge of the image). After background subtraction,
we estimate the total number of counts in the extracted MOS1, MOS2 and PN spectra of
G348.5$-$0.0 over the energy range of 0.5 to 10.0 keV to be
68, 26 and 31, respectively. These numbers of counts are unfortunately not
sufficient for a sophisticated spectral analysis. 
We placed an upper limit on X-ray flux and luminosity
of G348.5$-$0.0 over this energy range using the following method. We used the standard
program COLDEN\footnote{COLDEN is a standard tool in the CIAO software package that may
be accessed on-line at http://cxc.harvard.edu/toolkit/colden.jsp.} to determine that the 
nominal Galactic column density $N$$_H$ seen toward this SNR is 
$\sim$1.7$\times$10$^{22}$ cm$^{-2}$. Adopting half of this value as the column seen
toward G348.5$-$0.0 (because the value returned by COLDEN corresponds to the value
seen all the way through the Galactic disk) and assuming a Raymond-Smith thermal plasma 
model \citep{Raymond77} with solar abundances and a temperature of $kT$ $\sim$ 0.5 keV, 
we used the 
Portable, Interactive Multi-Mission Simulator (PIMMS\footnote{PIMMS may be accessed
on-line at http://cxc.harvard.edu/toolkit/pimms.jsp}) Version 4.2 to estimate a corresponding 
unabsorbed flux of $\sim$6.9$\times$10$^{-13}$ ergs cm$^{-2}$ sec$^{-1}$ for G348.5$-$0.0. 
Assuming a distance to G348.5$-$0.0 of 6.3 kpc, we estimate an upper limit for the unabsorbed
X-ray luminosity of the luminous radio rim to be 
$L$$_X$ $\sim$3$\times$10$^{33}$ ergs sec$^{-1}$ (over the energy range of 0.5-10.0 keV). The
true X-ray luminosity of G348.5$-$0.0 may be far below this value. Unfortunately, we still have not
clearly detected X-ray emission from this SNR: new X-ray observations conducted with
greater sensitivity may be more fruitful in detecting such emission. 

\section{Plasma Conditions of the SNRs}
\label{PlasmaSection}

We now discuss the properties of each SNR in turn. We  
derive and comment upon such physical properties of the SNRs as age ($t$), hydrogen nuclei density 
$n$$_H$, electron 
density $n$$_e$ (we will assume that $n$$_e$ = 1.2 $n$$_H$), pressure $P$ and swept-up 
X-ray-emitting mass $M$$_X$ for each SNR. We pattern the discussion in the present paper with 
the discussion we presented in a previous
work \citep{Pannuti10} where we derived and commented upon these properties for two other
Galactic MM SNRs, HB 21 (G89.0$+$4.7) and CTB 1 (G116.9$+$0.2). We use the following equations
for our analysis. Values for $n$$_e$ and $n$$_H$ may be calculated from the value obtained for the
normalization of thermal components in fits to the extracted X-ray spectra of the SNRs based 
on the equation
\begin{equation}
\mbox{Normalization}~(\mbox{cm$^{-5}$}) = \frac{10^{-14}}{4 \pi d^2} \int n_e n_H~dV, 
\label{NormEqn}
\end{equation}
where $d$ is the distance to the SNR in cm and $V$ = $\int$ $d$$V$ is the total volume of the SNR 
(we
will adopt uniform densities for $n$$_e$ and $n$$_H$ -- expressed here in cm$^{-3}$ -- and spherical 
volumes for the SNRs). We estimated the remnant age t$_{rad}$ assuming the SNRs are in the 
snowplow 
stage of evolution \citep[see][]{Rho02} using the following relation 
\begin{equation}
t_{rad} \mbox{(10$^4$ yr)} = 0.3~\mbox{$R$/$V$$_s$} = 30~\mbox{$R$$_s$(pc)/$V$$_s$(km/s)},
\label{RadAgeEqn}
\end{equation}
where $V$$_s$ is the shock velocity of the SNR. Values of $V$$_s$ were measured from ionic lines
detected in the infrared spectra by \citet{Hewitt09} and \citet{Andersen11} of the five SNRs 
considered in this paper (all of these SNRs are in the radiative phase of evolution). 
We also consider the ratio $P$/$k$ (where $k$ is Boltzmann's constant) and calculate this ratio 
using the equation 

\begin{equation}
P/k~\mbox{(K cm$^{-3}$)} = 2~n_e T
\label{PressureEqn}
\end{equation}

Lastly $M$$_X$ is calculated from the equation

\begin{equation}
M_X = f m_H n_H V 
\label{MassEqn}
\end{equation}

where $f$ is the volume filling factor of the X-ray emitting plasma (assumed to be unity in this
paper) and $m$$_H$ is the mass of a hydrogen nucleus. We discuss these calculated properties
for each SNR below: in Table \ref{ComparisonThreeSNRsTable} we present our calculated values for 
each property in the cases of the SNRs Kes 17, G346.6$-$0.2 and CTB 37A. For comparison, we 
include other published values for these same properties of these SNRs (we compare our results with 
these other publications in the text below). In Table \ref{SNRXraysTable} we present a summary of the 
properties 
of the diffuse emission of the SNRs considered in this paper (including the values of $V$$_s$ used in 
our calculations). We will discuss Table \ref{SNRXraysTable} in more detail in Section 
\ref{ComparisonSection}. 
\par 
{\it Kes 17:} Based on the derived parameters from our X-ray spectral fits (namely the 
PHABS$\times$APEC fits to the extracted MOS1+MOS2+PN spectra of Kes 17), we have applied 
Equations \ref{NormEqn}, \ref{RadAgeEqn},  
\ref{PressureEqn} and \ref{MassEqn} to calculate an electron density of the X-ray-emitting 
plasma to be $n$$_e$ = 0.42 cm$^{-3}$, a hydrogen density $n$$_H$ = 0.35 cm$^{-3}$, an age 
$t$$_{rad}$=2.3$\times$10$^4$,
a pressure $P$/$k$ = 
5.6$\times$10$^6$ K cm$^{-3}$ and a swept-up mass $M$$_X$ of 52 $M$$_{\odot}$. 
\par
{\it G311.5$-$0.3:} Through inspection of our spectral fit values and using the equations 
above, we calculate the following values for the properties of G311.5$-$0.3: $n$$_e$=0.20 cm$^{-3}$, 
$n$$_H$=0.17 cm$^{-3}$, $t$$_{rad}$=2.5-4.2$\times$10$^4$ yr,
$P$/$k$ = 4.4$\times$10$^6$ K and $M$$_X$ = 13 $M$$_{\odot}$.
\par
{\it G346.6$-$0.2:}
Using the values of the fit parameters that we derived for the GIS spectra of G346.6$-$0.0 using
the PHABS$\times$(NEI + Power Law) model (specifically where we froze the values of $\tau$
= 10$^{11}$ cm$^{-3}$ s and $\Gamma$=0.5), we calculate $n$$_e$ = 0.24 cm$^{-3}$, 
$n$$_H$ = 0.20 cm$^{-3}$, $t$$_{rad}$ = 4.7$\times$10$^4$ yr, 
$P$/$k$ = 
1.7$\times$10$^7$ K cm$^{-3}$ and $M$$_X$ = 84 $M$$_{\odot}$.  
We note that this is the only SNR considered in this paper which 
features X-ray spectra that indicate that the SNR is far from ionization equilibrium. 
\par
{\it CTB 37A:} Based on the fit parameters derived through analysis of the extracted 
MOS1+MOS2+PN spectra for the whole SNR,
we estimate the following properties for CTB 37A: $n$$_e$ = 0.77 cm$^{-3}$, $n$$_H$ = 0.64 
cm$^{-3}$, $t$$_{rad}$ = 3.2-4.2$\times$10$^4$ yr, $P$/$k$ =8.2$\times$10$^6$ K cm$^{-3}$ and 
$M$$_X$ = 76 $M$$_{\odot}$. All of these parameters are calculated based on spectral analysis of
the X-ray bright portion of the northern half of the SNR.  We emphasize
that the angular extent of this SNR as defined by the extended radio
emission includes a southwestern portion (a possible ``blow-out" region): no X-ray emission has
been detected from this southwestern portion. This region may in fact feature X-ray emission which
has remained undetected up to this point by pointed X-ray observations of the SNR. We therefore
view our estimate of the X-ray emitting mass associated with this SNR as a lower limit.

\section{Are the X-ray Emitting Plasmas of These SNRs Overionized?}
\label{OverionizationSection}

The phenomenon of overionization has been observed in the X-ray-emitting plasmas
associated with numerous Galactic MM SNRs (as well as other Galactic SNRs). Overionization
occurs when the electron temperature ($T$$_e$) is lower than the ionization temperature
($T$$_i$) of the plasma: in such cases, the recombination process is dominant. Examples of
MM SNRs that feature overionized X-ray-emitting plasmas are IC 443 (based on observations
made by {\it ASCA}, {\it XMM-Newton} and {\it Suzaku} -- see \citet{Kawasaki02}, \citet{Troja08}
and \citet{Yamaguchi09}), W28, W44, W49B and G359.1$-$0.5 (all based on observations made by
{\it Suzaku} -- see \citet{Sawada12}, \citet{Uchida12}, \citet{Ozawa09}, \citet{Ohnishi11}, respectively).
Reviews of the phenomenon of overionized X-ray-emitting plasmas associated with
SNRs is provided by \citet{Vink12}.

\par

Observationally, overionization of the X-ray-emitting plasma associated with an SNR is manifested
by significant residuals that appear when spectra of the plasmas are fitted with CIE and NEI
models. For example, in the case of W44, \citet{Uchida12} found significant residuals at 
$\sim$1.5 keV and $\sim$ 2 keV, which are attributed to the Mg-Ly$\alpha$ line and the 
Si-Ly$\alpha$ line, respectively. In the cases of W44
\citep{Uchida12}, IC 443 \citep{Yamaguchi09} and G359.1-0.5
\citep{Ohnishi11}, residuals that emerged after fitting with CIE models are seen at 2.7 and 3.5
keV: lines at these energies are associated with H-like Si (2.67 keV) and H-like S (3.48 keV),
respectively. Finally, in the case of W49B, \citet{Ozawa09} identified residuals at the energies of
5.6, 6.2, 6.7, 6.9, 7.9, 8.5 keV: these lines are attributed to Cr
K$\alpha$, Mn$\alpha$ and four Fe lines (of He$\alpha_{rec}$, Ly$\alpha$,
He$\beta_{rec}$ and He$\gamma$), respectively.

\par
We have examined the X-ray spectra of the SNRs in our sample to search for evidence of 
overionized plasmas associated with these sources. Unfortunately, in the case of CTB 37A, the 
extracted {\it Chandra} spectra do not have sufficiently high signal-to-noise to reveal significant 
residuals (if present): therefore, there is no clear evidence that the X-ray emitting plasma associated 
with this SNR is overionized. Recently, \citet{Yamauchi13} analyzed {\it Suzaku} spectra of 
G346.6$-$0.2 and argued that the plasma associated with this SNR is overionized. Unfortunately the 
signal-to-noise of our {\it ASCA} spectra of G346.6$-$0.2 is insufficient for a study of the overionization 
of this SNR. Inspection of the {\it ASCA} spectra reveals a hint of residual dips at 1.5 keV (note the 
residuals between 1.5-2.5 keV as shown in Figure \ref{G346Spectra}). The residual at 1.5 keV (which 
is consistent with the Mg Ly$\alpha$ line) could be evidence of over-ionization as seen in W44 
\citep{Uchida12} but the scatter is too large to quantify the overionization. Lastly, the signal-to-noise of 
our {\it ASCA} spectra of G311.5$-$0.3 is too poor to identify any lines, while G348.5$-$0.0 was not 
clearly detected by any of the X-ray observations considered here.
\par
Examination of the MOS1+MOS2+PN spectra of Kes 17 (as shown in Figure 
\ref{Kes17_MOS1MOS2PN_SPECTRA}) reveals evidence for residuals at 1.5 keV (a positive dip)
and 1.9 keV (a negative dip). Based on the presence of these residuals, we have investigated 
whether or not the X-ray-emitting plasma associated with Kes 17 is
indeed overionized. We conducted a spectral analysis patterned after the analysis presented
by \citet{Yamauchi13} where those authors determined that the X-ray-emitting plasma associated
with G346.6$-$0.2 is overionized. Using the SPEX 
software package \citep{Kaastra96}, those authors fitted the extracted {\it Suzaku} 
XIS spectra of the SNR using the {\tt neij} model. As described by \citet{Yamauchi13}, this model 
describes an X-ray emitting plasma where initially the ion temperature $kT$$_Z$ and the original 
electron temperature $kT$$_{e1}$ were the same (that is, $kT$$_Z$ = $kT$$_{e1}$) and the 
plasma itself was in collisional ionization equilibrium. Over time, the electron temperature drops
to $kT$$_{e2}$ due to rapid electron cooling. Such a model can reflect accurately a rapid 
electron cooling as predicted in models of SNR evolution.
In addition, the {\tt neij} model was multiplied by the {\tt absm} model to account for interstellar
absorption (also consistent with the work of \citet{Yamauchi13}).
\par
We first used the SPEX (Version 2.03.03) tool {\tt trafo} 
(Version 1.02) to convert the extracted {\it XMM-Newton} spectral files for the whole SNR into 
{\tt .spo} files suitable for analysis by the software package. We were unable to generate a
statistically-acceptable fit with the {\tt neij} model that is superior (in other words, with an improved
value for $\chi$$^2$$_{\nu}$) to the fits obtained with the thermal models described above. 
The residuals of the fits obtained with the {\tt neij} model originate from inconsistencies between
the different datasets and because of these limitations the overionization model cannot improve
the fit. Thus, the poor signal-to-noise of the datasets prevent us from testing sufficiently whether
or not the X-ray-emitting plasma is overionized. A deeper X-ray observation is required to 
investigate more rigorously the possibility that Kes 17 belongs to the class of overionized MM SNRs.

\section{Comparison of Properties of the Sampled SNRs: Implications for SNR evolution}
\label{ComparisonSection}

In Table \ref{SNRXraysTable} we present a summary of the properties of the diffuse emission
associated with the four SNRs examined in this paper that were detected in the X-ray. We find 
that the X-ray properties of three of these SNRs -- Kes 17, G311.5$-$0.3 and CTB 37A -- are 
similar to those of such archetypal MM SNRs as 3C 391 and W44 \citep{Rho94, Rho96, Shelton04}.
Specifically, the X-ray temperatures of these three SNRs range from $kT$$\sim$0.7 keV to 
$kT$$\sim$1 keV, the abundances of the diffuse emission associated with each SNR are 
approximately solar and all of the SNRs are middle-aged (that is, the ages of all of the SNRs are
$t$$_{rad}$$\sim$10$^4$ yr old). Lastly, all three of these SNRs are known to produce infrared
emission that originates from shocked molecular hydrogen (that is, warm H$_2$): this indicates
that these SNRs are interacting with dense (10$^{4}$ - 10$^{5}$ cm$^{-3}$) molecular clouds
\citep{Hewitt09, Andersen11}. Therefore, we conclude these three particular SNRs in our sample
are similar to the archetypal MM SNRs listed above and others \citep[see][]{Rho96}.
The X-ray properties of G346.6$-$0.2 differ somewhat from both those of these three MM SNRs
as well as those of the archetypical MM SNRs. 
\par
While G346.6$-$0.2 does feature a
center-filled thermal X-ray morphology with a contrasting well-defined shell-like radio morphology
that typifies MM SNRs, the X-ray spectrum of this SNR is harder with a higher temperature
and the presence of a second harder spectral component is implied.
The morphology shows center-filled X-ray emission within a well-defined radio shell like those in 
other MM SNRs. However, the X-ray spectrum is harder than these other MM SNRs (with a higher 
temperature, $kT$$\sim$2.5-3.4 keV) and the fitted ionization timescale is rather lower. We do 
note however
that an age estimate based on the observed infrared line emission from this SNR -- recall that
the shock velocity used in our analysis is based on detected ionic infrared emission -- is similar to the 
ages estimated
for Kes 17, G311.5$-$0.3, CTB 37A and the archetypal MM SNRs mentioned above. We do not have
a clear idea why G346.6$-$0.2 shows harder X-ray emission, but perhaps it is somehow related
to a particular property of this SNR: the detected H$_2$ emission from G346.6$-$0.2 differs from 
the H$_2$ emission detected from the other MM SNRs in that the detected emission from
G346.6$-$0.2 requires a high velocity J-shock component or a high-velocity C-shock component
coupled with a slow-velocity C-shock component to be fit properly. In contrast, the emission 
detected from the other SNRs requires two slow C-shock components. It is possible that the X-ray
emission from G346.6$-$0.2 is associated with the forward shock of this SNR, which is necessarily
a J-shock because C-shocks do not shock-heat the plasmas of SNRs. In addition, or perhaps
as an alternate explanation, C-shocks are associated with lower shock velocities: this might 
indicate that G346.6$-$0.2 is dynamically a younger SNR. Therefore, this would be an indirect
connection between the shock and the X-ray emission seen from the interior of this SNR. More
X-ray observations of this intriguing SNR are clearly warranted. 
\par
Observations made with the {\it Fermi} telescope have detected many MM SNRs at $\gamma$-ray 
energies, including W28, W44, W49B, W51C and IC 443 \citep{Abdo09, Abdo10a, Abdo10b, Abdo10c, 
Abdo10d, Uchiyama10}, among others \citep[also see][]{Brandt12,Hewitt12}. The $\gamma$-ray 
emission detected from these sources may be explained as hadronic particle acceleration by the MM 
SNRs expanding into a dense environment. Of the five SNRs considered in this paper, two -- Kes 17 and 
CTB 37A -- have been detected at $\gamma$-ray energies: Kes 17 was detected by {\it Fermi}  
\citep{Wu11},
while CTB 37A was detected with very high energy (VHE) $\gamma$-ray telescopes 
\citep{Aharonian08} and {\it Fermi} \citep{Castro10}. Our present research has helped to confirm that
Kes 17 and CTB 37A are MM SNRs that feature center-filled X-ray emission. We argue that additional
$\gamma$-ray observations of MM SNRs are needed to investigate the inter-relationship between
X-ray and $\gamma$-ray emission from these particular sources and mechanisms of 
particle acceleration.
\par
We note that infrared observations of the five SNRs considered in this paper
have revealed that the infrared emission detected from these SNRs is radiative and produced
by shocks with velocities $V$$_s$$\sim$70 to 150 km s$^{-1}$. It appears that each of these SNRs 
is currently in the snowplow stage of evolution and the X-ray emission detected from Kes 17,
G311.5$-$0.3, G346.6$-$0.2 and CTB 37A is fossil radiation that was produced during an
earlier evolutionary stage of the SNRs. The X-ray emission from several MM SNRs -- as observed by
{\it Suzaku} -- often exhibits multiple ionization states which come from elemental differences in
ionization and recombination timescales \citep[e.g.][]{Sawada12}. The rarefaction of the
recombining plasma is likely due to continuous fast cooling of the plasma
due to radiative shocks encountering dense clouds (as inferred by the characteristics of the 
detected infrared emission from these SNRs). Our sample of four X-ray-detected MM SNRs in this
paper help confirm that a strong correlation exists between infrared 
shocked H$_2$ emitting SNRs and MM SNRs.
\par
We also note that {\it Chandra} observations of other MM SNRs -- such as W44 and
3C 391 -- have confirmed the presence of uniform temperature profiles in X-ray emission from these 
sources \citep{Chen04,Shelton04}. In addition, wide ranges of plasma conditions and metal
enrichment have also been revealed by {\it ASCA} and {\it Chandra} X-ray observations of MM 
SNRs, such 
as an over-ionized plasma in IC 443 \citep{Kawasaki02}, a suggested
abundance gradient at the bright projected center of W44 \citep{Shelton04}
and iron-rich ejecta in the MM SNRs 3C 397 and Sgr A East \citep{SafiHarb05,Maeda02}. Finally, 
evidence for X-ray emission from an ejecta-dominated plasma associated with IC 443
as detected by {\it XMM-Newton} was presented by \citep{Troja08}. Simulations have shown
that the center-filled thermal X-ray emission as observed in MM SNRs can be produced through 
anisotropic thermal conduction. Specifically, SNRs expanding into denser environments tend to be 
smaller, making it easier for thermal conduction to dictate large changes in the temperatures of
their expanding hot gas bubbles. Therefore, thermal conduction plays a prominent role in 
determining the observed X-ray morphology of MM SNRs \citep{Tilley06}. Magnetohydrodynamic
(MHD) modeling of SNRs expanding
through an inhomogeneous ISM also confirms that X-ray emission detected from MM SNRs
can be reproduced when both thermal conduction and the reverse shock of the SNR are included in
the modeling \citep{Orlando09}. After the reverse shock has reached the center of the SNR, a 
maximum
in the X-ray emission is seen toward the center of the SNR and the morphology is centrally brightened:
such a morphology is a defining characteristic of MM SNRs. Therefore, it is evident that thermal
conduction plays a crucial role in producing SNRs of this class.
\par
Follow-up X-ray observations of the MM SNRs in our sample are vital to investigate further their
properties. While these SNRs have already been observed with such current
X-ray observatories as {\it XMM-Newton}, {\it Chandra} and {\it Suzaku}, we argue that a deeper
observing campaign of these SNRs (as well as other known MM SNRs) is crucial and will permit 
detailed comparisons between these SNRs to be made. Analysis of the X-ray spectra obtained
from these observations will allow us to investigate the line emission of elements present in the
X-ray emitting plasma of these sources, specifically their ionization states and temperatures. Such
analysis may also reveal the presence of enhanced elemental abundances in the ejecta emission:
this phenomenon has been seen in other MM SNRs. Finally, $\gamma$-ray observations of the
MM SNRs in our sample which have not yet been detected at such high energies -- such as 
G311.5$-$0.3 and G346.6$-$0.2 -- as well as other MM SNRs are clearly warranted. The detection
of $\gamma$-ray emission from such sources will provide exciting 
opportunities to advance our understanding of particle acceleration by these sources.

\section{Conclusions}
\label{ConclusionSection}

The conclusions of this paper may be summarized as follows: 
\par
1) Using archival X-ray observations made with {\it ASCA}, {\it XMM-Newton} and {\it Chandra},
we have conducted a spectroscopic analysis of the X-ray properties of five SNRs (Kes 17, G311.5$-$0.3,
G346.6$-$0.2, CTB 37A and G348.5$-$0.0) detected in the
infrared by the {\it Spitzer} GLIMPSE survey. Four of the five SNRs (namely Kes 17, G311.5$-$0.3, 
G346.6$-$0.2 and CTB 37A) are clearly detected by these observations and we have placed an upper limit 
on the X-ray luminosity of G348.5$-$0.0). Our work represents the first published detection of X-ray
emission from G311.5$-$0.3.
\par
2) The four X-ray detected SNRs all feature a center-filled X-ray morphology with a contrasting
shell-like radio morphology: therefore, these sources may be classified as MM SNRs. The X-ray 
emission from each SNR appears to be thermal in origin: however, when the spectral properties of the 
X-ray-emitting plasmas of the SNRs are compared, significant variations are seen from one SNR to the
other. For example, the plasma associated with Kes 17 may be overabundant in magnesium, 
suggesting that the plasma may be ejecta-dominated. In addition, evidence exists for
the presence of hard components in the X-ray emission from G346.6$-$0.2.
We also note that we 
have identified a hard discrete X-ray source seen in projection toward CTB 37A (and distinct from
the extended emission with a hard spectrum seen toward the northwest rim of the SNR) and suggest
that it may be a neutron star associated with this SNR.
\par 
3) We have discussed the plasma conditions of the four X-ray detected SNRs and estimated such 
properties of the plasma as $n$$_e$, $t$$_{rad}$ and $M$$_X$: these values range from 
$\sim$0.4-0.8 cm$^{-3}$, $\sim$2.3-4.2$\times$10$^4$ yr and $\sim$13-88 $M$$_{\odot}$, 
respectively. These values are similar to those observed for other MM SNRs.
\par
4) We discuss the implications of our results for studies of SNR evolution. Because all four of the X-ray
detected SNRs feature lines produced by shocked $H$$_2$ in their infrared spectra, we argue that 
each of them is interacting with a nearby molecular cloud.  We believe that our results help strengthen 
the link between MM SNRs and interactions between SNRs and molecular clouds: this may help to 
explain the origin of the center-filled thermal X-ray morphologies of these sources.

\acknowledgments

We thank the anonymous referee for many useful comments that helped to improve greatly the 
quality of this paper. T.G.P. also thanks Koji Mukai for assistance with analyzing the {\it ASCA} datasets
considered in this paper and Nicholas Lee for assistance with analyzing the {\it Chandra} observation
of CTB 37A. Finally, T.G.P. also thanks Jelle de Plaa for assistance with using the SPEX
software package. This research has made use NASA's Astrophysics Data System as well as 
data obtained from the High Energy 
Astrophysics Science Archive Research Center (HEASARC), provided by 
NASA's Goddard Space Flight Center.

\clearpage

\begin{landscape}
\begin{deluxetable}{lccccccc}
\tablecaption{Properties of Supernova Remnants\label{SNRPropsTable}}
\tabletypesize{\scriptsize}
\label{snrpropertiestable}
 \tablewidth{0pt}
 \tablehead{&  \colhead{Kes 17 (G304.6+0.1)} & \colhead{G311.5-0.3} &
 \colhead{G346.6-0.2} & \colhead{CTB 37A (G348.5$+$0.1)} & \colhead{G348.5-0.0}  & \colhead{References}}
 \startdata
RA (J2000.0)            & 13 05 59 & 14 05 38 & 17 10 19 & 17 14 06 & 17 15 26 & (1)  \\
Dec (J2000.0)           & $-$62 42 & $-$61 58 & $-$40 11 & $-$38 32 & $-$38 28 & (1) \\
Angular Diameter (arcmin)    & 8 & 5  & 8  & 15$^a$ & 5$^b$ & (1)  \\
distance (kpc)     & $\ge$9.7 & 12.5  & 11 & 6.3-9.5$^c$ &  $\le$6.3  & (2), (3), (4), (5) \\
physical size (pc)              &$\ge$23   & 18 & 26 & 21 & $\le$10 & (6) \\
$N$$_H$ (10$^{22}$ cm$^{-2}$) & $\sim$3.9 & $\sim$3.0 & $\sim$1.7 & $\sim$3.4 & \nodata & (6)  \\
$A$$_V$ (mag)                      & 19.3     & 11.4  & 14.3     & 15       &  15     & (6) 
\enddata
\tablecomments{References: (1) \citet{Green09a,Green09b}, (2) \citet{Caswell75}, (3) 
\citet{Shaver70c}, (4) \citet{Koralesky98}, (5) \citet{Tian12}, (6) This paper.}
\tablenotetext{a}{This size estimate includes an apparent extension of radio emission seen toward
the southwest which resembles a "blow-out" region and may not necessarily be associated with this
SNR. In the present paper, we concentrate on the X-ray emission detected from the bright radio
shell seen toward the north which is approximately 8 arcminutes in diameter. X-ray emission is
clearly detected from this northern portion of the SNR but it is not clearly detected from the 
southwestern portion. See Section \ref{CTB37ADiscussSubSection}.}
\tablenotetext{b}{This is the approximate size of the bright concave radio rim that defines this
SNR.}
\tablenotetext{c}{These are the lower and upper bounds on the distance to CTB 37A based
on analysis of the HI absorption profile toward these source \citep{Tian12}. For this paper, we
adopt a mean distance of 8 kpc to this SNR.} 
\end{deluxetable}

\begin{deluxetable}{lcccccccccc}
\tabletypesize{\scriptsize}
\tablecaption{Summary of {\it ASCA} GIS and SIS Observations of the 
SNRs\tablenotemark{a}\label{ASCAObsSummaryTable}}
\tablewidth{0pt}
\tablehead{
& & & & & \colhead{GIS2} & & \colhead{GIS3} & & \colhead{SIS0}\\
& & & & & \colhead{Effective} & \colhead{GIS2} & \colhead{Effective} & \colhead{GIS2} &
\colhead{Effective} & \colhead{SIS0} \\
& & & \colhead{Right} && \colhead{Exposure} & \colhead{Count} & \colhead{Exposure} 
& \colhead{Count} & \colhead{Exposure} & \colhead{Count} \\
\colhead{Sequence} & & \colhead{Observation} & \colhead{Ascencion}  & \colhead{Declination}
& \colhead{Time} &  \colhead{Rate} & \colhead{Time} & \colhead{Rate}
& \colhead{Time}  & \colhead{Rate} \\
\colhead{Number} & \colhead{SNR} & \colhead{Date} & \colhead{(J2000.0)}
& \colhead{(J2000.0)} & \colhead{(s)} & \colhead{(counts s$^{-1}$)} & \colhead{(s)}
& \colhead{(counts s$^{-1}$)} & \colhead{(s)} & \colhead{(counts s$^{-1}$)} 
}
\startdata
57013000 & Kes 17 & 1999 February 12 & 13 06 44.2 & $-$62 40 47 & 24656 & 
6.5$\times$10$^{-2}$ & 24656 & 7.8$\times$10$^{-2}$ & 23328 & 5.4$\times$10$^{-2}$ \\
56047000 & G311.5$-$0.3 & 1998 March 2 & 14 06 18.7 & $-$61 53 02 & 17858
& 4.2$\times$10$^{-3}$ & 17870 & 6.2$\times$10$^{-3}$ & \nodata & \nodata \\
54004090 & G346.6$-$0.2 & 1996 September 3 & 17 09 34.3 & $-$39 59 38 & 8986
& 7.4$\times$10$^{-2}$ & 8986 & 6.3$\times$10$^{-2}$ 
& \nodata & \nodata
\enddata
\tablecomments{The units of Right Ascension are hours, minutes and seconds while the units of
Declination are degrees, arc minutes and arcseconds. Count rates are for the energy range 
0.7-10.0 keV.}
\tablenotetext{a}{No SIS1 data with sufficient signal-to-noise for detailed spectral analysis was 
available for any of these observations.}
\end{deluxetable}

\begin{deluxetable}{ccccccccccc}
\tabletypesize{\scriptsize}
\tablecaption{Summary of {\it XMM-Newton} MOS1, MOS2 and PN Observations of the 
SNRs\label{XMMObsSummaryTable}}
\tablewidth{0pt}
\tablehead{
& & & & & \colhead{MOS1} & & \colhead{MOS2} & & & \colhead{PN}\\
& & & & & \colhead{Effective} & \colhead{MOS1} & \colhead{Effective} & \colhead{MOS2} &
\colhead{PN} & \colhead{Effective}\\
& & & \colhead{Right} && \colhead{Exposure} & \colhead{Count} & \colhead{Exposure} 
& \colhead{Count} & \colhead{Exposure} & \colhead{Count} \\
\colhead{Sequence} & & \colhead{Observation} & \colhead{Ascencion}  & \colhead{Declination}
& \colhead{Time} &  \colhead{Rate} & \colhead{Time} & \colhead{Rate}
& \colhead{Time}  & \colhead{Rate} \\
\colhead{Number} & \colhead{SNR} & \colhead{Date} & \colhead{(J2000.0)} 
& \colhead{(J2000.0)} & \colhead{(s)} & \colhead{(counts s$^{-1}$)} & \colhead{(s)} & 
\colhead{(counts s$^{-1}$)} & \colhead{(s)} & \colhead{(counts s$^{-1}$)}
}
\startdata
0303100201 & Kes 17 & 2005 August 25 & 13 05 40.7 & $-$62 41 36 & 20176 & 
1.3$\times$10$^{-2}$ & 19895 & 1.1$\times$10$^{-2}$ & 16151 & 9.7$\times$10$^{-3}$  \\
0306510101 & CTB 37A & 2006 March 1 & 17 14 41.0 & $-$38 30 53 & 3090 & 
7.7$\times$10$^{-2}$ & 2857 & 6.5$\times$10$^{-2}$ & 939 & 1.1$\times$10$^{-3}$ \\
0306510101 & G348.5$-$0.0 & 2006 March 1 & 17 14 41.0 & $-$38 30 53 & 3090 &
$<$2.2$\times$10$^{-2}$ & 2857 & $<$9.1$\times$10$^{-3}$ & 939 &
$<$3.3$\times$10$^{-2}$
\enddata
\tablecomments{The units of Right Ascension are hours, minutes and seconds while the
units of Declination are degrees, arcminutes and arcseconds. Count rates are for the
energy range 0.6-10.0 keV.} 
\end{deluxetable}
\clearpage
\end{landscape}

\begin{landscape}
\begin{deluxetable}{ccccccc}
\tabletypesize{\scriptsize}
\tablecaption{Summary of {\it Chandra} ACIS-I Observation of CTB 
37A\label{ChandraObsTable}}
\tablewidth{0pt}
\tablehead{
& & & & & \colhead{ACIS-I} & \colhead{ACIS-I} \\
\colhead{Sequence} & & \colhead{Observation} & \colhead{R.A.} & \colhead{Decl.} & 
\colhead{Effective Exposure} & \colhead{Count} \\
\colhead{Number} & \colhead{ObsID} & \colhead{Date} & \colhead{(J2000.0)} & 
\colhead{(J2000.0)} & \colhead{Time (s)} & \colhead{Rate (s)}
}
\startdata
500668 & 6721 & 2006 October 7 & 17 14 35.80 & $-$38 31 25 & 19656 & 3.0$\times$10$^{-2}$ 
\enddata
\tablecomments{The units of Right Ascension are hours, minutes and seconds while the
units of Declination are degrees, arcminutes and arcseconds. Count rates are for the
energy range 0.3-10.0 keV.} 
\end{deluxetable}

\begin{deluxetable}{lccccccccc}
\tabletypesize{\scriptsize}
\tablewidth{0pt}
\tablecaption{Summary of Fits to {\it ASCA} GIS and SIS Spectra of Kes 17
\label{GISSISSpectraSummaryKes17}}
\tablehead{& & & & & & &  \colhead{Absorbed} & \colhead{Unabsorbed} 
& \colhead{Unabsorbed} \\
& \colhead{$N$$_H$} & \colhead{$kT$} & \colhead{$\tau$} & & & \colhead{$\chi$$_{\nu}^2$} 
& \colhead{Flux\tablenotemark{b}} & \colhead{Flux\tablenotemark{b}} & 
\colhead{Luminosity\tablenotemark{b}} \\
\colhead{Model} & \colhead{(10$^{22}$ cm$^{-2}$)} & \colhead{(keV)} & 
\colhead{(10$^{12}$ cm$^{-3}$ s)} & \colhead{Normalization\tablenotemark{a}} & 
\colhead{Abundance} & 
\colhead{($\chi$$^2$/DOF)} & \colhead{(ergs cm$^{-2}$ s$^{-1}$)} & 
\colhead{(ergs cm$^{-2}$ s$^{-1}$)} & \colhead{(ergs s$^{-1}$)} 
}
\startdata
PHABS$\times$APEC & 3.48$^{+0.46}_{-0.30}$ & 0.70$^{+0.08}_{-0.11}$ & 
\nodata & 3.58$\times$10$^{-2}$ & 1.0 (frozen) & 1.18 (355.08/302) & 
2.00$\times$10$^{-12}$ & 7.96$\times$10$^{-11}$ & 8.96$\times$10$^{35}$  \\
PHABS$\times$APEC & 3.56$^{+0.48}_{-0.38}$ & 0.63$^{+0.09}_{-0.12}$ & 
\nodata & 1.30$\times$10$^{-1}$ & 0.20$^{+0.24}_{-0.12}$ & 1.12 (338.07/301) &
2.01$\times$10$^{-12}$ & 7.09$\times$10$^{-11}$ & 7.98$\times$10$^{35}$  \\
PHABS$\times$NEI & 3.72$^{+0.38}_{-0.37}$ & 0.64$^{+0.10}_{-0.08}$ & 9.60 ($>$0.5) &
5.34$\times$10$^{-2}$ & 1.0 (frozen) & 1.20 (359.74/301) & 1.96$\times$10$^{-12}$ 
& 1.15$\times$10$^{-10}$ & 1.29$\times$10$^{36}$ \\
PHABS$\times$NEI & 3.58$^{+0.42}_{-0.38}$ & 0.63$\pm$0.08 & 9.63 ($>$0.02) & 
1.26$\times$10$^{-1}$ & 0.22$^{+0.22}_{-0.13}$ & 1.13 (340.38/300) & 2.00$\times$10$^{-12}$
& 7.53$\times$10$^{-11}$ & 8.48$\times$10$^{35}$ \\
PHABS$\times$VAPEC\tablenotemark{c} & 4.22$^{+1.28}_{-1.02}$ & 0.60$^{+0.15}_{-0.14}$ 
& \nodata & 6.41$\times$10$^{-2}$ & Mg=2.38$^{+4.32}_{-1.58}$ & 1.17 (352.69/301) &
1.89$\times$10$^{-12}$ & 1.43$\times$10$^{-10}$ & 1.61$\times$10$^{36}$  \\
PHABS$\times$VAPEC\tablenotemark{c} & 4.27$^{+0.53}_{-0.52}$ & 
0.56$^{+0.10}_{-0.11}$ & \nodata & 1.13$\times$10$^{-1}$ & Si=0.47$^{+0.27}_{-0.19}$, & 
1.11 (334.12/300) & 1.99$\times$10$^{-12}$ & 2.32$\times$10$^{-10}$ & 2.61$\times$10$^{36}$ 
\\
& & & & & S=0.21($<$0.56)
\enddata
\tablecomments{All quoted error bounds correspond to the 90\% confidence level.}
\tablenotetext{a}{In the cases of the PHABS$\times$APEC, PHABS$\times$NEI and 
PHABS$\times$VAPEC models, the normalization is defined in units
of (10$^{-14}$/4$\pi$$d$$^2$)$\times$$\int$$n$$_e$$n$$_p$~$dV$, where $d$ is the 
distance to the source (in units of cm), $n$$_e$ and $n$$_p$ are the number densities 
of electrons
and hydrogen nuclei, respectively (in units of cm$^{-3}$) and finally $\int$~$dV$ is the
integral over the entire volume of the X-ray-emitting plasma (in units of cm$^{-3}$).}
\tablenotetext{b}{For all the considered energy range is 0.7 to 10.0 keV. All luminosities
have been calculated assuming a distance of 9.7 kpc to Kes 17.}
\tablenotetext{c}{In the first VAPEC fit listed here, the abundance of magnesium was allowed
to vary while the abundances of the other elements were frozen to solar values. In the 
second VAPEC fit listed here, the abundances of silicon and sulfur were allowed to vary while
the abundances of the other elements were frozen to solar values.}
\end{deluxetable}
\clearpage
\end{landscape}

\begin{landscape}
\begin{deluxetable}{lccccccccc}
\tabletypesize{\scriptsize}
\tablewidth{0pt}
\tablecaption{Summary of Fits to {\it XMM-Newton} MOS1, MOS2 and PN Spectra of
Diffuse X-ray Emission from Kes 17\tablenotemark{a}\label{XMMSpectraSummaryKes17}}
\tablehead{& & & & & & &  \colhead{Absorbed} & \colhead{Unabsorbed} 
& \colhead{Unabsorbed} \\
& \colhead{$N$$_H$} & \colhead{$kT$} & \colhead{$\tau$} & & & \colhead{$\chi$$_{\nu}^2$} 
& \colhead{Flux\tablenotemark{b}} & \colhead{Flux\tablenotemark{b}} & 
\colhead{Luminosity\tablenotemark{b}} \\
\colhead{Model} & \colhead{(10$^{22}$ cm$^{-2}$)} & \colhead{(keV)}
& \colhead{(10$^{11}$ cm$^{-3}$ s)} & \colhead{Normalization\tablenotemark{a}} & 
\colhead{Abundance} & 
\colhead{($\chi$$^2$/DOF)} & \colhead{(ergs cm$^{-2}$ s$^{-1}$)} & 
\colhead{(ergs cm$^{-2}$ s$^{-1}$)} & \colhead{(ergs s$^{-1}$)} 
}
\startdata
PHABS$\times$APEC & 3.15$^{+0.39}_{-0.18}$ & 0.92$^{+0.09}_{-0.15}$ & \nodata &
1.67$\times$10$^{-2}$ & 1.0 (frozen) & 1.10 (308.77/281) & 1.81$\times$10$^{-12}$  & 
3.35$\times$10$^{-11}$ & 3.77$\times$10$^{35}$ \\
PHABS$\times$APEC &  3.29$^{+0.33}_{-0.51}$ & 0.82$^{+0.17}_{-0.10}$ & \nodata & 
3.17$\times$10$^{-2}$ & 0.53$^{+0.77}_{-0.23}$ & 1.09 (306.48/280) &
1.76$\times$10$^{-12}$ & 3.95$\times$10$^{-11}$ & 4.45$\times$10$^{35}$\\
PHABS$\times$NEI &  3.14$^{+0.34}_{-0.38}$ & 1.10$^{+0.26}_{-0.20}$ & 
3.70$^{+6.30}_{-1.70}$ &  1.32$\times$10$^{-2}$ & 1.0 (frozen) & 1.08 (303.10/286) &
1.95$\times$10$^{-12}$ & 2.56$\times$10$^{-11}$ & 2.88$\times$10$^{35}$ \\
PHABS$\times$VAPEC\tablenotemark{c} & 3.69$^{+0.61}_{-0.65}$ & 
0.79$^{+0.17}_{-0.09}$ & \nodata &
2.31$\times$10$^{-2}$ & Mg=2.75$^{+2.20}_{-1.70}$ & 1.08 (303.11/280) & 
1.77$\times$10$^{-12}$ & 5.28$\times$10$^{-11}$ & 5.95$\times$10$^{35}$ \\
PHABS$\times$VAPEC\tablenotemark{c} & 3.47$^{+0.37}_{-0.55}$ & 
0.80$^{+0.25}_{-0.10}$ & \nodata &
2.56$\times$10$^{-2}$ & Si=0.79$^{+0.51}_{-0.27}$, & 1.08 (314.50/291) & 
1.77$\times$10$^{-12}$ & 5.43$\times$10$^{-11}$ & 6.11$\times$10$^{35}$\\
& & & & & S=0.71$^{+0.49}_{-0.29}$\\
PHABS$\times$VAPEC\tablenotemark{c} & 3.46$^{+0.33}_{-0.62}$  & 0.90$^{+0.08}_{-0.10}$ 
& \nodata & 1.82$\times$10$^{-2}$ & Mg=2.76$^{+1.22}_{-1.04}$, & 1.09 (302.70/278) & 
1.81$\times$10$^{-12}$ & 5.34$\times$10$^{-11}$ & 6.02$\times$10$^{35}$\\
& & & & & Si=1.08$^{+0.21}_{-0.18}$, \\
& & & & & S=0.88$^{+0.12}_{-0.24}$ \\
PHABS$\times$(VAPEC & 3.23$^{+0.46}_{-0.58}$ & 0.84$^{+0.10}_{-0.12}$ & \nodata & 
3.01$\times$10$^{-3}$, & Mg=9.49($>$1.65), & 1.07 (292.50/274) & 2.05$\times$10$^{-12}$ & 
5.40$\times$10$^{-11}$ & 6.09$\times$10$^{35}$ \\
+ POWER LAW)\tablenotemark{c} & & $\Gamma$=3.98$^{+0.24}_{-1.63}$ & & 
8.01$\times$10$^{-3}$ & Si=5.17($>$0.85), \\
& & & & & S=4.88($>$0.44) \\
\hline
Reference Fits\tablenotemark{d,e}\\
\hline
WABS$\times$(PSHOCK & 3.09 (frozen) & 0.75 (frozen) & 21 (frozen) &
1.92$\times$10$^{-2}$, & 0.8 (frozen) & 1.21 (338.16/280) & 2.23$\times$10$^{-12}$ &
3.83$\times$10$^{-11}$ & 4.31$\times$10$^{35}$ \\
+ POWER LAW)\tablenotemark{d} & & $\Gamma$=2.4 (frozen) & & 9.12$\times$10$^{-4}$  \\
WABS$\times$(VMEKAL & 3.77 (frozen) & 0.78 (frozen) & \nodata & 2.66$\times$10$^{-2}$, &
Mg = 1.1 (frozen), & 1.11 (312.68/252) &
1.93$\times$10$^{-12}$ & 5.77$\times$10$^{-11}$ & 6.50$\times$10$^{35}$ \\
+ POWER LAW)\tablenotemark{e} & & $\Gamma$=1.5 (frozen) & & 5.32$\times$10$^{-5}$ & 
Si = 0.97 (frozen), \\
& & & & & S = 0.50 (frozen)\\

WABS$\times$(PSHOCK & 3.67$^{+0.52}_{-0.55}$ & 0.75$^{+0.20}_{-0.12}$ & 1667 ($>$10) & 
3.05$\times$10$^{-2}$, & 0.61$^{+0.75}_{-0.44}$ & 1.11 (305.83/275) & 1.88$\times$10$^{-12}$
& 4.62$\times$10$^{-11}$ & 5.76$\times$10$^{35}$ \\
+ POWER LAW)\tablenotemark{d} & & $\Gamma$=1.58$^{+2.24}_{-0.25}$ & & 
1.17$\times$10$^{-4}$  \\

WABS$\times$(VMEKAL & 3.73$^{+0.63}_{-0.54}$ & 0.79$^{+0.22}_{-0.16}$ & \nodata & 
1.33$\times$10$^{-2}$, & Mg=4.96($>$1.34), & 1.08 (298.93/276) & 1.92$\times$10$^{-12}$ & 
5.61$\times$10$^{-11}$ & 6.12$\times$10$^{35}$ \\
+ POWER LAW)\tablenotemark{e} & & $\Gamma$=4.16$^{+1.22}_{-2.12}$ & & 
8.08$\times$10$^{-3}$ & Si=2.18 ($>$0.64), \\
& & & & & S=1.95 ($>$0.32) \\

\enddata
\tablecomments{All quoted error bounds correspond to the 90\% confidence level.}
\tablenotetext{a}{In the cases of the PHABS$\times$APEC, PHABS$\times$NEI,
PHABS$\times$VAPEC, PHABS$\times$(VPSHOCK+Power Law) and 
PHABS$\times$(VMEKAL+Power Law) models, the normalization of the thermal component
is defined in units
of (10$^{-14}$/4$\pi$$d$$^2$)$\times$$\int$$n$$_e$$n$$_p$~$dV$, where $d$ is the 
distance to the source (in units of cm), $n$$_e$ and $n$$_p$ are the number densities 
of electrons
and hydrogen nuclei, respectively (in units of cm$^{-3}$) and finally $\int$~$dV$ is the
integral over the entire volume of the X-ray-emitting plasma (in units of cm$^{-3}$). Where
a power law is present, the normalization is defined in the units of photons keV$^{-1}$ cm$^{-2}$
s$^{-1}$ at 1 keV.}
\tablenotetext{b}{For all the considered energy range is 0.7 to 10.0 keV. All luminosities
have been calculated assuming a distance of 9.7 kpc to Kes 17.}
\tablenotetext{c}{In the first VAPEC fit listed here, the abundance of magnesium was allowed
to vary while the abundances of the other elements were frozen to solar values. In the 
second VAPEC fit listed here, the abundances of silicon and sulfur were allowed to vary while
the abundances of the other elements were frozen to solar values. In the third VAPEC fit
listed here, the abundances of magnesium, silicon and sulfur were allowed to vary while the
abundances of the other elements were frozen to solar values. In the VAPEC$\times$POWER LAW
fit listed here, the abundances of magnesium, silicon and sulfur were allowed to vary while the
abundances of the other elements were frozen to solar values.}
\tablenotetext{d}{Fit presented by \citet{Combi10} in the analysis of 
MOS1+MOS2+PN spectra of the SNR. Specific fit values for
each parameter given here correspond to means for parameter values obtained for fits to three
different regions of the SNR. The value of the photon index $\Gamma$ of the power law component 
is given: the second listed normalization corresponds to the normalization of this component.} 
\tablenotetext{e}{Fit presented by \citet{Gok12} in the analysis of {\it Suzaku} XIS spectra of the
SNR. Specific fit values for each parameter given here correspond to means for parameter values
obtained for fits to three different regions of the SNR. The value of the photon index $\Gamma$ of 
the power law component 
is given: the second listed normalization corresponds to the normalization of this component.}
\end{deluxetable}

\begin{deluxetable}{lcccccccc}
\tabletypesize{\scriptsize}
\tablewidth{0pt}
\tablecaption{Summary of Fits to {\it ASCA} GIS Spectra of 
G311.5$-$0.3\label{GISSpectraSummaryG311}}
\tablehead{& & & & & &  \colhead{Absorbed} & \colhead{Unabsorbed} 
& \colhead{Unabsorbed} \\
& \colhead{$N$$_H$} & \colhead{$kT$} & & & \colhead{$\chi$$_{\nu}^2$} 
& \colhead{Flux\tablenotemark{b}} & \colhead{Flux\tablenotemark{b}} & 
\colhead{Luminosity\tablenotemark{b}} \\
\colhead{Model} & \colhead{(10$^{22}$ cm$^{-2}$)} & \colhead{(keV)} & \colhead{$\Gamma$}
& \colhead{Normalization\tablenotemark{a}} & 
\colhead{($\chi$$^2$/DOF)} & \colhead{(ergs cm$^{-2}$ s$^{-1}$)} & 
\colhead{(ergs cm$^{-2}$ s$^{-1}$)} & \colhead{(ergs s$^{-1}$)} 
}
\startdata
PHABS$\times$Power Law & 3.31$^{+6.09}_{-1.91}$ & \nodata & 4.82$^{+3.78}_{-2.32}$ 
& 4.49$\times$10$^{-3}$ & 0.79 (5.56/7) & 2.25$\times$10$^{-13}$ & 
6.95$\times$10$^{-12}$ & 1.30$\times$10$^{35}$\\ 
PHABS$\times$APEC\tablenotemark{c} & 2.69$^{+1.51}_{-1.59}$ & 
0.98$^{+1.02}_{-0.22}$ & \nodata & 1.58$\times$10$^{-3}$ & 0.57 (3.97/7) & 
2.23$\times$10$^{-13}$ & 3.00$\times$10$^{-12}$ & 5.61$\times$10$^{34}$ 
\enddata
\tablecomments{All quoted error bounds correspond to the 90\% confidence level.}
\tablenotetext{a}{In the case of the PHABS$\times$Power Law model, the normalization is
defined in the units of photons keV$^{-1}$ cm$^{-2}$ s$^{-1}$ at 1 keV. In the case of the
PHABS$\times$APEC model, the normalization is defined in units
of (10$^{-14}$/4$\pi$$d$$^2$)$\times$$\int$$n$$_e$$n$$_p$~$dV$, where $d$ is the 
distance to the source (in units of cm), $n$$_e$ and $n$$_p$ are the number densities 
of electrons
and hydrogen nuclei, respectively (in units of cm$^{-3}$) and finally $\int$~$dV$ is the
integral over the entire volume of the X-ray-emitting plasma (in units of cm$^{-3}$).}
\tablenotetext{b}{Measured over the energy range of 0.7 to 5.0 keV.}
\tablenotetext{c}{Abundance parameter frozen to solar values.}
\end{deluxetable}
\clearpage
\end{landscape}

\clearpage
\begin{landscape}
\begin{deluxetable}{lcccccccccc}
\tabletypesize{\scriptsize}
\tablewidth{0pt}
\tablecaption{Summary of Fits to {\it ASCA} GIS Spectra of G346.6$-$0.2
\label{GISSpectraSummaryG346}}
\tablehead{& & & & & & & &  \colhead{Absorbed} & \colhead{Unabsorbed} 
& \colhead{Unabsorbed} \\
& \colhead{$N$$_H$} & \colhead{$kT$} & & \colhead{$\tau$} & & & \colhead{$\chi$$_{\nu}^2$} 
& \colhead{Flux\tablenotemark{b}} & \colhead{Flux\tablenotemark{b}} & 
\colhead{Luminosity\tablenotemark{b}} \\
\colhead{Model} & \colhead{(10$^{22}$ cm$^{-2}$)} & \colhead{(keV)} & \colhead{$\Gamma$}
& \colhead{(10$^9$ cm$^{-3}$ s)} & \colhead{Normalization\tablenotemark{a}} & 
\colhead{Abundance} & 
\colhead{($\chi$$^2$/DOF)} & \colhead{(ergs cm$^{-2}$ s$^{-1}$)} & 
\colhead{(ergs cm$^{-2}$ s$^{-1}$)} & \colhead{(ergs s$^{-1}$)} 
}
\startdata
PHABS$\times$Power Law & 1.54$^{+0.50}_{-0.40}$ & \nodata & 2.54$^{+0.40}_{-0.34}$ 
& \nodata & 7.68$\times$10$^{-3}$ & \nodata & 1.34 (66.78/50) & 9.07$\times$10$^{-12}$
& 2.35$\times$10$^{-10}$ & 3.40$\times$10$^{36}$  \\
PHABS$\times$APEC & 1.39$^{+0.33}_{-0.31}$ & 2.49$^{+0.51}_{-0.39}$ & \nodata &
\nodata & 1.34$\times$10$^{-2}$ & 1.0 (frozen) & 1.45 (72.25/50) & 8.43$\times$10$^{-12}$
& 1.80$\times$10$^{-10}$ & 2.61$\times$10$^{36}$\\
PHABS$\times$APEC & 1.08$^{+0.34}_{-0.30}$ & 3.44$^{+1.30}_{-0.84}$ & \nodata & \nodata 
& 1.46$\times$10$^{-2}$ & $<$0.4 & 1.22 (59.88/49) & 8.68$\times$10$^{-12}$ & 
1.52$\times$10$^{-10}$ & 2.20$\times$10$^{36}$ \\
PHABS$\times$NEI & 2.13$^{+0.40}_{-0.68}$ & 2.80$^{+0.80}_{-0.50}$ & 
\nodata & 7.27$^{+6.23}_{-3.77}$ & 1.77$\times$10$^{-2}$ & 1.0 (frozen) & 
1.07 (52.52/49) & 8.66$\times$10$^{-12}$ & 1.77$\times$10$^{-10}$ & 
2.56$\times$10$^{36}$ \\
PHABS$\times$NEI & 1.97$^{+0.55}_{-0.77}$ & 2.82$^{+1.14}_{-0.54}$ & \nodata &
9.84$^{+46}_{-6.44}$ & 1.82$\times$10$^{-2}$ & 0.49$^{+1.11}_{-0.45}$
& 1.07 (51.38/48) & 8.62$\times$10$^{-12}$ & 1.06$\times$10$^{-10}$ &
1.53$\times$10$^{36}$ \\
PHABS$\times$(NEI+Power Law) & 1.77$^{+0.40}_{-0.66}$ & 1.57$^{+0.92}_{-0.40}$ & 
0.5 (frozen) & 80 (frozen)  & 1.81$\times$10$^{-2}$,
& 0.25$^{+1.10}_{-0.10}$ & 1.17 (63.06/54) & 1.43$\times$10$^{-11}$ & 3.59$\times$10$^{-11}$ & 
5.20$\times$10$^{35}$ \\
& & & & & 3.43$\times$10$^{-4}$\\
PHABS$\times$(NEI+Power Law) & 1.76$^{+0.52}_{-0.70}$ & 0.97$^{+1.05}_{-0.37}$ & 2.0 (frozen) 
& 80 (frozen) & 2.42$\times$10$^{-3}$,
& 2.67 ($>$0.56) & 1.16 (62.37/54) & 1.10$\times$10$^{-11}$ & 4.46$\times$10$^{-11}$ & 
6.46$\times$10$^{35}$ \\
& & & & & 4.13$\times$10$^{-3}$ \\
PHABS$\times$(NEI+Power Law) & 1.74$^{+0.48}_{-0.68}$ & 1.50$^{+1.00}_{-0.58}$ & 
0.5 (frozen) & 100 (frozen) & 
1.88$\times$10$^{-2}$, & 0.23$^{+1.20}_{-0.05}$ & 1.19 (64.21/54) & 1.45$\times$10$^{-11}$ 
& 3.37$\times$10$^{-11}$ & 4.88$\times$10$^{35}$ \\
& & & & & 3.57$\times$10$^{-4}$\\
PHABS$\times$(NEI+Power Law) & 1.79$^{+0.56}_{-0.72}$ & 0.87$^{+0.98}_{-0.33}$ & 
2.0 (frozen) & 100 (frozen) & 
5.46$\times$10$^{-3}$, & 1.32 ($>$0.42) & 1.16 (62.82/54) & 1.09$\times$10$^{-11}$ & 
4.50$\times$10$^{-11}$ & 6.52$\times$10$^{35}$ \\
& & & & & 4.09$\times$10$^{-3}$ \\
\hline
Reference Fits\tablenotemark{c,d,e}\\ 
\hline
WABS$\times$(VMEKAL & 2.0 (frozen) & 0.97 (frozen) & 0.5 (frozen) & \nodata & 
2.29$\times$10$^{-2}$, & Varied\tablenotemark{c} & 1.16 (65.02/56) 
& 1.63$\times$10$^{-11}$ & 4.91$\times$10$^{-11}$ & 7.11$\times$10$^{35}$ \\
+Power Law)\tablenotemark{c} & & & & & 4.5$\times$10$^{-4}$ \\
WABS$\times$(VNEI & 2.1 (frozen) & 1.22 (frozen) & 0.5 (frozen) & 292 (frozen) & 
2.29$\times$10$^{-2}$, & Varied\tablenotemark{d} & 1.16 (65.05/56) 
& 1.63$\times$10$^{-11}$ & 4.91$\times$10$^{-11}$ & 7.11$\times$10$^{35}$ \\
+Power Law)\tablenotemark{d} & & & & & 4.5$\times$10$^{-4}$ \\
WABS$\times$(VPSHOCK & 2.1 (frozen) & 1.3 (frozen) & 0.6 (frozen) & 692 (frozen) & 
2.29$\times$10$^{-2}$, & Varied\tablenotemark{e} & 1.15 (64.50/56) 
& 1.63$\times$10$^{-11}$ & 4.91$\times$10$^{-11}$ & 7.11$\times$10$^{35}$ \\
+Power Law)\tablenotemark{e} & & & & & 4.5$\times$10$^{-4}$ \\
\enddata
\tablecomments{All quoted error bounds correspond to the 90\% confidence level.}
\tablenotetext{a}{In the case of the PHABS$\times$Power Law model, the normalization is
defined in the units of photons keV$^{-1}$ cm$^{-2}$ s$^{-1}$ at 1 keV. In the cases of the
PHABS$\times$APEC and PHABS$\times$NEI models, the normalization is defined in units
of (10$^{-14}$/4$\pi$$d$$^2$)$\times$$\int$$n$$_e$$n$$_p$~$dV$, where $d$ is the 
distance to the source (in units of cm), $n$$_e$ and $n$$_p$ are the number densities 
of electrons
and hydrogen nuclei, respectively (in units of cm$^{-3}$) and finally $\int$~$dV$ is the
integral over the entire volume of the X-ray-emitting plasma (in units of cm$^{-3}$). In the cases
of the PHABS$\times$(NEI+Power Law) model and the reference fits, the first listed normalization 
is for the NEI 
component and the second listed normalization is for the Power Law component: these
normalizations are defined as above.}
\tablenotetext{b}{For all the considered energy range is 0.6 to 10.0 keV. All luminosities
have been calculated assuming a distance of 11 kpc to G346.6$-$0.2.}
\tablenotetext{c}{These fit parameters are from the fit obtained by \citet{Sezer11b} to extracted
{\it Suzaku} spectra of this SNR. The fitted variable elemental abundances in that fit are as  
follows: Mg = 0.65, Si = 0.44, S = 0.70, Ca = 3.4 and Fe = 0.4. All of these elemental abundances
(as well as the other fit parameters except for the normalizations) 
were frozen as we applied these spectral fit parameters to the extracted GIS data.}
\tablenotetext{d}{These fit parameters are from the fit obtained by \citet{Sezer11b} to extracted
{\it Suzaku} spectra of this SNR. The fitted variable elemental abundances in that fit are as  
follows: Mg = 0.51, Si = 0.47, S = 0.70, Ca = 2.3 and Fe = 0.6. All of these elemental abundances
(as well as the other fit parameters except for the normalizations) 
were frozen as we applied these spectral fit parameters to the extracted GIS data.}
\tablenotetext{e}{These fit parameters are from the fit obtained by \citet{Sezer11b} to extracted
{\it Suzaku} spectra of this SNR. The fitted variable elemental abundances in that fit are as  
follows: Mg = 0.76, Si = 0.56, S = 0.8, Ca = 2.3 and Fe = 0.6. Here, the given value for 
$\tau$ corresponds to the upper limit on the ionization timescale as derived by those authors.
All of these elemental abundances
(as well as the other fit parameters except for the normalizations) 
were frozen as we applied these spectral fit parameters to the extracted GIS data.}

\end{deluxetable}

\begin{deluxetable}{lccccccccc}
\tabletypesize{\scriptsize}
\tablewidth{0pt}
\tablecaption{Summary of Fits to {\it XMM-Newton} MOS1, MOS2 and PN Spectra of
Diffuse X-ray Emission from CTB 37A\tablenotemark{a}\label{XMMSpectraSummaryCTB37A}}
\tablehead{& & & & & & &  \colhead{Absorbed} & \colhead{Unabsorbed} 
& \colhead{Unabsorbed} \\
& \colhead{$N$$_H$} & \colhead{$kT$} & \colhead{$\tau$} & & & \colhead{$\chi$$_{\nu}^2$} 
& \colhead{Flux\tablenotemark{c}} & \colhead{Flux\tablenotemark{c}} & 
\colhead{Luminosity\tablenotemark{c}} \\
\colhead{Model} & \colhead{(10$^{22}$ cm$^{-2}$)} & \colhead{(keV)}
& \colhead{(10$^{11}$ cm$^{-3}$ s)} & \colhead{Normalization\tablenotemark{b}} & 
\colhead{Abundance} & 
\colhead{($\chi$$^2$/DOF)} & \colhead{(ergs cm$^{-2}$ s$^{-1}$)} & 
\colhead{(ergs cm$^{-2}$ s$^{-1}$)} & \colhead{(ergs s$^{-1}$)} 
}
\startdata
PHABS$\times$APEC & 3.15$^{+0.75}_{-0.40}$ & 0.73$^{+0.15}_{-0.16}$ & \nodata &
3.37$\times$10$^{-2}$ & 1.0 (frozen) & 1.11 (207.36/192) & 2.32$\times$10$^{-12}$ &
7.43$\times$10$^{-11}$ & 5.69$\times$10$^{35}$ \\
PHABS$\times$APEC & 3.26$^{+0.66}_{-0.56}$ & 0.64$^{+0.18}_{-0.14}$ & \nodata &
1.03$\times$10$^{-1}$ & 0.29$^{+0.43}_{-0.15}$ & 1.08 (200.04/186) & 2.28$\times$10$^{-12}$ 
& 7.73$\times$10$^{-11}$ & 5.92$\times$10$^{35}$ \\
PHABS$\times$NEI & 3.53$^{+0.80}_{-0.58}$ & 0.62$^{+0.21}_{-0.16}$ & 357 ($>$3) & 
6.29$\times$10$^{-2}$ & 1.0 (frozen) & 1.11 (207.13/186) & 2.24$\times$10$^{-12}$ &
1.35$\times$10$^{-10}$ & 1.03$\times$10$^{36}$  \\
PHABS$\times$NEI & 3.40$\pm$0.60 & 0.64$^{+0.16}_{-0.14}$ & 6.73 ($>$0.52) & 
1.05$\times$10$^{-1}$ & 0.35$^{+0.50}_{-0.20}$ & 1.09 (201.19/185) & 2.29$\times$10$^{-12}$
& 9.37$\times$10$^{-11}$ & 7.18$\times$10$^{35}$ \\
PHABS$\times$VAPEC\tablenotemark{d} & 3.71$^{+0.71}_{-0.56}$ & 0.55$^{+0.16}_{-0.11}$ 
& \nodata & 9.29$\times$10$^{-2}$ & Si=0.48$^{+0.26}_{-0.23}$ & 1.05 (198.98/186) &
2.24$\times$10$^{-12}$ & 1.91$\times$10$^{-10}$ & 1.46$\times$10$^{36}$ \\
PHABS$\times$VNEI\tablenotemark{d} & 3.69$^{+0.61}_{-0.49}$ & 0.57$^{+0.13}_{-0.11}$ &
43 ($>$1) & 9.36$\times$10$^{-2}$ & Si=0.55$^{+0.30}_{-0.27}$ & 1.07 (198.82/185) &
2.25$\times$10$^{-12}$ & 1.95$\times$10$^{-10}$ & 1.49$\times$10$^{36}$ \\
\hline
Reference Fits\tablenotemark{e} &&&\\
\hline
PHABS$\times$(VMEKAL & 2.90 (frozen) & 0.63 (frozen) & \nodata & 3.14$\times$10$^{-2}$, &
1.0 (frozen) & 1.13 (214.80/190) & 3.20$\times$10$^{-12}$ & 7.41$\times$10$^{-11}$ & 
5.68$\times$10$^{35}$ \\
+Power Law)\tablenotemark{e} & & and $\Gamma$=1.6 (frozen) & & 3.40$\times$10$^{-4}$\\
\enddata
\tablecomments{All quoted error bounds correspond to the 90\% confidence level.}
\tablenotetext{a}{X-ray emission from the known hard X-ray source has been excluded
when spectra have been extracted.}
\tablenotetext{b}{In the cases of the PHABS$\times$APEC, PHABS$\times$NEI, 
PHABS$\times$VAPEC, PHABS$\times$VNEI and PHABS$\times$(VMEKAL+Power Law)
models, the normalization of the thermal component is defined in units
of (10$^{-14}$/4$\pi$$d$$^2$)$\times$$\int$$n$$_e$$n$$_p$~$dV$, where $d$ is the 
distance to the source (in units of cm), $n$$_e$ and $n$$_p$ are the number densities 
of electrons
and hydrogen nuclei, respectively (in units of cm$^{-3}$) and finally $\int$~$dV$ is the
integral over the entire volume of the X-ray-emitting plasma (in units of cm$^{-3}$). In the case
of the PHABS$\times$(VMEKAL+Power Law) model, the normalization of the power law component
is defined in units of photons keV$^{-1}$ cm$^{-2}$ s$^{-1}$ at 1 keV.}
\tablenotetext{c}{For the considered energy range 0.7 to 10.0 keV. All luminosities have
been calculated assuming a distance of 8 kpc to CTB 37A.}
\tablenotetext{d}{For these fits, the abundances for all elements were frozen to solar except for
silicon, which was allowed to vary.}
\tablenotetext{e}{These fit parameters are from the fit obtained by \citet{Sezer11a} to extracted
{\it Suzaku} spectra of this SNR. All elemental abundances were frozen to solar values and 
all of the fit parameters (except for the normalizations)  
were frozen as we applied these spectral fit parameters to the extracted GIS data.}
\end{deluxetable}
\clearpage
\end{landscape}

\begin{landscape}
\begin{deluxetable}{lcccccccccc}
\tabletypesize{\scriptsize}
\tablewidth{0pt}
\tablecaption{Summary of Fits to {\it Chandra} ACIS-I Spectra of CTB 37A}
\tablehead{& & & & & & & &  \colhead{Absorbed} & \colhead{Unabsorbed} 
& \colhead{Unabsorbed} \\
& \colhead{$N$$_H$} & \colhead{$kT$} & & \colhead{$\tau$} & & & \colhead{$\chi$$_{\nu}^2$} 
& \colhead{Flux\tablenotemark{b}} & \colhead{Flux\tablenotemark{b}} & 
\colhead{Luminosity\tablenotemark{b}} \\
\colhead{Model} & \colhead{(10$^{22}$ cm$^{-2}$)} & \colhead{(keV)}
& \colhead{$\Gamma$}
& \colhead{(10$^{11}$ cm$^{-3}$ s)} & \colhead{Normalization\tablenotemark{a}} & 
\colhead{Abundance} & 
\colhead{($\chi$$^2$/DOF)} & \colhead{(ergs cm$^{-2}$ s$^{-1}$)} & 
\colhead{(ergs cm$^{-2}$ s$^{-1}$)} & \colhead{(ergs s$^{-1}$)} 
}
\startdata
\multicolumn{11}{c}{CXOU J171419.8$-$383023} \\
\hline
PHABS$\times$Power Law & 4.08$^{+1.32}_{-0.88}$ & \nodata & 1.18$^{+0.42}_{-0.33}$ 
& \nodata & 3.97$\times$10$^{-4}$ & \nodata & 0.84 (91.97/109) & 2.99$\times$10$^{-12}$ & 
4.49$\times$10$^{-12}$ & 3.43$\times$10$^{34}$ \\
\hline
\multicolumn{11}{c}{Northeast Region} \\ 
\hline
PHABS$\times$APEC & 3.07$^{+0.23}_{-0.27}$ & 0.80$^{+0.13}_{-0.08}$ & \nodata
& \nodata & 1.31$\times$10$^{-2}$ & 1.0 (frozen) & 1.09 (144.42/132)  & 1.12$\times$10$^{-12}$
& 1.88$\times$10$^{-11}$ & 1.44$\times$10$^{35}$ \\ 
PHABS$\times$APEC & 3.03$\pm$0.25 & 0.78$^{+0.11}_{-0.08}$ & \nodata & \nodata
& 2.30$\times$10$^{-2}$ & 0.45$^{+0.45}_{-0.10}$ & 1.06 (138.81/131) & 
1.13$\times$10$^{-12}$ & 1.64$\times$10$^{-11}$ & 1.26$\times$10$^{35}$ \\
PHABS$\times$NEI & 3.30$^{+0.24}_{-0.28}$ & 0.83$^{+0.20}_{-0.14}$ & \nodata &
4.15 ($>$2.55) & 1.53$\times$10$^{-2}$ & 1.0 (frozen) & 1.07 (140.38/131) &
1.16$\times$10$^{-12}$ & 2.42$\times$10$^{-11}$ & 1.85$\times$10$^{35}$ \\
PHABS$\times$NEI & 3.24$^{+0.40}_{-0.24}$ & 0.80$^{+0.13}_{-0.10}$ & \nodata &
2.44 ($>$0.25) & 2.63$\times$10$^{-2}$ & 0.43$^{+0.37}_{-0.15}$ & 1.02 (132.56/130) & 
1.16$\times$10$^{-12}$ & 2.14$\times$10$^{-11}$ & 1.64$\times$10$^{35}$\\
\hline
\multicolumn{11}{c}{Southeast Region} \\
\hline
PHABS$\times$APEC & 3.03$^{+0.51}_{-0.33}$ & 0.57$\pm$0.12 & \nodata 
& \nodata & 1.79$\times$10$^{-2}$ & 1.0 (frozen) & 0.95 (95.32/100) & 7.11$\times$10$^{-13}$
& 1.55$\times$10$^{-11}$ & 1.19$\times$10$^{35}$ \\
PHABS$\times$APEC & 3.02$^{+0.52}_{-0.34}$ & 0.57$\pm$0.12 & \nodata 
& \nodata & 2.30$\times$10$^{-2}$ & 0.73$^{+5.49}_{-0.43}$ & 0.96 (95.02/99) & 
7.12$\times$10$^{-13}$ & 1.51$\times$10$^{-11}$ & 1.16$\times$10$^{35}$ \\
PHABS$\times$NEI & 3.31$^{+0.43}_{-0.36}$ & 0.55$^{+0.11}_{-0.06}$ & \nodata
& 4.23($>$0.86) & 2.76$\times$10$^{-2}$ & 1.0 (frozen) & 0.94 (93.08/99) & 
7.36$\times$10$^{-13}$ & 2.51$\times$10$^{-11}$ & 1.92$\times$10$^{35}$ \\
PHABS$\times$NEI & 3.32$^{+0.36}_{-0.30}$ & 0.55$\pm$0.13 & \nodata & 
3.34($>$0.52) & 2.79$\times$10$^{-2}$ & 0.86 ($>$0.34) & 0.95 (93.04/98) &
7.37$\times$10$^{-13}$ & 2.55$\times$10$^{-11}$ & 1.95$\times$10$^{35}$ 
\enddata
\tablecomments{All quoted error bounds correspond to the 90\% confidence level.}
\tablenotetext{a}{In the case of the PHABS$\times$Power Law model, the normalization is
defined in the units of photons keV$^{-1}$ cm$^{-2}$ s$^{-1}$ at 1 keV. In the cases of the
PHABS$\times$APEC and PHABS$\times$NEI models, the normalization is defined in units
of (10$^{-14}$/4$\pi$$d$$^2$)$\times$$\int$$n$$_e$$n$$_p$~$dV$, where $d$ is the 
distance to the source (in units of cm), $n$$_e$ and $n$$_p$ are the number densities 
of electrons
and hydrogen nuclei, respectively (in units of cm$^{-3}$) and finally $\int$~$dV$ is the
integral over the entire volume of the X-ray-emitting plasma (in units of cm$^{-3}$).}
\tablenotetext{b}{For CXOU J171319.8$-$383023, the northeast region and the southeast
region, the considered energy ranges are 0.8 to 10.0 keV, 0.9 to 5.0 keV and 0.9 to 4.0 
keV, respectively. All luminosities have been calculated assuming a distance of 8 kpc to
CTB 37A.}
\label{ACISSpectraSummaryCTB37A}
\end{deluxetable}
\clearpage
\end{landscape}

\begin{deluxetable}{lcccc}
\tabletypesize{\scriptsize}
\tablecaption{Comparison of Derived X-ray Properties with Previously-Published 
Results\label{ComparisonThreeSNRsTable}}
\tablewidth{0pt}
\tablehead{ & \colhead{Properties} & \colhead{Properties} \\
\colhead{SNR} & \colhead{(This Paper)} & \colhead{(Other Papers)} & \colhead{References} 
& \colhead{Notes}}
\startdata
Kes 17\tablenotemark{a} (G304.6$+$0.1) & $n$$_e$ = 0.42 cm$^{-3}$ & $n$$_e$ = 0.9-2.3 cm$^{-3}$ & (1) 
& (1) calculated age from $t$ = $\tau$/$n$$_e$.\\
& $\tau$ $\sim$ 10$^{12}$ s cm$^{-3}$ & $\tau$ = 1-3$\times$10$^{12}$ s cm$^{-3}$\\ 
& $t$$_{rad}$ = 2.3$\times$10$^4$ yr  & $t$ = 2.8-6.4$\times$10$^4$ yr \\
& $M$$_X$ = 52 $M$$_{\odot}$ & $M$$_X$ = 8-15 $M$$_{\odot}$ \\
& & $n$$_e$$\sim$2.3 $f$$^{-1/2}$ & (2) & (2) assumed distance of 10 kpc \\
& & $\tau$$\sim$10$^{12}$ s cm$^{-3}$ & & and adopted $f$ as filling factor. \\ 
& & $t$$\sim$1.4$\times$10$^4$$f$$^{1/2}$ yr \\
& & $M$$_X$$\sim$380$f$$^{1/2}$ $M$$_{\odot}$ \\
G346.6$-$0.2\tablenotemark{b} & $n$$_e$ = 0.24 cm$^{-3}$ & $n$$_e$ = 0.82 cm$^{-3}$ & (3) 
& (3) assumed distance of 8.3 kpc \\
& $\tau$ $\sim$ 10$^{11}$ s cm$^{-3}$ & 
$\tau$ = 2.92$\times$10$^{11}$ s cm$^{-3}$ & & and calculated age from $t$=$\tau$/$n$$_e$.\\ 
& $t$$_{rad}$ = 4.7$\times$10$^4$ yr  & $t$ = 1.1$\times$10$^4$ yr \\
& $M$$_X$ = 84 $M$$_{\odot}$ & & & \\
CTB 37A\tablenotemark{c} (G348.5$+$0.1) & $n$$_e$ = 0.77 cm$^{-3}$ & $n$$_e$ = 1 cm$^{-3}$ 
& (4) 
& (3) assumed distance of 11.3 kpc \\
& $\tau$ $\geq$10$^{12}$ s cm$^{-3}$ & $\tau$ = $\geq$10$^{12}$ s cm$^{-3}$ & & and a 
filling factor 
$f$=1. Age  \\
&  $t$$_{rad}$ = 3.2-4.2$\times$10$^4$ yr  & $t$ = 3$\times$10$^4$ yr & & calculated from 
$t$=$\tau$/$n$$_e$.\\
& $M$$_X$=76 $M$$_{\odot}$ & $M$$_X$ = 530 $M$$_{\odot}$ \\
\enddata
\tablecomments{References: (1) -- \citet{Combi10}, (2) -- \citet{Gok12}, (3) -- \citet{Sezer11b}, 
(4) -- \citet{Sezer11a}.}
\tablenotetext{a}{Properties calculated based on fits to extracted {\it XMM-Newton} MOS1+MOS2+PN
spectra obtained using PHABS$\times$VAPEC model (with Mg or Si and S thawed). The ionization
timescale quoted here for comparison purposes was obtained from a PHABS$\times$NEI fit to the
same spectra (see Table \ref{GISSISSpectraSummaryKes17}): this resulting fit suggests that the
X-ray emitting plasma associated with Kes 17 is close to ionization equilibrium.}
\tablenotetext{b}{Properties calculated based on fits to extracted {\it ASCA} GIS2+GIS3
spectra obtained using PHABS$\times$NEI model.}
\tablenotetext{c}{Properties calculated based on fits to extracted {\it XMM-Newton} MOS1+MOS2+PN
spectra obtained using PHABS$\times$VAPEC model (with Mg or Si and S thawed). The estimate of the 
X-ray-emitting mass associated with CTB 37A that we provide here should be treated 
as a lower limit: see Section \ref{CTB37ADiscussSubSection}.} 
\end{deluxetable}


\begin{deluxetable}{lcccc}
\tablecaption{Properties of the Diffuse Emission from X-ray-Detected SNRs in Present Sample
\tablenotemark{a}}
\tabletypesize{\scriptsize}
\setlength{\tabcolsep}{0.03in}
\tablewidth{0pt}
\tablehead{& \colhead{Kes 17 (G304.6+0.1)} & \colhead{G311.5-0.3} &
\colhead{G346.6-0.2} & \colhead{CTB 37A (G348.5$+$0.1)}}
\startdata
Radius  (pc) & $\ge$11 & 9 & 13 & 10 \\
$V$$_s$ (km s$^{-1}$, radiative)\tablenotemark{b}  & 150 & 100, 170 & 70-100 & 75-100 \\
$t$$_{rad}$ (10$^4$ yr) & 2.3 & 2.5-4.2 & 3.3-4.7 & 3.2-4.2 \\
X-ray Mechanism  & Thermal  & Likely Thermal & Thermal & Thermal \\
$kT$ (keV) &  0.79-1.10 & 0.98 & 2.49-3.44 & 0.55-0.83 \\
$\tau$  (cm$^{-3}$ s)  & 3.7$\times$10$^{11}$ & $\sim$$>$10$^{12}$ & 7.3$\times$10$^{9}$ 
&$>$10$^{12}$ \\ 
$n$$_e$  (cm$^{-3}$)   & 0.42 & $>$0.40 & 0.25 & 0.77 \\ 
$M$$_X$  ($M$$_{\odot}$)  & 52 & 13 & 88 & 76\tablenotemark{c} \\
Morphology  & Center-Filled & Center-Filled & Center-Filled & Center-Filled\\
SNR classification & MM SNR  & MM SNR & MM SNR & MM SNR \\
Evidence of Molecular Cloud interaction\tablenotemark{b} & Yes (H$_2$) & Yes (H$_2$)  &  Yes (H$_2$) & 
Yes (H$_2$) \\
$\gamma$-ray detection & Yes & No & No & Yes
\enddata
\tablenotetext{a}{For the sake of completeness, we note that the fifth SNR considered in this
paper (G348.5$-$0.0) was not detected in the X-ray but -- similarly to the four SNRs listed in this table
that were detached in the X-ray -- evidence exists that G348.5$-$0.0 is interacting with a molecular
cloud.} 
\tablenotetext{b}{See \citet{Hewitt09} and \citet{Andersen11}.}
\tablenotetext{c}{This estimate of the X-ray-emitting mass associated with CTB 37A should be treated 
as a lower limit: see Section \ref{CTB37ADiscussSubSection}.} 
\label{SNRXraysTable}
\end{deluxetable}

\clearpage

\clearpage

\newpage

\clearpage

\end{document}